\documentclass[%
 reprint,
 amsmath,amssymb,
 aps,
floatfix,
]{revtex4-2}

\usepackage{graphicx}% Include figure files
\usepackage{dcolumn}% Align table columns on decimal point
\usepackage{bm}% bold math
\usepackage{chngcntr}
\usepackage{hyperref}
\usepackage{xcolor}
\usepackage{float}
\usepackage{cancel}
\usepackage{comment}
\usepackage{array}
\colorlet{importantColor}{red}

\begin{document}

\setlength{\abovedisplayskip}{4pt plus 0pt minus 0pt}
\setlength{\belowdisplayskip}{4pt plus 0pt minus 0pt}
\setlength{\abovedisplayshortskip}{0pt}
\setlength{\belowdisplayshortskip}{0pt}

\setlength{\lineskiplimit}{-3pt}
\setlength{\lineskip}{2pt plus 0.5pt minus 0.5pt}

\preprint{APS/123-QED}

\title {\textbf {Cell-type-specific synaptic motifs generate coupled mean-variability dynamics in recurrent neural networks} 
}% 

\author{Meiyi Zhang}%
\affiliation{%
 Center for Quantitative Biology, Academy for Advanced Interdisciplinary Studies, Peking University, Beijing, China
}
\author{Jinjian Yu}%
\affiliation{%
 School of Physics, Peking University, Beijing, China
}%
\author{Louis Tao}
 \affiliation{
 Center for Bioinformatics, National Laboratory of Protein Engineering and Plant Genetic Engineering, School of Life Sciences
 and
 Center for Quantitative Biology, Academy for Advanced Interdisciplinary Studies,Peking University, Beijing, China
}
\author{Yuxiu Shao}
\thanks{Yuxiu.SHAO@univ-cotedazur.fr}
\affiliation{CNRS, LJAD and Institute NeuroMod, Université Côte d’Azur, Nice, France
}

\date{\today}% It is always \today, today,
             %  but any date may be explicitly specified
\begin{abstract}
Synaptic-resolution network connectomics has revealed that brain circuits feature fine-scale structural connectivity, such as pairs of correlated synaptic couplings known as second-order motifs. Large-scale recordings of neuronal activity in networks containing nonlinear neurons reveal macroscopic heterogeneous population dynamics throughout the brain. 
These findings rekindle the inquiry into this intriguing question: Can microscale synaptic structures contribute to macroscopic heterogeneous dynamics and computations in ways that canonical brain circuit models cannot?
To answer this question, we construct random RNNs with various cell types, nonlinear non-negative neural responses, and arbitrary marginal and second-order correlated synaptic statistics. We derive low-rank mean-field equations for \(P\)-population networks in which the pre- and postsynaptic neuronal population identities determine the synaptic and motif strengths. Our framework requires \(2P\) latent dynamic variables with \(P\) variables describing mean population activity and \(P\) variables capturing within-population variability. Theoretical and numerical results demonstrate that chain motifs induce correlations in synaptic variability, coupling within-population variability to population mean dynamics through nonlinear gain.  We use this framework to construct network models constrained by experimentally observed motif statistics that reproduce key population activity patterns in the mouse primary visual cortex. By explicitly linking synaptic organization to coupled mean-variability dynamics, our results provide a testable framework for studying the relationship between fine-scale connectivity, heterogeneous dynamics, and task-related responses.  
% \begin{description}
% % \item[Usage]
% % Secondary publications and information retrieval purposes.
% % \item[Structure]
% % You may use the \texttt{description} environment to structure your abstract;
% % use the optional argument of the \verb+\item+ command to give the category of each item. 
% \end{description}
\end{abstract}

%Use showkeys class option if keyword
                              % display desired
\maketitle

\section{\label{sec:Intro}Introduction}
\begin{figure*}[t]
\centering
\includegraphics[width=0.9\linewidth]{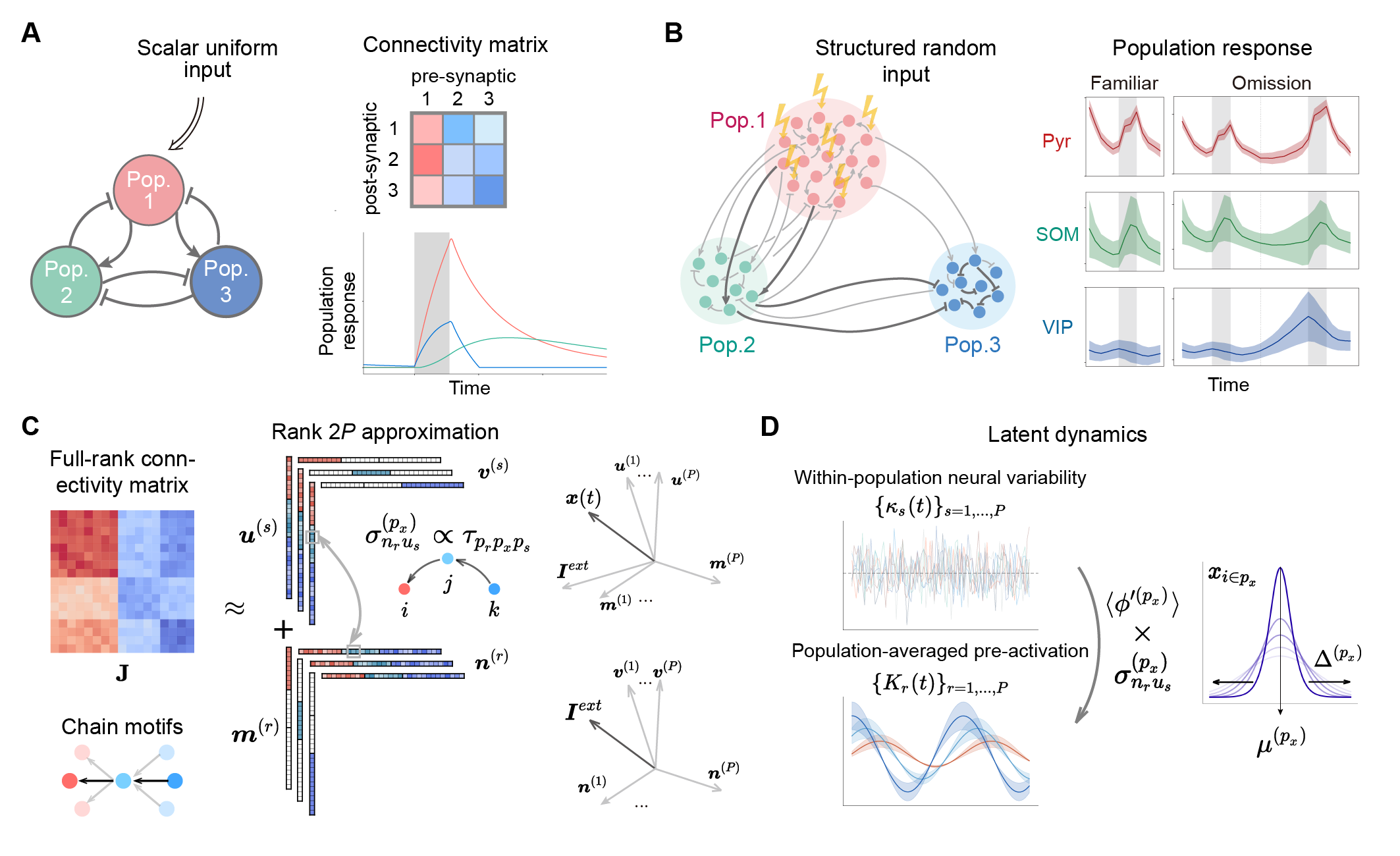}
\caption{{\bf Schematics of classical neural models and low-rank mean-field networks.} (A) Schematic of the multi-population neural circuit model. Each population is treated as a single computational unit with emphasis on the average connectivity between populations. 
(B) Schematic of a multiscale neural network model incorporating local connectivity patterns within populations and between-population structures. Externally structured inputs interact with recurrent connectivity to produce heterogeneous neural population responses.
(C) The low-rank approximation (Rank \(2P\)) of the full-rank connectivity matrix $\mathbf{J}$ with \(P\) populations and chain motifs. The connectivity vectors $\boldsymbol{u}^{(1)}...\boldsymbol{u}^{(P)}$ correspond to the $P$ incoming random patterns, and the connectivity vectors $\boldsymbol{n}^{(1)}... \boldsymbol{n}^{(P)}$ correspond to the $P$ outgoing random patterns. The strength of the second-order chain motifs $\tau_{p_rp_xp_s}$ reflects the population-specific correlation coefficient between these two sets of patterns $\boldsymbol{u}^{(s)}$ and $\boldsymbol{n}^{(r)}$ on population $p_x$. 
Neural activity \(\boldsymbol{x}\) is constrained in the space spanned by the left connectivity vectors and the input vector. External inputs interact with recurrent connectivity to shape the resulting dynamics.
(D) The $P$ latent quantities $\{\kappa_r\}_{r=1...P}$ along directions $\boldsymbol{u}^{(r)}$ reflect the within-population neural variability, and the $P$ latent quantities $\{K_r\}_{r=1...P}$ along directions $\boldsymbol{m}^{(r)}$ reflect the population-averaged pre-activations.
Microscopic variability is scaled up to influence population-level heterogeneity via the correlated recurrent connectivity.
}
\label{fig:fig_schematic}
\end{figure*}
Synaptic-level couplings in cortical circuits are neither independent nor identically distributed. Cellular- and synaptic-resolution connectomics studies have shown nonrandom local wiring patterns, including second-order motifs formed by pairs of correlated synapses: reciprocal, convergent, divergent, and chain motifs\cite{campagnola2022local,perin2011synaptic,gal2017rich,rieubland2014structured,reimann2024specific,song2005highly,dahmen_strong_2026}. 
Chain motifs, which involve triplets of neurons, are distinctive because they correlate the incoming and outgoing connectivities associated with an intermediate neuron, and their statistics are heterogeneously distributed across cell types \cite{rieubland2014structured,reimann2024specific}.
Functionally, advances in large-scale neural recordings have enabled researchers to characterize network dynamics using data analysis and other computational approaches \cite{steinmetz2018challenges,steinmetz2021neuropixels,trautmann2025large,international2025brain,schneider2023learnable,gosztolai2025marble}. Recent studies have revealed how the cerebral cortex operates across different dynamical regimes, and have identified computational principles underlying information processing through coordinated activity and communication across neuron populations, cortical circuits and brain areas \cite{panichello2021shared,dahmen2025heterogeneity,meissner2025computational,luo2025transitions}.
Yet it remains unclear whether and how cell-type-specific synaptic motifs contribute to this population-level heterogeneity beyond what can be explained by average connectivity between neuronal classes.

Random recurrent neural networks provide a natural framework for relating connectivity statistics to collective dynamics to functional computations.  Networks with independently and identically distributed~(i.i.d.) Gaussian connectivity have been used to characterize fixed points, bifurcations, chaos, learning, and memory \cite{sompolinsky1988chaos,kadmon2015transition,brunel2000dynamics,stubenrauch2025fixed,van2021large}.
Because of their statistical tractability in the presence of nonlinearity, related Gaussian random network models have also become important in theoretical machine learning, where they have been used to study learning, memory, and optimization \cite{clark2024theory,clark2025transient,kusmierz2020edge,liu2024connectivity,mastrovito2024transition}.
% clark2025connectivity,
Advanced extensions incorporating symmetric connectivity disorder, corresponding to reciprocal motifs, have shown they, even by themselves, can substantially enrich nonlinear network dynamics relative to the independent case \cite{marti2018correlations,clark2023dimension,clark2025connectivity}. 
However, a thorough treatment is still lacking for the other second-order motif types.
In parallel, large-scale recordings in freely behaving animals \cite{steinmetz2021neuropixels,hong2019novel,steinmetz2018challenges} have revealed low-dimensional activity manifolds in neuronal populations\cite{perich2025neural,langdon2023unifying,gallego2017neural}, while low-rank recurrent networks have provided an interpretable framework for generating the observed population dynamics and implementing task-related computations, such as decision-making, working memory, and related cognitive functions \cite{mastrogiuseppe2018linking,beiran2021shaping,pals2024trained,valente2022extracting,dubreuil2022role,beiran2023parametric,pellegrino2023low}.
In these low-rank models, global correlations among connectivity vectors, whether predefined or learned, determine the latent dynamical modes\cite{dubreuil2022role,barbosa2023early}. What remains unresolved is whether such global structures can arise from cell-type-specific local connectivity statistics, particularly chain motifs, and how these motif-generated structures interact with neuronal and network nonlinearities.

Previous work showed that chain motifs can generate isolated eigenmodes and change linear stability and steady-state responses \cite{shao2025impact}. These results established a link between local synaptic correlations and global network dynamics but did not examine how motif-induced structures interact with neuronal nonlinearities or how within-population heterogeneity participates in setting up nonlinear network states. Here we extend this line of inquiry to biologically motivated, non-negative nonlinearities and a rich repertoire of nonlinear dynamics that emerge beyond the linear regime. Addressing these questions requires both theoretical and corresponding computational advances, retaining not only the cell-type-specific population structures but also the structured neuron-to-neuron heterogeneity generated by synaptic motifs.

Here we develop a biologically grounded low-rank mean-field framework building upon the classical Gaussian-mixture low-rank framework. 
On the one hand, we introduce a novel low-rank approximation model that accommodates recurrent connectivity with arbitrary incoming- and outgoing-degree distributions, manifesting as heterogeneous second-order chain-motif correlations across cell types, the resulting rank structure thus reflects multiple sources of cell-type-specific heterogeneity. Unlike mean connectivity approximation models \cite{hertag2020learning,palmigiano2020common,murphy2009balanced}, where rank is solely determined by the number of coarse-graining population blocks, our approximation includes additional rank components arising from correlated incoming- and outgoing-degrees \cite{nykamp2017mean}, which capture microscopic connectivity heterogeneity of individual units as well~(Fig.~\ref{fig:fig_schematic}A-C).

On the other hand, this low-rank approximation gives rise to low-dimensional nonlinear dynamics that carry neural computations. The latent quantities associated with individual ranks naturally separate into two sets: one capturing dynamical heterogeneity across populations, and the other reflecting microscopic variability among individual rate units. 
Beyond providing these biologically interpretable latent quantities, our framework yields mathematical equations linking the two sets of quantities, articulating a principle of information processing whereby neuronal-level variability influences, and is in turn shaped by, coarse-grained population-level mean activity~(Fig.~\ref{fig:fig_schematic}D). 
Theoretical and numerical examinations and dynamical systems analyses of simplified Excitatory-Inhibitory (EI) neural networks reveal distinct, identifiable autonomous attractor states and rich network dynamics in response to structured input.
Finally, relating these dynamical phenotypes to chain-motif statistics from the mouse visual cortex connectome, we construct a motif-informed multipopulation network model that successfully recapitulates the empirically observed dynamical heterogeneity across populations and stimulus conditions.

The paper is organized as follows. In Sec.~\ref{sec:Method}, we provide a brief recap of two recurrent models used here. In Sec.~\ref{sec:relationship_between_motifs_and_lowrank}, we derive the mathematical relationship between circuit models and the corresponding low-rank models.
In Secs.~\ref{sec:toy-examples} and \ref{secs:exp-application}, we use the framework to examine the comprehensive pathway from local circuits to global structures and the emergent multiscale dynamics, drawing on both simplified theoretical models and experiment-based datasets.

\vspace{-8pt}
\section{\label{sec:Method}Classic models}
\subsection{\label{sec:NeuralCircuit}Neural circuit model}
\vspace{-8pt}
We consider a circuit model of \(N\) recurrently connected neurons. 
To capture the heterogeneity observed in brain connectivity, we partition these neurons into \(P\) nonoverlapping populations, \(p_1\ldots p_P\), satisfying \(\sum_{x=1}^P N_{p_x}=N\).
Recurrent connectivity is represented by a weight matrix \(\mathbf{J}\), where an element \(J_{ij}\) denotes the synaptic weight from a presynaptic neuron \(j\) to a postsynaptic neuron \(i\). 
We assume that synaptic weights between neurons from the same pair of populations share identical statistics. Consequently, the network connectivity matrix \(\mathbf{J}\) exhibits a block structure (\(P^2\) blocks), with each block corresponding to connections between a specific pair of pre- and postsynaptic neuron populations.
This structure emphasizes cross-population heterogeneity, which is the conventional explanatory framework used in circuit models.

A random weight matrix can always be decomposed as 
\begin{equation}\label{eq:circuit_decomposition_main}
    \mathbf{J}=\mathbf{J^0}+\mathbf{Z},
\end{equation} 
where \(\mathbf{J}^0\) is the mean connectivity and \(\mathbf{Z}\) is a zero-mean random matrix \cite{thibeault2024low}. 
Among the properties of \(\mathbf{J}\), we focus on first- and second-order statistics: marginal statistics -- the mean \(J_{pq}^0\) and variance \(\sigma_{pq}^2\), for population \(p\) and \(q\), and correlation coefficients
between synaptic weights.
Correlation coefficients are defined for pairs of synapses sharing a common neuron (two common neurons in reciprocal motifs), forming structures known as second-order motifs. 
We primarily examine chain motifs, in which an intermediate neuron~\(j\) (in population~\(q\)) receives input from neuron~\(k\) (in population~\(s\)) and projects to neuron~\(i\) (in population~\(p\)), where \(i \neq j \neq k\). The strength of chain motifs is quantified by the correlation coefficient \(\tau_{pqs}^{\text{chain}}=\tau_{ijk}^{\text{chain}}\) between the two synaptic weights.
Importantly, the marginal statistics form low-rank block \textit{matrices}, while the pairwise correlation coefficients constitute a low-rank \textit{tensor}.

We describe neural dynamics using a rate model with positive, nonlinear activation. The temporal evolution of the pre-activation variable, $x_i$, of neuron~\(i\) is described by
\begin{equation}\label{eq:rate_neural_network_main}
\dot{x}_i(t) = -x_i(t) + \sum_{q}^{p_1\ldots p_P}\sum_{j\in q} J_{ij}\phi(x_j(t)) + I_i^{ext}u(t),
\end{equation}
where \(\phi(x)=b+\tanh{(x-\theta)}\) is the nonlinear activation function and \(I_i^{ext}u(t)\) represents the external input to neuron \(i\).
The network's response to external input perturbations is characterized using linear response theory \cite{miller2020generalized,chau2025exact,hu2018feedback,tsodyks1997paradoxical}. 
Consider an input perturbation with a spatial pattern \(\boldsymbol{h}\) (normalized such that \(\Vert\boldsymbol{h}\Vert^2=1\)) and a magnitude \(\delta_h\). This perturbation induces a deviation in the pre-activation of neuron~\(i\) from its equilibrium as \(\delta x_i\). The response function of neuron~\(i\) is \(\chi_{ih}^x = \delta x_i/\delta_h\).
When \(\boldsymbol{h}\) applies only to the population \(q\) and exhibits no randomness or only i.i.d. randomness, the population-level averaged response is calculated as \(\langle\chi_{ih}^x\rangle_{i\in p}\), representing the mean response of neurons in the population \(p\) to a uniform input to neurons in population \(q\), as \(\chi_{pq}^x\) \cite{shao2025impact}. Details for the circuit model are provided in Appendix~\ref{app:appendix_neuralcircuit}.

\subsection{\label{sec:lrRNN_model}Low-rank recurrent neural network model}
Recurrent connectivity \(\mathbf{J}\) denotes, in general, a matrix of full rank \(N\). 
However, real world networks, including brain networks, gene regulatory networks, social networks, and others, often exhibit approximately low-rank structures.
In neuroscience, low-rank network connectivity can emerge from the organization of neurons into distinct cell types (i.e. in \(\mathbf{J^0}\)) or functional ensembles with shared computational roles. 
The low-rank connectivity matrix takes the following form
\begin{equation}\label{eq:gen_synapse_lowrank_main}
\mathbf{J} = \frac{1}{N}\sum_{r=1}^R\boldsymbol{m}^{(r)}\boldsymbol{n}^{(r)} = \frac{1}{N}\mathbf{M}\mathbf{N}^{\intercal},
\end{equation}
where rank \(R\ll N\), \(\boldsymbol{m}^{(r)}\) and \(\boldsymbol{n}^{(r)}\) are the \(r^{\rm th}\) columns of matrices \(\mathbf{M}\) and \(\mathbf{N}\in \mathbb{R}^{N\times R}\). 
We similarly characterize the first- and second-order statistics of connectivity vectors for population \(p\): marginal statistics including the mean \(a^{\smash{(p)}}_{\smash{x_r}}\) and variance \(\sigma_{\smash{x_r^2}}^{(p)}\) for \(x\in\{m,~n\}\), and pairwise covariance \(\sigma\smash{_{n_r m_s}^{(p)}}\), where \(r,s=1\ldots R\).
We then substitute this low-rank decomposition into Eq.~\eqref{eq:rate_neural_network_main} to examine corresponding network dynamics.

Under the low-rank structure, population neural activity evolves within the subspace spanned by columns of \(\mathbf{M}\) and the orthogonalized input vector \(\boldsymbol{I}^{ext}_{\perp}\), expressed as
\begin{equation}\label{eq:lowdim_subspace_main}
\boldsymbol{x}(t) = \sum_{r=1}^R\kappa_r(t)\boldsymbol{m}^{(r)} + \nu(t)\boldsymbol{I}^{ext}_{\perp}.
\end{equation}
We thus focus on the low-dimensional dynamics of the latent variables \(\kappa_r\) and \(\nu\) emerging from the low-rank connectivity structure rather than the high-dimensional dynamics of individual neuronal activations (Eq.~\eqref{eq:rate_neural_network_main}). 
Specifically, the temporal evolution of \(\kappa_r\) is given by
\begin{small}
\begin{equation}\label{eq:general_latent_rec_dynamics_main}
\begin{aligned}
\dot{\kappa}_r(t) = &-\kappa_r(t) + \sum_{p}\alpha_p\Bigg[a_{n_r}^{(p)}\langle\phi^{(p)}\rangle+\bigg(\sigma^{(p)}_{n_rI_{\perp}^{ext}}\nu(t)\\
&+\sum_{s=1}^R\sigma_{n_rm_s}^{(p)}\kappa_s(t)\bigg)\langle\phi'^{(p)}\rangle\Bigg] +\hat{\boldsymbol{m}}^{(r)\intercal}\boldsymbol{I}^{ext}u(t),
\end{aligned}
\end{equation}
\end{small}
where \(\alpha_p = N_p/N\) is the proportion of neurons in the population \(p\),
\(\hat{\boldsymbol{x}} = \boldsymbol{x}/\Vert\boldsymbol{x}\Vert^2\) represents the scaled vector,
and \(\langle f^{(p)}\rangle = \int_{z\in D}p(z)f(z)\mathcal{D}z\) denotes the population average with respect to the population-specific activity distribution \(D = \mathcal{N}(\mu^{(p)},\Delta^{(p)})\).
Similar to \(\sigma_{n_r m_s}^{(p)}\), \(\sigma_{n_rI_{\perp}^{ext}}^{(p)}\) is the covariance between the associated connectivity vector and external input.

Focusing on low-dimensional dynamics, we define the latent response function as the response function of latent dynamical quantities to input perturbations
\begin{equation}
\chi_{rh}^{\kappa}=\delta\kappa_r/\delta_h,\,\,\,\chi_{h}^{\nu}=\delta \nu/\delta_h.
\end{equation}
Unlike the population-level response function \(\chi_{pq}^x\) defined in the circuit model, the latent response function offers a distinctive advantage in relating external inputs to global random modes that conduct computations \cite{clark2025connectivity, ostojic2024computational}. 
% Latent response functions capture the structural variability and computational modes beyond the population-level average responses; a
Analyzing them provides insight into how inputs interact with connectivity to regulate computational modes. Details for the low-rank neural network model and latent response function are provided in Appendices~\ref{app:appendix_lrRNN} and \ref{app:response_func}.

\section{\label{sec:relationship_between_motifs_and_lowrank}Relationship between circuit and low-rank network models}
We now investigate how these two models relate to each other in the presence of local chain motifs.
Specifically, we examine how connectivity statistics -- both for marginal mean connectivity and correlated random fluctuations -- jointly shape the nonlinear dynamics of the network.
To simplify notation, we omit the superscript in $\tau^{\rm chain}_{pqs}$ for chain motifs unless stated otherwise.

\subsection{\label{subsec:novel_connectivity}Low-rank connectivity approximation}
We characterize a single synaptic weight \(J_{ij}\) from a presynaptic neuron \(j\) in population \(q\) to a postsynaptic neuron \(i\) in population \(p\), following the network construction approach in \cite{shao2025impact,hu2018feedback,dahmen_strong_2026} (Sec.~\ref{sec:NeuralCircuit}). 
Beyond the mean connectivity \(J^0_{pq}\), we characterize the connectivity fluctuations through two complementary parts that share underlying random components. 
The strength of their correlation is quantified by the correlation coefficient of chain motifs.
The correlated incoming (converging) connectivity fluctuation is the summation over \(P\) populations
\begin{equation}\label{eq:incoming_comp}
\frac{\sigma_{pq}}{\sqrt{N}} \left(\sum_{x=1}^P{\rm sgn}(\tau_{p_xpq})\sqrt{\vert\tau_{p_xpq}\vert}\eta_i^{(p_xq)}\right),
\end{equation}
and the correlated outgoing (diverging) connectivity fluctuation aggregates over \(P\) populations
\begin{equation}\label{eq:outgoing_comp} 
\frac{\sigma_{pq}}{\sqrt{N}} \left(\sum_{x=1}^P\sqrt{\vert\tau_{pqp_x}\vert}\eta_j^{(pp_x)}\right).
\end{equation}
Overall, the synaptic weight is 
\begin{small}
\begin{equation}\label{eq:synapse_construction}
    \begin{aligned}
J_{ij}^{pq} =& J^0_{pq} + \frac{\sigma_{pq}}{\sqrt{N}} \Big(\sum_{x=1}^P{\rm sgn}(\tau_{p_xpq})\sqrt{\vert\tau_{p_xpq}\vert}\eta_i^{(p_xq)}\\
&+\sum_{x=1}^P\sqrt{\vert\tau_{pqp_x}\vert}\eta_j^{(pp_x)}+c_{pq}y_{ij}\Big),
    \end{aligned}
\end{equation}
\end{small}
where \(\boldsymbol{\eta}^{(pq)}\in\mathbb{R}^N\) and \(\mathbf{Y}\in\mathbb{R}^{N\times N}\) are vectors and matrix with standard normal distributed random elements. 
The incoming fluctuation Eq.~\eqref{eq:incoming_comp} contributes to chain motifs when \(J_{ij}\) is the first synapse and neuron \(i\) the intermediate neuron. While the outgoing fluctuation Eq.~\eqref{eq:outgoing_comp} contributes when \(J_{ij}\) is the second synapse and neuron \(j\) the intermediate one.
The decomposition in Eq.~\eqref{eq:synapse_construction} shares the same structure as in the circuit model Eq.~\eqref{eq:circuit_decomposition_main}.

How does this construction, Eq.~\eqref{eq:synapse_construction}, relate to the low-rank connectivity?
The key idea is that the incoming component Eq.~\eqref{eq:incoming_comp} corresponds to the connectivity structure with a set of random left vectors $\{\boldsymbol{u}^{(r)}\}$ and deterministic right vectors $\{\boldsymbol{v}^{(r)}\}$, where $r=1\dots P$.
The deterministic right connectivity vector is expressed as 
\begin{equation}\label{eq:incoming_right}
% \boldsymbol{v}^{(k)}=\frac{1}{N}\left(1_{\{i\in p^k\}}\right)_{i=1}^N,
% \boldsymbol{v}^{(r)}=\left(1_{\{i\in p_r\}}\right)_{i=1}^N,
\boldsymbol{v}^{(r)}=\{\mathbf{1}_{p_r}(i)\}_{i=1}^N
\end{equation}
where \(\mathbf{1}_A(x)\) is the indicator function, i.e., \(\mathbf{1}_A(x) = 1\) if \(x\in A\), otherwise zero. 
The left vector is 
\begin{small}
\begin{equation}\label{eq:incoming_left}
\begin{aligned}
% \boldsymbol{u}^{(k)} = &N\sum_{x=1}^P\beta_{\rm sgn}^{x,k}\odot\boldsymbol{\eta}^{(p_xp_k)},\\
\boldsymbol{u}^{(r)} = &N\sum_{x=1}^P\boldsymbol{\beta}_{\rm sgn}^{x,r}\odot\boldsymbol{\eta}^{(p_xp_r)},\\
\boldsymbol{\beta}_{\rm sgn}^{x,r}
=&\Big[\frac{\sigma_{p_1 p_r}}{\sqrt{N}}{\rm sgn}(\tau_{p_x p_1 p_r})\sqrt{\vert\tau_{p_x p_1 p_r}\vert}\boldsymbol{1}_{N_{p_1}};\dots;\\
&\frac{\sigma_{p_P p_r}}{\sqrt{N}}{\rm sgn}(\tau_{p_x p_P p_r})\sqrt{\vert\tau_{p_x p_P p_r}\vert}\boldsymbol{1}_{N_{p_P}}\Big],
\end{aligned}
\end{equation}
\end{small}
where \(\odot\) represents the Hadamard~(elementwise) product.
Note that we match individual rank indices to population indices and fix this \emph{one-to-one mapping} in this study.

The outgoing component Eq.~\eqref{eq:outgoing_comp}, together with the mean connectivity \(\mathbf{J^0}\), associates with the structure with a set of deterministic left vectors $\{\boldsymbol{m}^{(r)}\}$ and random right vectors $\{\boldsymbol{n}^{(r)}\}$, again with $r=1\dots P$.
The left vector is expressed as 
\begin{equation}\label{eq:outgoing_left}
\boldsymbol{m}^{(r)}=\{\mathbf{1}_{p_r}(i)\}_{i=1}^N,
% \boldsymbol{m}^{(r)}=\left(1_{\{i\in p_r\}}\right)_{i=1}^N,
\end{equation}
and the right vector is expressed as the summation:
\begin{equation}\label{eq:outgoing_right}
\begin{aligned}
\boldsymbol{n}^{(r)} =& N\sum_{x=1}^P\boldsymbol{\beta}^{r,x}\odot\boldsymbol{\eta}^{(p_r p_x)}+N\boldsymbol{J}^0_{p_r p_1:p_P},\\
\boldsymbol{\beta}^{r,x}=&
\Big[\frac{\sigma_{p_r p_1}}{\sqrt{N}}\sqrt{\vert\tau_{p_r p_1 p_x}\vert}\boldsymbol{1}_{N_{p_1}};\dots ;\\
&\frac{\sigma_{p_r p_P}}{\sqrt{N}}\sqrt{\vert\tau_{p_r p_P p_x}\vert}\boldsymbol{1}_{N_{p_P}}\Big],\\
\boldsymbol{J}^0_{p_r[p_1:p_P]}&=\left[J^0_{p_r p_1}\boldsymbol{1}_{N_{p_1}};\dots; J^0_{p_r p_P}\boldsymbol{1}_{N_{p_P}}\right].
\end{aligned}
\end{equation}
Overall, we have
\begin{equation}\label{eq:synapse_lowrank}
J_{ij} = \frac{1}{N}\sum_{r=1}^P\left( u_i^{(r)}v_j^{(r)} + m_i^{(r)}n_j^{(r)}\right)+\frac{c_{pq}\sigma_{pq}}{\sqrt{N}}y_{ij},
\end{equation}
revealing a rank $2P$ structure (Sec.~\ref{sec:lrRNN_model}, Eq.~\eqref{eq:gen_synapse_lowrank_main}). 
The connectivity matrix presented in the low-rank and i.i.d. random components format, as Eq.~\eqref{eq:synapse_lowrank}, is equivalent to that of Eq.~\eqref{eq:synapse_construction}, which means that we converted the matrix given by the locally canonical neural circuit incorporating heterogeneous chain motifs into the form of low-rank decomposition.
In the following, we consider only the first term within the bracket on the right-hand side of Eq.~\eqref{eq:synapse_lowrank} as the low-rank approximation model.

We moreover examine and list the statistics of connectivity vectors, for those with non-zero means
\begin{equation}\label{eq:conn_stats_0}
a_{v_r}^{(p_k)} = \delta_{rk},\,\,\,a_{m_r}^{(p_k)} = \delta_{rk},\,\,\,a_{n_r}^{(p_k)} = NJ^0_{p_rp_k},
\end{equation}
for those with non-zero variances
\begin{small}
\begin{equation}\label{eq:conn_stats_1}
\begin{aligned}
\sigma_{u_r^2}^{(p_k)} &= N\sum_{x=1}^P\sigma_{p_kp_r}^2\vert\tau_{p_xp_kp_r}\vert,\\
\sigma_{n_r^2}^{(p_k)} &= N\sum_{x=1}^P\sigma_{p_rp_k}^2\vert\tau_{p_rp_kp_x}\vert,
\end{aligned}
\end{equation}
\end{small}
and for those with non-zero covariances (Fig. \ref{fig:fig_schematic}C)
\begin{equation}\label{eq:conn_stats_2}
\sigma_{n_ru_{r'}}^{(p_k)} = N\sigma_{p_rp_k}\sigma_{p_kp_{r'}}\tau_{p_rp_kp_{r'}}.
\end{equation}

\noindent{\em Eigenvalue outliers.}
As shown in previous studies \cite{schuessler2020dynamics,shao2025impact}, the eigenvalue outliers of $\mathbf{J}$, in the form of Eq.(\ref{eq:circuit_decomposition_main}), can be computed analytically using the Matrix Determinant Lemma \cite{schuessler2020dynamics} and the Sherman-Morrison Formula.
The averaged eigenvalue outliers \([\lambda]\) across realizations are the solutions of 
\begin{equation}\label{eq:characteristic_polynomia}
\begin{aligned}
\det\left(\mathbf{I}\lambda - \frac{1}{N}\mathbf{N}_0^{\top}\left(\mathbf{I}-\frac{[\mathbf{Z}^2]}{\lambda^2}\right)^{-1}\mathbf{M}_0\right)=0
\end{aligned}
\end{equation}
where \(\mathbf{J^0}=\mathbf{M_0}\mathbf{N_0}^{\top}/N\), and \([\mathbf{Z}^2]\) denotes the second-order moment matrix of the random connectivity.
The elements of \([\mathbf{Z}^2]\) are 
\begin{equation}\label{eq:Z_second_order_moment}
    \left[\mathbf{Z}^2\right]_{ij} = \sum_{x=1}^{P} \alpha_{p_x}\sigma_{pp_x}\sigma_{p_xq}{\rm sgn}(\tau_{pp_xq})|\tau_{pp_xq}| 
\end{equation}
where neurons \(i,j\) belong to populations \(p,q\).
Therefore, \([\mathbf{Z}^2]\) is also a block-structured low-rank matrix with rank at most \(P\). Exploiting this low-rank structure with some linear algebra, we obtain the trial-averaged eigenvalues.
The details for computing the eigenvalue outliers are provided in Appendix~\ref{app:outliers}.

\subsection{Low-dimensional network dynamics}
By comparing Eqs.~\eqref{eq:gen_synapse_lowrank_main} and \eqref{eq:synapse_lowrank} and examining the statistics of the connectivity vectors, we demonstrate that the set of connectivity vectors \(\{\boldsymbol{u}^{(r)},\boldsymbol{m}^{(r)}\}_{r=1\dots P}\) constitutes a new orthogonal basis with rank \(R= 2P\).
This new set of vectors, along with the external input vector, characterize the low-dimensional embedding space \cite{ostojic2024computational}.
Accordingly, the temporal dynamics of the synaptic current of neuron \(i\) evolve in the embedding space as follows~(Eq.~\eqref{eq:lowdim_subspace_main})
\begin{equation}\label{eq:latent_dynamics_x}
x_i(t) = \sum_{r=1}^P\left(\kappa_r(t)u_i^{(r)}+K_r(t)m_i^{(r)}\right)+\nu(t)I_{\perp,i}^{ext}
\end{equation}
where \(\{\kappa_r(t),K_r(t)\}_{r=1\dots P}\) correspond to the latent quantities evolving within the recurrent subspace along \(\boldsymbol{u}^{(r)},~\boldsymbol{m}^{(r)}\) respectively, and \(\nu(t)\) along the external input direction.
Analogous to Eq.~\eqref{eq:general_latent_rec_dynamics_main}, we 
combine connectivity-vector statistics Eqs.~\eqref{eq:conn_stats_0}-\eqref{eq:conn_stats_2} and obtain temporal dynamics of the latent quantities as
\begin{equation}\label{eq:low_dyn_quantities}
    \begin{aligned}
    \dot{\kappa}_r(t) =& -\kappa_r(t) 
    + \alpha_{p_r}\langle\phi^{(p_r)}\rangle
    +\hat{\boldsymbol{u}}^{(r)\intercal}\boldsymbol{I}^{ext}u(t),\\
    \dot{K}_r(t) =& -K_r(t) 
    + \sum_{x=1}^P\alpha_{p_x}\Bigg[a_{n_r}^{(p_x)}\langle\phi^{(p_x)}\rangle+\Big(\sigma^{(p_x)}_{n_rI_{\perp}^{ext}}\nu(t)
    \\
    &+\sum_{r'=1}^P\sigma_{n_ru_{r'}}^{(p_x)}\kappa_{r'}(t)\Big)\langle\phi'^{(p_x)}\rangle\Bigg] 
    +\hat{\boldsymbol{m}}^{(r)\intercal}\boldsymbol{I}^{ext}u(t),\\
    \dot{\nu}(t) =& -\nu(t) 
    + \hat{\boldsymbol{I}}_{\perp}^{ext\intercal}\boldsymbol{I}^{ext}u(t).
    \end{aligned}
\end{equation}
The high-dimensional network dynamics \(\{x_i\}_{i=1\dots N}\) are fully characterized by the low-dimensional dynamic quantities~\(\{\kappa_1\dots\kappa_P, K_1\dots K_P,\nu\}\).

Noticing that the orthogonalized connectivity basis vectors have specific statistical properties: \(\sigma_{m_r^2}^{(p_k)} = 0\), \(\forall r,k\). Consequently, only the latent quantities \(\{\kappa_r\}\) contribute to neural activity variability:
\begin{equation}
    \Delta^{(p_k)}(t)=\sum_{r=1}^P\sigma_{u_r^2}^{(p_k)}\left(\kappa_r(t)\right)^2 + \sigma_{I_{\perp}^{ext}{}^2}^{(p_k)}\left(\nu(t)\right)^2.
\end{equation}
\begin{small}
\renewcommand{\arraystretch}{1.5} % 
\begin{table*}[t]
\centering
\caption{Connectivity Statistics of EI Networks}
\label{tab:EI_connectivity_vectors}
\begin{tabular}{c|cccccc|cccccccc}
\hline\hline
Stats.
 & \(a_{v_1}^{(E)}\) & \(a_{v_1}^{(I)}\) & \(a_{m_1}^{(E)}\) & \(a_{m_1}^{(I)}\) & \(a_{n_1}^{(E)}\) & \(a_{n_1}^{(I)}\) & \(\sigma_{u_1^2}^{(E)}\)
 & \(\sigma_{u_1^2}^{(I)}\) & \(\sigma_{n_1^2}^{(E)}\) & \(\sigma_{n_1^2}^{(I)}\) & \(\sigma_{n_1u_1}^{(E)}\) & \(\sigma_{n_1u_1}^{(I)}\) & \(\sigma_{n_1u_2}^{(E)}\) & \(\sigma_{n_1u_2}^{(I)}\) \\
\hline\hline
Homo. & 1 & 1 & 1 & 1 & \(NJ^0\) & \(-NgJ^0\) & \(N\sigma^2|\tau|\) & \(N\sigma^2|\tau|\) & \(N\sigma^2|\tau|\) & \(N\sigma^2|\tau|\) & \(N\sigma^2\tau\) & \(N\sigma^2\tau\) & NAN  & NAN \\\hline
Hetero.~1 & 1 & 0 & 1 & 0 & \(NJ^0\) & \(-NgJ^0\)  & \(N\sigma^2\sum_{p}^{E,I}|\tau_p|\) & \(N\sigma^2\sum_{p}^{E,I}|\tau_p|\) & \(2N\sigma^2|\tau_E|\) & \(2N\sigma^2|\tau_E|\) & \(N\sigma^2\tau_E\) & \(N\sigma^2\tau_E\) & \(N\sigma^2\tau_E\) & \(N\sigma^2\tau_E\) \\\hline
Hetero.~2 & NAN & NAN & 1 & 0 & \(NJ^0\) & \(-Ng_EJ^0\)  & NAN & NAN & 0 & 0 & NAN & NAN & 0 & 0 \\
\hline\hline
Stats.
 & \(a_{v_2}^{(E)}\) & \(a_{v_2}^{(I)}\) & \(a_{m_2}^{(E)}\) & \(a_{m_2}^{(I)}\) & \(a_{n_2}^{(E)}\) & \(a_{n_2}^{(I)}\) & \(\sigma_{u_2^2}^{(E)}\)
 & \(\sigma_{u_2^2}^{(I)}\) & \(\sigma_{n_2^2}^{(E)}\) & \(\sigma_{n_2^2}^{(I)}\) & \(\sigma_{n_2u_1}^{(E)}\) & \(\sigma_{n_2u_1}^{(I)}\) & \(\sigma_{n_2u_2}^{(E)}\) & \(\sigma_{n_2u_2}^{(I)}\) \\
 \hline\hline
Homo. &  &  & NAN &  &  &  &  &  &  & NAN &  &  &  &  \\\hline
Hetero.~1 & 0 & 1 & 0 & 1 & \(NJ^0\) & \(-NgJ^0\) & \(N\sigma^2\sum_{p}^{E,I}|\tau_p|\)  & \(N\sigma^2\sum_{p}^{E,I}|\tau_p|\) & \(2N\sigma^2|\tau_I|\) & \(2N\sigma^2|\tau_I|\) & \(N\sigma^2\tau_I\) & \(N\sigma^2\tau_I\) & \(N\sigma^2\tau_I\) & \(N\sigma^2\tau_I\)  \\\hline
Hetero.~2 & 0 & 1 & 0 & 1 & \(NJ^0\) & \(-Ng_IJ^0\) & 0  & \(N\sigma^2|\tau_{III}|\) & 0 & \(N\sigma^2|\tau_{III}|\) & NAN & NAN & 0 & \(N\sigma^2\tau_{III}\)  \\
\hline\hline
\end{tabular}
\end{table*}
\end{small}
More interestingly, \(a_{u_r}^{(p_k)} = 0\), \(\forall r,k\), while only \(a_{m_r}^{(p_k)} = 1\) for a specific neuron population where \(k=r\).
Thus, each latent quantity \(K_r\) associates with mean activities of neurons specifically in the mapped population \(p_r\)
\begin{equation}
    \mu^{(p_r)}(t) = K_r(t)+a_{I_{\perp}^{ext}}^{(p_r)}\nu(t).
\end{equation}
Accordingly, the \(2P\)-rank structure and its associated \(2P\)-dimensional latent variables admit specific \emph{physical interpretations}: the set \(\{\kappa_r\}_{r=1\dots P}\) parameterizes the heterogeneity of within-population activity variability, whereas the set \(\{K_r\}_{r=1\dots P}\) encodes the heterogeneity of mean activity across populations. This framework thereby separates microscopic variability from population-level mean activity heterogeneity.

Despite this distinction, the two sets of latent dynamics are not independent. The variability variable \(\kappa_{r'}\) directly influences the mean variable \(K_{r}\) through the covariance \(\sigma_{n_{r}u_{r'}}\) and population gain modulation. Moreover, they interact nonlinearly with each other through the population-averaged activation \(\langle\phi^{(p)} \rangle\) (Fig. \ref{fig:fig_schematic}D). Consequently, our framework provides a principled and quantitative characterization of the mean-variability coupling.

\section{\label{sec:toy-examples}Rich dynamics in two-population EI networks}
\subsection{\label{subsec:homogeneousGauss}Self-excitation in nonlinear networks via positively correlated chain motifs}
\begin{figure*}
\centering
\includegraphics[width=1.0\linewidth]{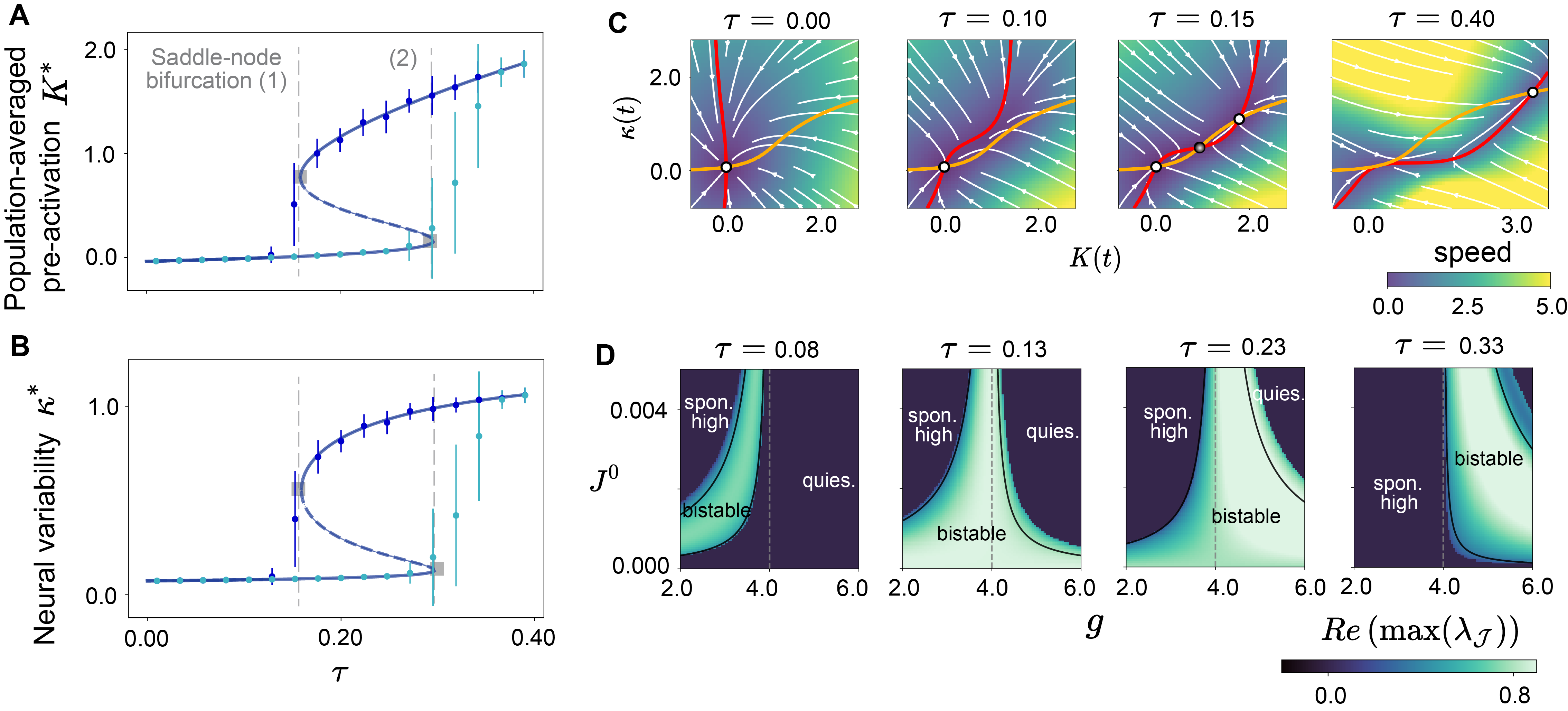}
\caption{{\bf Impact of homogeneous chain motifs on dynamics in EI networks. } (A,~B) Saddle-node bifurcation of latent quantities $K$ and $\kappa$, induced by homogeneous chain motifs with strength $\tau$. 
Solid lines are theoretical predictions using Eq.~\eqref{eq:kappa_K_homo},
dots are numerical full-rank network simulations with high initial conditions (dark blue) and with low initial conditions (light blue).
(C) Velocity field on \(\kappa-K\) plane with increased strength $\tau$ of homogeneous chain motifs. The $\kappa$-nullcline is shown by the orange line, and the $K$-nullcline is shown by the red line.
White circles are stable steady states, and gray circle is the unstable steady state.
Parameters in (A-C): $N_E=2000$, $N_I=500$, $\sigma=0.1$, $J^0=0.001$, $\theta=1.6$, $g=5.0$. (D) Phase diagram in mean connectivity plane $g-J^0$, with $\tau$ increases. Parameters in (D): $N_E=2000$, $N_I=500$, $\sigma=0.1$, $\theta=1.6$.
}
\label{fig:homo_fig}
\end{figure*}
We first focus on the simple network architecture, in which only the mean connectivity depends on the presynaptic neuron population \cite{brunel2000dynamics}, that is \(J^0_{EE} = J^0_{IE} = J^0\) and \(J^0_{EI} = J^0_{II} = -gJ^0\). Here, we set the ratio of mean inhibition to mean excitation is denoted by \(g\) and the ratio of the inhibitory to the excitatory neuron number is \(\gamma = \alpha_I/\alpha_E\). 
Both the magnitude of the connectivity fluctuations and the correlation coefficient are homogeneous and independent of the pre- and postsynaptic neuron populations~(\(\sigma\) and \(\tau\) respectively).

According to Eq.~\eqref{eq:synapse_construction}, the connectivity matrix of this simple homogeneous EI network can be constructed as 
\begin{small}
\begin{equation}\label{eq:simpleEI_homo}
J_{ij} = \frac{\sigma}{\sqrt{N}}{\rm sgn}(\tau)\sqrt{\vert\tau\vert}\eta_i+\left(\frac{\sigma}{\sqrt{N}}\sqrt{\vert\tau\vert}\eta_j+J^0_{pq}\right) + \frac{\sigma c y_{ij}}{\sqrt{N}}
\end{equation}
\end{small}
where neurons \(i,~j\) belong to populations \(p,~q\) respectively, \(\eta_i\) is a standard Gaussian-distributed random variable, and \(c=\sqrt{1-2|\tau|}\).
So for this two-population network, instead of being rank-4, the homogeneous property allows for a rank-2 structure with connectivity vectors
\begin{small}
\begin{equation}\label{eq:simpleEI_homo_connvec}
\begin{aligned}
\boldsymbol{u} &=  N\frac{\sigma}{\sqrt{N}}{\rm sgn}(\tau)\sqrt{\vert\tau\vert}\boldsymbol{\eta},\,\,\,\boldsymbol{v} = \boldsymbol{1},\\
\boldsymbol{m} &= \boldsymbol{1},\,\,\,\boldsymbol{n} = N\left(\frac{\sigma}{\sqrt{N}}\sqrt{\vert\tau\vert}\boldsymbol{\eta}+\boldsymbol{J^0}\right)
\end{aligned}
\end{equation}
\end{small}
where \(\boldsymbol{J^0} = [J^0\dots,-gJ^0\dots]^{\intercal}\) is the vector representing the mean connectivity.
The statistics of the entries on these recurrent connectivity vectors are listed in Table~\ref{tab:EI_connectivity_vectors} -- Homo.

Following Eqs.~\eqref{eq:characteristic_polynomia} and \eqref{eq:Z_second_order_moment}, we obtain the expression for the eigenvalue outliers of this connectivity matrix as 
\begin{equation}\label{eq:homo_outlier}
\lambda = \frac{\lambda_0\mp\sqrt{\lambda_0^2+4\Delta^2}}{2}
\end{equation}
where the eigenvalue perturbation strength
\begin{equation}
\Delta^2 = \sigma^2\tau(N-1).
\end{equation}
As $\tau$ increases, the two real eigenvalue outliers deviate from $\lambda_0$ in the positive and negative directions \cite{shao2025impact}.

Next, we examine the dynamics of this simple homogeneous EI network, focusing on the autonomous dynamics where no external input is applied (\(u(t)=0\), \(\forall t\)).
For the rank-two orthogonalized connectivity vectors \(\boldsymbol{u}\) and \(\boldsymbol{m}\), we have two corresponding low-dimensional latent quantities defined as \(\kappa(t),~K(t)\) (Eq.~\eqref{eq:low_dyn_quantities}), which evolve according to
\begin{small}
\begin{equation}\label{eq:homo_dynamics}
\begin{aligned}
    \dot{\kappa}&= -\kappa + \sum_p^{E,I}\alpha_p\langle\phi^{(p)}\rangle,\\
    \dot{K} &= -K + \sum_p^{E,I}\alpha_p\left(a_n^{(p)}\langle\phi^{(p)}\rangle+\sigma_{nu}^{(p)}\langle\phi'^{(p)}\rangle\kappa\right),
\end{aligned}
\end{equation}
with \(\langle\phi^{(p)}\rangle=\langle\phi(\mu^{(p)},\Delta^{(p)})\rangle,~\langle\phi'^{(p)}\rangle=\langle\phi'(\mu^{(p)},\Delta^{(p)})\rangle\) and
\begin{equation}
\begin{aligned}
    \mu^{(p)} = K,\,\,\,\Delta^{(p)} = \kappa^2N\sigma^2\vert\tau\vert.\,\,% {\rm for}\,\,p\in E,I.
\end{aligned}
\end{equation}
\end{small}
Noticing that the collective dynamics of neurons are identical across neuron populations, we set the population-averaged activity as \(\langle\phi\rangle\) and population-averaged gain as \(\langle\phi'\rangle\), independent across populations.

Using Eq.~\eqref{eq:homo_dynamics}, we fully characterize the dynamical landscape of the macroscopic latent quantities \(\kappa\) and \(K\) in the embedding space spanned by the basis vectors \(\boldsymbol{u}\) and \(\boldsymbol{m}\). 
The steady-state of the autonomous dynamics satisfies the following self-consistent equations:
\begin{small}
\begin{equation}\label{eq:kappa_K_homo}
\kappa^* = \langle\phi^*\rangle,\,\,\,K^* = N\alpha_E(1-g\gamma)J^0\langle\phi^*\rangle+N\sigma^2\tau\langle\phi'{}^*\rangle\kappa^*
\end{equation}
where \(\langle f^*\rangle=\langle f(\mu^*,\Delta^*)\rangle\) with \(\mu^* = K^*\), \(\Delta^* = (\kappa^*)^2N\sigma^2\vert\tau\vert\).
\end{small}

\vspace{12pt}
We then examine the effects of the homogeneous positive strength of the chain motifs \(\tau\) on network dynamics in networks specifically with inhibition-dominant mean connectivity.
Simulations using full-rank connectivity reveal self-excited dynamics with increasing \(\tau\), that is, the autonomous population-averaged neural activity of excitatory and inhibitory neurons achieves a relatively high state with a strong \(\tau\). 
This dynamic self-excitation can be precisely captured by our theoretical predictions of the latent quantity \(K\) (Eq.~\eqref{eq:kappa_K_homo}, Fig.~\ref{fig:homo_fig}A).

Specifically, both latent quantities \(\kappa\) and \(K\) initially exhibit a single stable low steady-state. Gradually increasing \(\tau\) reveals a saddle-node bifurcation~(1), yielding a stable high steady-state and an unstable saddle point at intermediate activity levels.
This bifurcation enables self-excitation to arise in the autonomous dynamics of both excitatory and inhibitory neurons and creates a bistable regime~(coexistence of low and high steady states).
Further increasing \(\tau\) destabilizes the low steady-state through a second saddle-node bifurcation~(2), restoring the network to a monostable regime with a single stable high-activity steady-state~(Fig.~\ref{fig:homo_fig}A,~B).
The velocity field on the two-dimensional \(\kappa-K\) plane~(Fig.~\ref{fig:homo_fig}C) demonstrates that increase in $\tau$ causes the nullcline \(\dot{K}=0\)
%\ys{(red lines)???} 
to progressively curve, generating two intersection points -- an unstable saddle point and a stable high-activity steady-state. Further increase of $\tau$ causes the intermediate saddle point and the low steady-state to come close and collide.
These latent dynamical properties are not specific to the EI network with our chosen nonlinearity but generalize to networks with other saturating positive nonlinearities, as demonstrated in Supp.~Fig.~\ref{supp_fig:homo_general_nonlinearity}.

Beyond examining the impact of motifs in isolation, we analytically investigate how mean connectivity and chain motif components jointly regulate dynamical bifurcations through stability analysis of the network's steady-state. 
For a given motif strength \(\tau\), the steady-state solution and its stability (measured by \(\det(\mathcal{J})\), where \(\mathcal{J}\) is the Jacobian of the system around the steady-state)
% ${\rm det}(\mathbf{J}_{\rm jacobian})$) 
result in a set of self-consistent equations for the three variables \((\kappa^*,~K^*,~\lambda_0)\)
\begin{equation}\label{eq:solve_transition_homogeneous}
\begin{aligned}
\kappa^* &= \langle\phi^*\rangle,\\
K^* &= \lambda_0\langle\phi^*\rangle + N\sigma^2\langle\phi^{\prime*}\rangle\kappa^*,\\
\lambda_0 &= \frac{1}{\langle\phi'^{*}\rangle}
(1-C\langle\phi''^{*}\rangle\kappa^*-|C|\langle\phi''^{*}\rangle\kappa^* \\
& \ \ +\text{sgn}(\tau)(C\langle\phi''^{*}\rangle\kappa^*)^2-C(\langle\phi'^{*}\rangle)^2) ,
\end{aligned}
\end{equation}
where \(\phi^* = \phi(K^*,(\kappa^*\sigma)^2N|\tau|)\) and coefficient $C=\sigma^2N\tau$. Solving these self-consistent equations yields two solution sets corresponding to the bifurcation transitions from the quiescent regime to the bistable regime and from the bistable regime to the high-activity monostate regime.

We here focus on two solutions for \(\lambda_0\), denoted as \(\{\lambda_0^{+},~\lambda_0^{-}\}_{(\tau)}\), to explore how mean connectivity structure regulates the dynamical bifurcations.
For a fixed network size \(N\) and EI population ratio \(\gamma\), the relationship \(\lambda_0 = N\alpha_E(1-g\gamma)J^0\) leads to two bifurcation transition lines in the \(g -J^0\) plane with the region between them representing the bistable regime.
When $\tau\to 0$, the bistable regime only arises in an excitation-dominant regime (to the left of \(g=\gamma^{-1}=4\)) and requires moderate \(J^0\)~(Fig.~\ref{fig:homo_fig}D, \(\tau=0.08\)). 
As $\tau$ increases, both bifurcation transition lines shift towards the inhibition-dominant regime (to the right of \(g=\gamma^{-1}\)), indicating that homogeneous chain motifs enhance the bistability while reducing the required mean excitation~(Fig.~\ref{fig:homo_fig}D, upper boundary of \(g\) increases from left to right subpanels).
Moreover, with a larger \(J^0\), the two transition lines tangentially converge along \(g=\gamma^{-1}\) and the bistable regime shrinks, ultimately vanishing. In this limit, the steady states of the latent dynamical variables as a function of \(g\) exhibit behavior characteristic of a cusp bifurcation (Supp.~Fig.~\ref{supp_fig:homo_gj0_saddle}C,~D).
For a comprehensive discussion of bifurcation and stability analysis, see Appendix~\ref{app:stability_analysis} for references.

Overall, the findings demonstrate that incorporating chain motifs generates rich bifurcation diagrams in network dynamics. The mathematical framework thus enables a mechanistic understanding of the network state transitions~(i.e. between bistable and monostable regimes) underlying diverse cognitive processes.
\begin{figure*}[htb]
\centering
\includegraphics[width=0.95\linewidth]{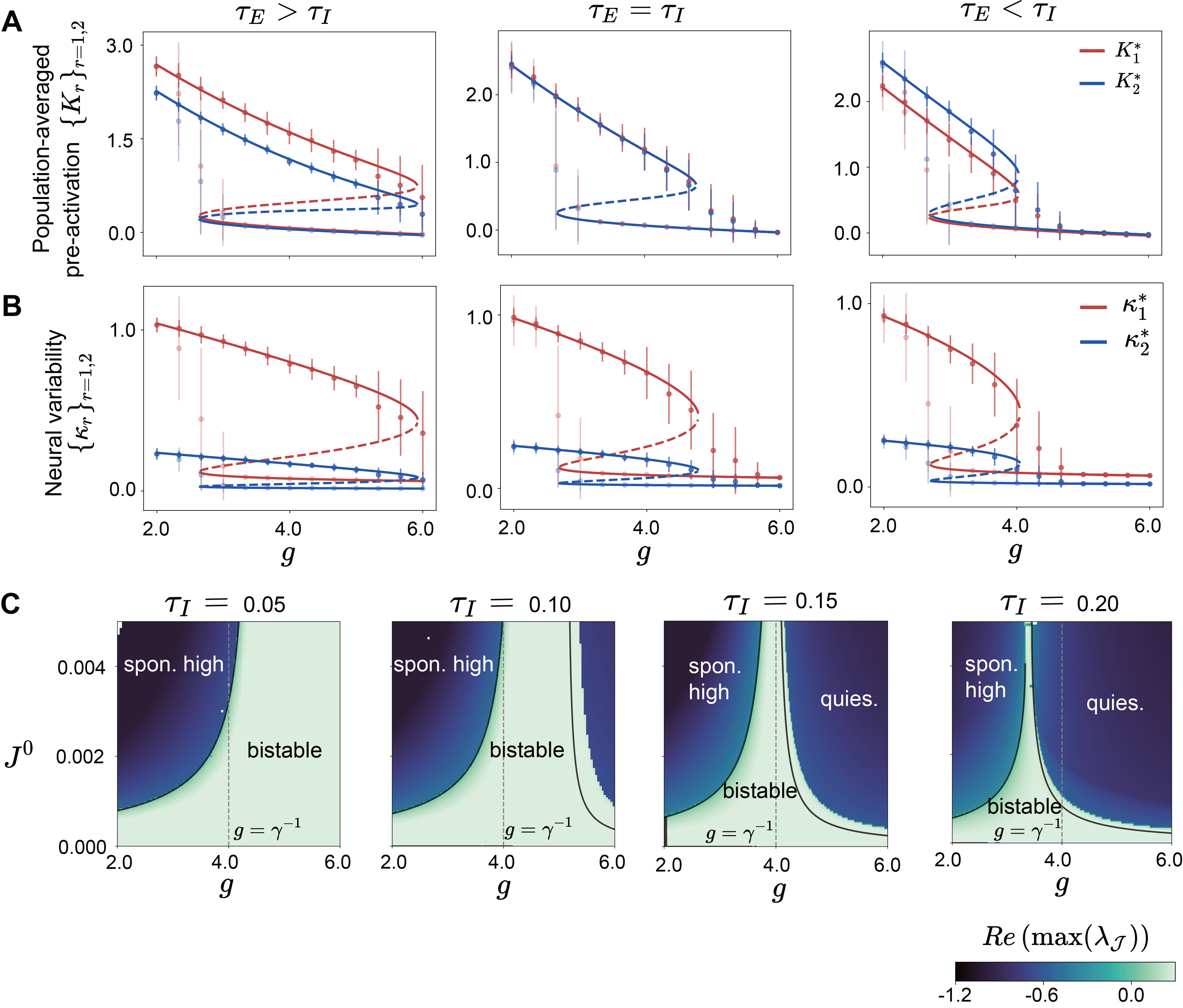}
\caption{{\bf Dependence of latent quantities steady states on increasing global inhibition strength $g$, across varying end-neuron-specific chain motifs $\tau_E$ and $\tau_I$ in EI network.} (A) Illustration of steady-state of $\{K_r\}_{r=1,2}$ corresponding to the population-averaged pre-activation of the excitatory population and the inhibitory population, respectively. When $g$ is small both latent quantities maintain a high steady-state. As $g$ increases, the system exhibits bistability. A decrease in $g$ destabilizes the high steady-state, leaving only the low steady-state stable. When $\tau_E>\tau_I$, the excitatory population has a higher steady-state. With $\tau_E=\tau_I$, two populations have identical steady-states. When $\tau_I<\tau_E$, the inhibitory population has a higher steady-state. (B) Illustration of steady-state of $\{\kappa_r\}_{r=1,2}$. 
Solid lines are theoretical predictions according to Eq.~\eqref {eq:latents_heter}, dots are numerical full-rank network simulations with different initializations (Light-colored dots represent simulations under low initial conditions, while dark-colored dots denote those under high initial conditions).
Parameters in (A) and (B): $\tau_E=0.15$. $\tau_I=0.10$ for the left, $\tau_I=0.15$ for the middle, and $\tau_I=0.20$ for the right. Fixing $J^0=0.001$. 
(C) Phase diagram in the mean connectivity plane $g-J^0$, at different level of $\tau_E$ and $\tau_I$. $\tau_E$ is fixed at $0.15$. Other parameters: $N_E = 2000$, $N_I = 500$, $J^0 = 0.001$, $\sigma = 0.1$, $\theta = 1.6$. 
}
\label{fig:hetero_auto_EI_fig}
\end{figure*}
\subsection{\label{subsec:heterogeneousGauss}Multi-scale heterogeneity induced by end-neuron-specific chain motifs}
\vspace{-8pt}
We then examine our framework under a more general condition -- a two-population EI network with heterogeneous, cell-type-specific chain motif strengths. 

We start with a simplification where the chain motifs are unique to the first index, meaning that the strengths depend on the end neuron, $\tau_{Epq}=\tau_E$, $\tau_{Ipq}=\tau_I$ for
 \(p,q\in {E,I}\).  Other network settings are the same as those used in the homogeneous EI network. 
The network connectivity $J_{ij}$ is expressed as 
\begin{equation}\label{eq:heterogeneous_EI}
\begin{aligned}
J_{ij} =& J_{pq}^0 +\frac{\sigma}{\sqrt{N}}\Big(\sum_s^{E,I}{\rm sgn}(\tau_{s})\sqrt{|\tau_{s}|}\eta_i^{(sq)}\\
&+\sum_s^{E,I}\sqrt{|\tau_{p}|}\eta_j^{(ps)}+c_{pq}y_{ij}\Big)
\end{aligned}
\end{equation}
where \(i,j\) belong to population \(p,q\) respectively.
By analogous method in \cite{shao2025impact}, the eigenvalue outliers of connectivity matrix with end-neuron-specific chain motif $\tau_E$ and $\tau_I$ satisfy the expression
\begin{equation}\label{eq:heteroEI_outlier}
\lambda = \lambda_0 + \frac{(1+\gamma)\sigma^2N_E^2(\tau_E-g\gamma\tau_I)}{\lambda^2-\sigma^2N_E(\tau_E+\gamma\tau_I)}J^0.
\end{equation}
In general, the third-order polynomial Eq.~\eqref{eq:heteroEI_outlier} yields three eigenvalue outliers. When \(\tau_E = \tau_I\), the eigenspectrum reduces to the homogeneous case, with outliers given by Eq.~\eqref{eq:homo_outlier}.
\begin{figure*}[htb]
\centering
\includegraphics[width=0.95\linewidth]{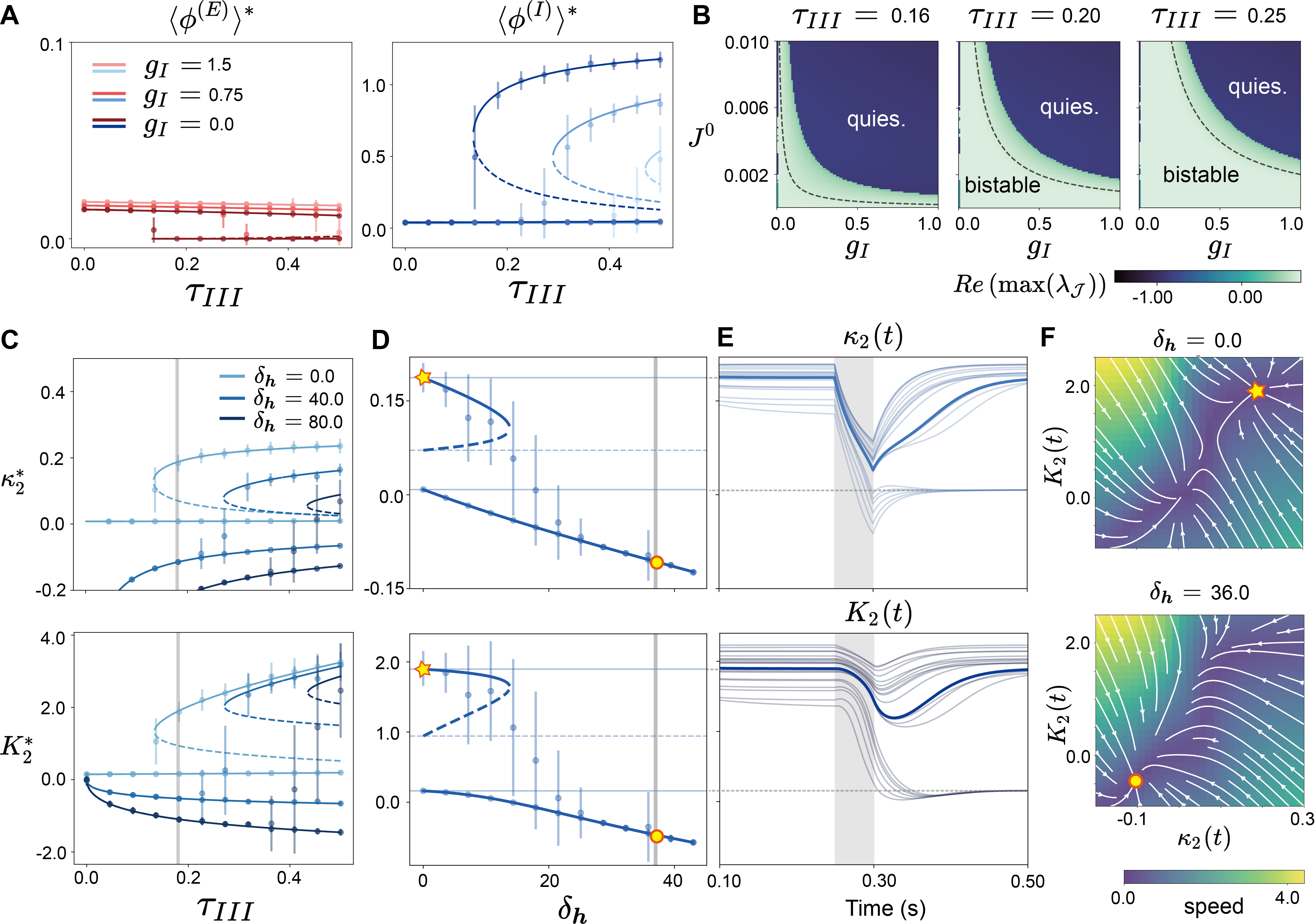}
\caption{{\bf Steady-state and transient activity of inhibitory neuron population modulated by chain motifs $\tau_{III}$.} (A) Bistable transition of population-averaged firing rates $\langle\phi^{(E)}$ and $\langle\phi^{(I)}\rangle$ under different global inhibitions $g_I$. With $g_I=1.5$, steady states $\langle\phi^{(E)}\rangle$ and $\langle\phi^{(I)}\rangle$ stay in the low state. 
(B) Phase diagram in the mean connectivity plane $g_{I}-J^0$.
(C) Bifurcation shift occurs with increasing magnitude of the patterned input $\boldsymbol{h} = -\boldsymbol{\eta}^{(II)} / \sqrt{N_I}$. 
(D) Transition from bistability to quiescence of latent quantities $\kappa_2, K_2$ with patterned input $\boldsymbol{h}$ that is parallel to $\boldsymbol{\eta}^{(II)}$. $\tau_{III}=0.18$. 
In panels (A,~C,~and D), solid lines are theoretical predictions of latent dynamics, dots are numerical results from full-rank network simulations.
(E) Transient activity before, during, and after the patterned input $\boldsymbol{h}$. $\tau_{III}=0.18$ (the gray line in panel (C)), $\delta_{\boldsymbol{h}} = 36.0$ (the gray line in panel (D)).  
The two darkest lines are theoretical results directly derived from the latent quantity dynamics, the lighter lines are results obtained from full-rank network simulations.
(F) Reduced velocity field in plane $\kappa_2-K_2$, under perturbation strength $\delta_{\boldsymbol{h}}=0.0$ (up) and $\delta_{\boldsymbol{h}}=36.0$ (down). Fixing $K_1$ at its steady-state under the autonomous scenario (up) and input-driven scenario (down). Parameters: $J^0 = 0.005$, $g_E = 5.0$, $\sigma = 0.6$, $\theta = 2.1$.}
\label{fig:tau_iii_reponse_fig}
\end{figure*}

According to our theoretical framework and the heterogeneous property of the network settings, the low-rank approximation model is of rank four~(\(2P=4\)), with \(\{\boldsymbol{u}^{(r)},\boldsymbol{v}^{(r)},\boldsymbol{m}^{(r)},\boldsymbol{n}^{(r)}\}_{r=1,~2}\). 
We set the one-to-one rank-population mapping between the neural population indices and the rank indices as \(r=1,~2\) and \(p_1=E,~p_2=I\).
Using the generic formats given in Eqs.~\eqref{eq:incoming_left}-\eqref{eq:outgoing_right}, the connectivity vectors \(\boldsymbol{u}^{(r)}\) and \(\boldsymbol{n}^{(r)}\) for \(r=1,2\) satisfy the following expressions:
\begin{small}
\begin{equation}
\begin{aligned}
    \boldsymbol{u}^{(r)} =& \sqrt{N}\sigma\left({\rm sgn}(\tau_E)\sqrt{|\tau_E|}\boldsymbol{\eta}^{(Ep_r)} +{\rm sgn}(\tau_I)\sqrt{|\tau_I|}\boldsymbol{\eta}^{(Ip_r)}\right),\\
    \boldsymbol{n}^{(r)} =& \sqrt{N}\sigma\left(\sqrt{|\tau_{p_r}|}\boldsymbol{\eta}^{(p_rE)}+\sqrt{|\tau_{p_r}|}\boldsymbol{\eta}^{(p_rI)}\right) \\
    &+ N[J^0_{p_rE}\dots J^0_{p_rI}]^{\intercal}.
\end{aligned}
\end{equation}
\end{small}
The statistics of the entries on all connectivity vectors are listed in Table~\ref{tab:EI_connectivity_vectors} -- Hetero.~1.
Observe that, in contrast to the homogeneous model, the rank (three) inferred from the number of eigenvalue outliers does not equal the rank (four) derived from our novel matrix decomposition approach under this more general, heterogeneous condition. 
The non-uniqueness of matrix decompositions may explain the rank mismatches \cite{kolda2009tensor}.

There are four latent dynamic quantities \(\kappa_1,~\kappa_2,~K_1,~K_2\) that correspond to the four rank recurrent connectivity vectors \(\boldsymbol{u}^{(1)},~\boldsymbol{u}^{(2)},~\boldsymbol{m}^{(1)},~\boldsymbol{m}^{(2)}\).
The four variables and these two sets evolve as follows:
\begin{equation}\label{eq:latents_heter}
\begin{aligned}
\dot\kappa_r &= -\kappa_r + \alpha_{p_r}\langle\phi^{(p_r)}\rangle, \\
\dot K_r &= -K_r + \bar K+N\sigma^2\tau_{p_r}\sum_{r'=1}^2\kappa_{r'}\sum_{s}^{E,I}\alpha_{s}\langle\phi'{}^{(s)}\rangle,
\end{aligned}
\end{equation}
where \(r=1,2\) and \(\bar K = N_EJ^0\langle\phi^{(E)}\rangle-N_IgJ^0\langle\phi^{(I)}\rangle\) 
representing the dynamical contributions from network mean connectivity.
%and \(p_1=E,~p_2=I\).
The means and variances of the population activity are 
\begin{equation}\label{eq:hetero_end_means_variance}
\mu_{p_r} = K_r,\,\,\,\Delta_{E/I} = (\kappa_1^2+\kappa_2^2)N\sigma^2(|\tau_E|+|\tau_I|).
\end{equation}

Turning again to autonomous dynamics, we discover that latent dynamical quantities $(\kappa_1, \kappa_2, K_1, K_2)$ exhibit saddle-node bifurcations, similar to those in networks with homogeneous chain motifs (Supp.~Fig.~\ref{suppfig:supp_fig_hetero_end_type}). 
Unlike the degenerate homogeneous case, heterogeneity in the strengths of motif produces distinct \(K_1,~K_2\) (resp. \(\kappa_1,~\kappa_2\))~(Fig.\ref{fig:hetero_auto_EI_fig}A,~B).
The distinction, particularly between \(K_1\) and \(K_2\) indicates that population-averaged neuronal activity varies across populations. 
When $\tau_E > \tau_I$, the excitatory population exhibits a higher steady-state mean activity level~(Fig.\ref{fig:hetero_auto_EI_fig}A, left panel).
Conversely, stronger $\tau_I$ drives higher mean activity in the inhibitory population~(Fig.\ref{fig:hetero_auto_EI_fig}A, right panel). 
The population-specific steady-state structure is directly and selectively reshaped by the end-neuron-specific chain motifs, since they determine how strongly the neural variability $\kappa_r$ contributes to the population-averaged pre-activation $K_r$ (Fig.\ref{fig:fig_schematic}D).

Furthermore, we identify regulations of combined mean and motif components on autonomous dynamics. The modulations of $\tau_p \ (p=E,I)$ interact with the effects of mean connectivity in a more intricate way. Specifically, increasing \(\tau_I\) shifts the right transition line of \(g\) leftward rather than rightward (as in the homogeneous case, Fig.~\ref{fig:homo_fig}D), thereby shrinking the bistable regime within the inhibition-dominant mean connectivity regime (Fig.~\ref{fig:hetero_auto_EI_fig}C from the left subpanels to the right).
The asymptotics as \(J^0\to+\infty\) are determined flexibly by strengths of chain motifs rather than being the exact value  \(g=\gamma^{-1}\). 
The relative levels of $\tau_E$ and $\tau_I$ thus critically determine network excitability within each mean connectivity regime.

So, in networks with homogeneous global inhibitory mean connectivity \(-gJ^0\), chain motifs ending in the inhibitory population suppress bistability, driving the system toward quiescence.
However, introducing heterogeneous mean inhibitions allows these inhibitory-ending chain motifs to also support bistable states (Supp. Fig.\ref{supp_fig:hetero_auto_tauI}). 
To examine this, we now focus on a special case of cell-type-specific heterogeneity in mean connectivity and chain motifs: \(J_{II}^0=0\) with \(\tau_{III}>0\) only (see  Table~\ref{tab:EI_connectivity_vectors} -- Hetero.~2).
This configuration moreover reflects the sparse and correlated couplings between interneurons in empirical investigations \cite{biswas2025inhibitory,fishell2020interneuron,pfeffer2013inhibition,lee2013disinhibitory,pi2013cortical}. 

In this special case, the inhibitory neuron population activity is characterized by the following self-consistent equations:
\begin{equation}\label{eq:auto_hetero_tauIII}
\begin{aligned}
\dot{\kappa}_2 &= -\kappa_2 +\alpha_I \langle\phi^{(I)}\rangle\\
\dot{K}_2 &= -K_2 + N_EJ^0\langle\phi^{(E)}\rangle + N\sigma^2\tau_{III} \kappa_2 \alpha_I\langle\phi'{}^{(I)}\rangle.
\end{aligned}
\end{equation}
The dynamics in Eq.~\eqref{eq:auto_hetero_tauIII} form a positive feedback loop, where increasing \(\tau_{III}\) induces a saddle-node bifurcation. Only the inhibitory population exhibits a high-firing state, while the excitatory population remains quiescent (Fig.\ref{fig:tau_iii_reponse_fig}A). 
Consequently, the second-order chain motif on its own can initiate self-excitation (analogous to the disinhibition effect) in the target neuron population upon the removal of global inhibition.
To mechanistically understand how the combined mean \(J_{II}^0\) and chain-motif strength \(\tau_{III}\) within interneurons impact network dynamics, we conduct the stability analysis in the plane \(J^0-g_I\), where \(g_I\) is specified for \(J_{II}^0\). Results elucidate that either maintaining a minimal mean \(J_{II}^0\to 0^+\) or increasing \(\tau_{III}\) ensures the existence of bistability~(Fig.~\ref{fig:tau_iii_reponse_fig}B, bottom-left lighter blue areas).

\vspace{6pt}
Beyond modulating autonomous dynamics by connectivity, we moreover tame neural population dynamics with structured external inputs.
We apply external inputs \(\boldsymbol{h}\delta_h\) with 
\small{\(h_i = -\eta_i^{(II)}/\sqrt{N_I}\)} for inhibitory neuron \(i\) only. 
Since \(\boldsymbol{h}\) aligns with the existing connectivity structure \(\boldsymbol{u}^{(2)}\)~(associated with \(\kappa_2\)), no additional orthogonal basis vector is introduced.
The new steady-state of \(\tilde{\kappa}_2\) accordingly becomes 
\begin{equation}
    \tilde{\kappa}_2 = \alpha_I\langle\phi(\tilde{\mu}_I,\tilde{\Delta}_I)\rangle - \delta_h/(N\sigma\sqrt{\alpha_I\tau_{III}}),
\end{equation}
leading to a negative latent response function 
\(\chi_{2h}^{\kappa}=(\tilde \kappa_2-\kappa_2^*)/\delta_h<0\).
This negative response function indicates a reduction in the activity variability within the inhibitory neuron population.
Beyond directly modulating the variability \(\kappa_2\), this external input indirectly impacts the mean activity of inhibitory neurons \(K_2\) through correlated chain motifs (Eq.~\eqref{eq:auto_hetero_tauIII}, third term in \(K_2\) dynamics).

A thorough investigation of the bifurcation diagram reveals that the aligned external input vector shifts the bifurcation point along the \(\tau_{III}\) axis.
As external input strength increases, stronger synaptic correlations within the inhibitory neurons are necessary to trigger the dynamical transition from the monostable state, a state with lower \(\kappa_2,~K_2\), to the bistable regime (Fig.\ref{fig:tau_iii_reponse_fig}C). 
Examination along the external input axis demonstrates that, in a network autonomously residing in the bistable regime, \(\kappa_2,~K_2\) for inhibitory neurons decrease as the magnitude of structural input increases. 
When sufficiently strong input is applied for a sufficiently long duration, the steady-state dynamics are pushed across the bifurcation transition, landing the network back in the monostable regime (Fig.\ref{fig:tau_iii_reponse_fig}D and F).
We additionally simulate temporal network dynamics in response to a transient pulse input. Upon delivery of the aligned input, the variable \(\kappa_2\) instantaneously decreases, followed by a decrease in the mean inhibitory neural activity \(K_2\). Once the input is withdrawn after a short duration, the inhibitory neural activity is effectively captured by the high activity basin, restoring the system to its original dynamics (Fig.\ref{fig:tau_iii_reponse_fig}E). Due to trial-to-trial variability, a small portion of simulations escape and are drawn to the low activity state.

Heterogeneous chain-motif correlations critically shape network dynamics. By establishing distinct synaptic pathways, these motifs differentially integrate and propagate microscale variability across scales, thereby sculpting the population-specific neural dynamics.
Beyond internal recurrent connectivity alone, external inputs fundamentally modulate these dynamics through their interaction with recurrent connectivity structure. This input-connectivity interplay might reflect how networks operate in naturalistic environments, where sensory inputs continuously interact with circuit structures. 
\vspace{-12pt}
\section{\label{secs:exp-application}Mean-variability interactions in experience-dependent neural population responses}
\vspace{-8pt}
\subsection{\label{subsec:biological_network}Biologically inspired multipopulation network model architecture}
\begin{table*}[htb]
\caption{Connectivity Vectors of the Multi-population Network}
\label{tab:connectivity_vectors}
\begin{tabular}{c | c | c | c}
\hline\hline
 & Pyr cell population & SOM cell population & VIP cell population\\
\hline\hline
\(u_i^{(1)}\)& \(\sigma\sqrt{N\vert\tau^r\vert}\eta_i^{(e)}\)&\(-\sigma\sqrt{N|\tau_{vse}|}\eta_i^{(ve)}\)&0\\
\hline
\(v_i^{(1)}\) & \(C_f(\sigma\sqrt{N|\tau^r|})\eta_i^{(e)}\)& 0 &0\\
\hline
\(u_i^{(2)}\) & 0 & 0& \(\sigma\sqrt{N\vert\tau_{vvs}\vert}\eta_i^{(vs)}\) \\
\hline
\(v_i^{(2)}\)  & 0& 1 & 0\\
\hline
\(u_i^{(3)}\)  & 0& \(\sigma\sqrt{N|\tau_{vsv}|}\eta_i^{(vv)}\)  & 0\\
\hline
\(v_i^{(3)}\)  & 0& 0 & 1\\
\hline
\(m_i^{(1)}\) & 1 &0 & 0 \\
\hline
\(n_i^{(1)}\) & \(NJ_{EE}^0\) &\(NJ_{ES}^0\)&\(NJ_{EV}^0\) \\
\hline
\(m_i^{(2)}\) & 0 &1 & 0 \\
\hline
\(n_i^{(2)}\) & \(NJ_{SE}^0\) &\(NJ_{SS}^0\)&\(NJ_{SV}^0\) \\
\hline
\(m_i^{(3)}\) & 0 &0 & 1 \\
\hline
\(n_i^{(3)}\) & \(NJ_{VE}^0\) & \(\sigma\sqrt{N}(\sqrt{|\tau_{vse}|}\eta_i^{(ve)} + \sqrt{|\tau_{vsv}|}\eta_i^{(vv)})+NJ_{VS}^0\)  & \(\sigma\sqrt{N|\tau_{vvs}|}\eta_i^{(vs)}+NJ_{VV}^0\) \\
\hline
\hline\hline
\end{tabular}
\end{table*}

\begin{figure*}[htb]
  \centering 
\includegraphics[width=0.95\linewidth]{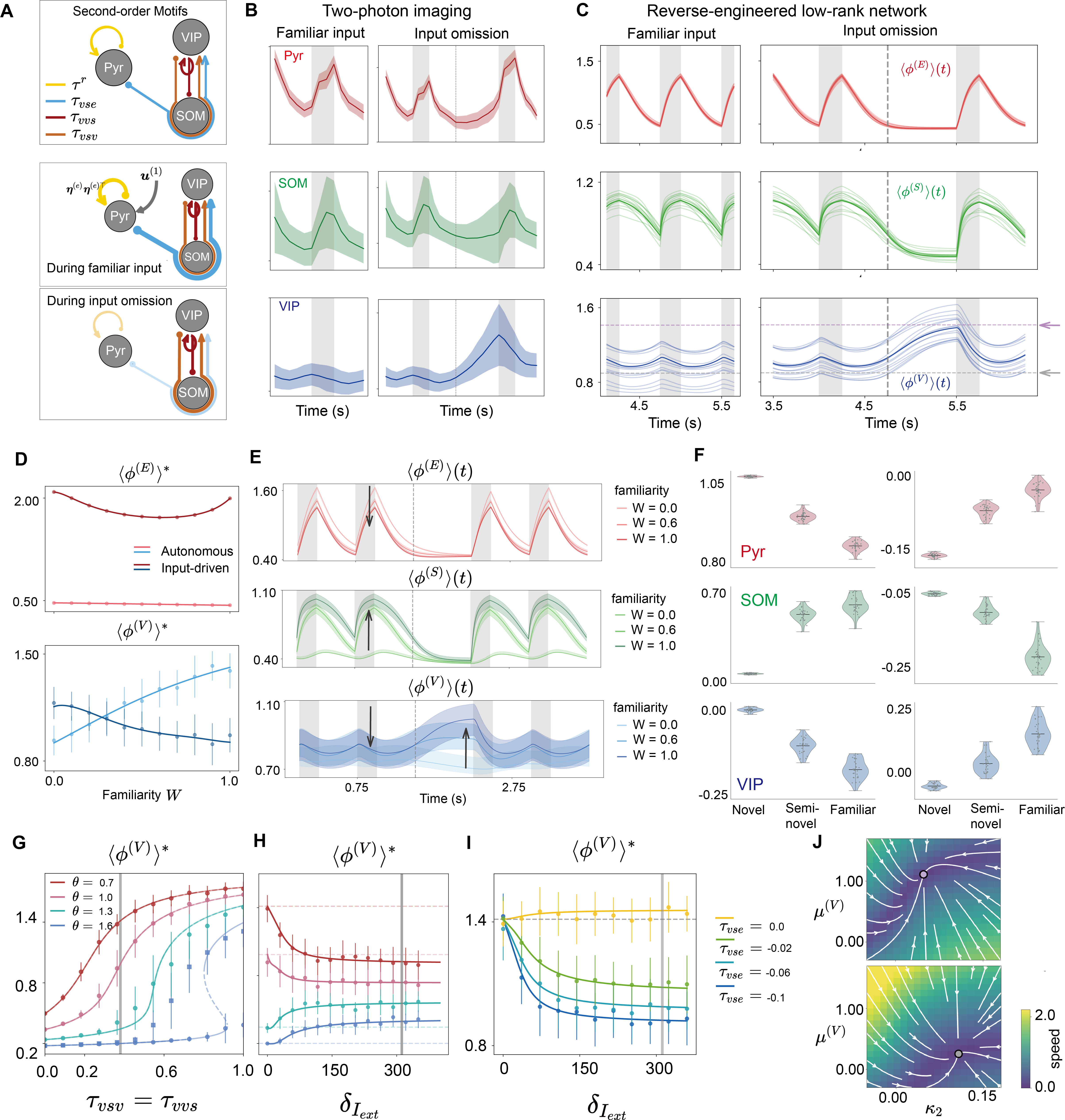}
\caption{{\bf Cell-type-specific motif statistics predict the experience-shaped population neuron dynamics.} (A) Upper panel: schematic of our V1 three-population low-rank recurrent neural networks, exhibiting the population-specific chain motifs $\tau_{vse}$, $\tau_{vsv}$ and $\tau_{vvs}$, and the reciprocal motif $\tau^r$ within Pyr neuron population. Bottom panel: contribution of distinct motifs. During familiar input, the recurrent-aligned patterned input enhances the preactivation of Pyr neurons and VIP neurons are suppressed through a negative $\tau_{vse}$. 
During input omission, positive $\tau_{vsv}$ and $\tau_{vvs}$ contribute to the ramping behavior of VIP neuron population. 
(B) Transient activity of Pyr, SOM and VIP neuron populations (two-photon imaging) during detection-of-change task under familiar input (left) and its omission (right).
(C) Population-averaged firing rate of Pyr, SOM and VIP neurons $\langle\phi^{(E)}\rangle(t)$, $\langle\phi^{(S)}\rangle(t)$ and $\langle\phi^{(V)}\rangle(t)$ are presented under conditions incorporating reciprocal motifs $\tau_r=0.001$, and chain motifs $\tau_{vse}=-0.1$, $\tau_{vsv}=0.38$, $\tau_{vvs}=0.38$. The input strength of the mean part $\mu_{I_{ext}} = 1.8$, and the input strength of the structured random part $\delta_{I_{ext}}=312$.
The purple dashed line represents the autonomous steady state and the gray dashed line represents the input-driven steady state, with input parallel to $\boldsymbol{u}^{(1)}$.
(D) Steady-state change of population-averaged firing rate $\langle\phi^{(E)}\rangle$ and $\langle\phi^{(V)}\rangle$ as a function of familiarity $W$. The input strength of the mean part $\mu_{I_{ext}} = 1.8$ and the input strength of the random part is fixed at $\delta_{I_{ext}}=312$.%
(Caption next page.)}
\label{fig:experience}
\end{figure*}

\addtocounter{figure}{-1}
\begin{figure*} [t!]
  \caption{(Previous page.)
  (E) Transient activity of Pyr, SOM and VIP neurons under different values of $W$. 
  (F) Illustration of the distributions of numerically obtained difference values for $\langle\phi^{(E)}\rangle$, $\langle\phi^{(S)}\rangle$, and $\langle\phi^{(V)}\rangle$.
  Left: differences are computed as the value at the stimulus onset time step minus the value at the stimulus offset time step. 
  Right: differences are computed as the value at the expected stimulus onset time step during omission minus the value at the subsequent actual stimulus onset time step. 
  In the violin plots, black horizontal lines indicate the mean, and gray upper‑lower lines indicate the maximum and minimum values across all simulations. Individual simulation data points are overlaid. Pairwise one‑tailed Mann–Whitney U tests with Bonferroni correction indicate significant directional differences for all group comparisons (adjusted $p < 0.001$).
  (G) Bifurcation of the steady-state firing rate $\langle\phi^{(V)}\rangle$ with simultaneous increases in $\tau_{vsv} = \tau_{vvs}$ analyzed under varying nonlinearity thresholds $\theta$. \(\tau_{vse}=-0.1,~\delta_{I_{ext}}=0\). The vertical gray line shows the $\tau_{vsv}=\tau_{vvs}=0.38$, which is adopted in panel (C).
  (H) Change in steady-state firing rate as a function of increasing patterned input strength $\delta_{I_{ext}}$ of the patterned input, under varying nonlinearity thresholds $\theta$. \(\tau_{vse}=-0.1,~\tau_{vsv}=\tau_{vvs}=0.38\). The vertical gray line shows the input strength of the structured random part $\delta_{I_{ext}}=312$, which is adopted in panel (C).
  (I) Same as in (H), but examined under varying values of $\tau_{vse}$. \(\tau_{vsv}=\tau_{vvs}=0.38,~\theta=0.7\). 
  The vertical gray line shows the input strength of the structured random part $\delta_{I_{ext}}=312$, which is adopted in panel (C).
  (J) Velocity field in plane $\kappa_2-\mu^{(V)}$ ($\kappa_2- K_3$). 
  Upper: velocity field without input. The purple dot represents the autonomous steady state (purple dashed line in panel (C)).
  Bottom: velocity field under patterned input parallel to $\boldsymbol{u}^{(1)}$.  The gray dot represents the input-driven steady state, with input parallel to $\boldsymbol{u}^{(1)}$ (gray dashed line in panel (C)).
  The darkest/solid lines in panels (C-E) and (G-I) are theoretical results derived directly from the latent quantity dynamics, the lighter lines, dots, and shaded regions are numerical results obtained from low-rank network simulations.
  Other parameters: $C_f=0.1$, $\sigma=0.4$, $b=1.1$ and $\theta=0.7$ in nonlinearity $\phi(\cdot) = b+\text{tanh}(x-\theta)$, $\alpha_E=0.8, \alpha_S=\alpha_V=0.1$. 
}
\end{figure*}

We next ask whether the mean–variability~(\(\kappa-K\)) mechanism developed above can provide a circuit-level account of heterogeneous population responses observed in the mouse primary visual cortex. Guided by experimentally observed cell-type-specific motif statistics, we construct a multipopulation network and examine whether the resulting connectivity pathways, consistent with the observed motif statistics, can reproduce key qualitative features of the recorded population activity. Here the aim is not to infer or even imply a unique underlying circuit from population activity, but to use our theoretical framework to identify a structurally interpretable mechanism linking fine-scale connectivity statistics to heterogeneous population dynamics.

We focus on pyramidal~(Pyr), vasoactive intestinal peptide-positive~(VIP), and somatostatin-positive~(SOM) neurons recorded in mouse VISp during a visual change-detection task~\cite{garrett2020experience,garrett2023stimulus}. Visual images were presented as familiar or unfamiliar stimulus sets, and the different cell populations exhibited distinct activity patterns across stimulus conditions. Here we concentrate on responses to the familiar image set and to stimulus omissions. In the experiments, Pyr cells show stimulus-evoked activity that ramps across repeated familiar presentations, whereas VIP cells exhibit a different temporal profile: their activity rises during inter-stimulus intervals and is suppressed during stimulus presentation. During stimulus omissions, Pyr responses are strongly reduced, while VIP activity continues to rise through the extended inter-stimulus interval. SOM cells are included both as an experimentally recorded population and as an intermediate population in the connectivity architecture developed below.

To examine whether these population-specific response patterns can arise through the mean–variability mechanism developed above, we model the circuit as three interacting populations -- Pyr, SOM, and VIP (\(P=3\)). Under our general framework, the corresponding structured connectivity has rank \(R=2 P=6\), with three pairs of connectivity directions, \(\{\boldsymbol{u}^{(r)},\boldsymbol{v}^{(r)},\boldsymbol{m}^{(r)},\boldsymbol{n}^{(r)}\}_{r=1,2,3}\), associated with the three cell populations. Guided by experimentally observed connectivity statistics, we incorporate reciprocal structure within the Pyr population and strong chain-type correlations involving the SOM and VIP populations. These structural components define the six connectivity-vector pairs used below~(Table~\ref{tab:connectivity_vectors} ). We will examine their dynamical roles separately and, subsequently, their combined effects on the recorded population responses.

\subsection{\label{subsec:adapt_pyrvip}Computational principles underlying adaptive Pyr and VIP activity}
\subsubsection{\label{subsubsec:tonic_familiar}Feedforward chain motifs underlie Pyr-VIP population responses to familiar stimuli}
We begin by elucidating the factor that underlies the nontrivial, yet distinct, Pyr and VIP population activities observed during the presentation of externally familiar stimuli \(\boldsymbol{I}^{ext}\), with uniform contrast magnitude \(\mu_{I_{ext}} = \langle I^{ext}_i\rangle_{i\in Pyr}\).
Motivated by the analytical findings in Sec.~\ref {subsec:heterogeneousGauss} (Fig.~\ref{fig:tau_iii_reponse_fig}), the extent of familiarity can be characterized by the prior knowledge ingrained in the internal network connectivity and dynamics. Here, we operationalize this concept by first assuming that the variability in the external input is aligned with the variability in the connectivity vectors: \(\boldsymbol{I}^{ext} -\mu_{I_{ext}}\boldsymbol{m}^{(1)}=\delta_{I_{ext}} \boldsymbol{u}^{(1)}/\Vert\boldsymbol{u}^{(1)}\Vert\).
% with \(\boldsymbol{u}^{(1)}/\Vert\boldsymbol{u}^{(1)}\Vert\) the normalized structural vector.
The dynamics of total synaptic inputs received by the neuron indexed \(i\) followed 
\begin{equation}\label{eq:LD_manifold_app}
x_i(t) = \sum_{r=1}^3\kappa_ru_i^{(r)} + \sum_{r=1}^3K_{r}m_i^{(r)}.
\end{equation}

Furthermore, we question the origin of this alignment between environmental information and connectivity variability. Motivated by the work of modern Hopfield networks on associative memory \cite{krotov2016dense,clark2025transient,whittington2025tale,shan2025graph,pachitariu2026critical}, and a series of experimental investigations of the symmetry of network connectivity in learning systems \cite{garrett2020experience,saxena2025enriched,horton2024excitatory}, we argue that the aligned connectivity structure \(\boldsymbol{u}^{(1)}\) arose from learning and plasticity\cite{miehl2023formation}. 
Consequently, we redesign \(\boldsymbol{v}^{(1)}\) to be symmetric to \(\boldsymbol{u}_{E}^{(1)}\) rather than using a uniform vector \cite{sherf2025complexity}.
With these presumptions, we characterize the temporal evolution of the latent quantity \(\kappa_1\) as 
\begin{equation}\label{eq:hblearn_var}
\dot{\kappa_1}= -\kappa_1 + \alpha_EC_fN\sigma^2\tau^r\langle\phi'^{(E)}\rangle\kappa_1 + \delta_{I_{ext}}/\Vert\boldsymbol{u}^{(1)}\Vert.
\end{equation}
The familiar external stimuli do not directly influence the mean activity of the population but instead enhance population variability through structural alignment, thereby reflecting the learnt information. 

To transfer this enhanced variability to influence population-averaged activities -- in this example, the mean activity of VIP cells \(K_3\) -- chain motifs are employed. 
We return to the VISp connectome dataset (publicly available connectivity data from Allen Institute for Brain Science  \href{https://brain-map.org/}{https://brain-map.org/}), and find that the strength of chain motifs, which begins at Pyr cells and relays at SOM cells, has a significant value\cite{dahmen_strong_2026} (Supp.~Fig.~\ref{suppfig:pop3_transient_mean}A). 
We accordingly assign a negative correlation \(\tau_{vse}<0\) to the over-representation of this kind of chain motifs that targets VIP cells~(begins at Pyr, relays at SOM and targets at VIP; \(J^{se}_{kj}>0\) followed by \(J_{ik}^{vs}<0\) leads to a negative correlation) (Fig.~\ref{fig:experience}A, Fig.~\ref{fig:motif_ablation}C,~D).
The mean–variability interaction between the variability~\(\kappa_1\) of Pyr and the mean activity \(K_3\) of VIP is governed by \(\tau_{vse}\)  via
\begin{equation}
-N_S\sigma^2|\tau_{vse}|\langle\phi'^{(S)}\rangle\kappa_1. 
\end{equation}
Therefore, aligned structured external inputs enhance Pyr population variability~\(\kappa_1\). This enhanced variability propagates through feedforward chain motifs to suppress VIP mean activity~
(Fig.~\ref{fig:experience}B and C, the left columns).
VIP inhibition of Pyr is then conveyed through mean connectivity.
Due to \(J_{SV}^0<0\), the diminished mean activation of VIP population
alleviates the inhibition on the SOM neuron population, 
which subsequently inhibits the Pyr cells through \(J_{ES}^0<0\) (Fig.~\ref{fig:experience}E, arrows from lighter to darker areas).

Therefore, during the presentation of the familiar tonic input signals, the aligned internal connectivity  processes the familiar signal, generating increased variability in Pyr population activity.  This variability is propagated to and suppresses VIP mean activity through mean-variability interaction regulated by chain motifs \(\tau_{vse}\).
Subsequently, mediated by the VIP-to-SOM-to-Pyr disinhibitory averaged connectivity circuit, reduced VIP inhibition leads to decreased Pyr mean population activity.

\subsubsection{\label{subsubsec:tonic_ITI} Recurrent chain motifs among inhibitory populations drive VIP ramping during stimulus intervals and omission}
\begin{figure*}[t]
\centering
\includegraphics[width=0.95\linewidth]{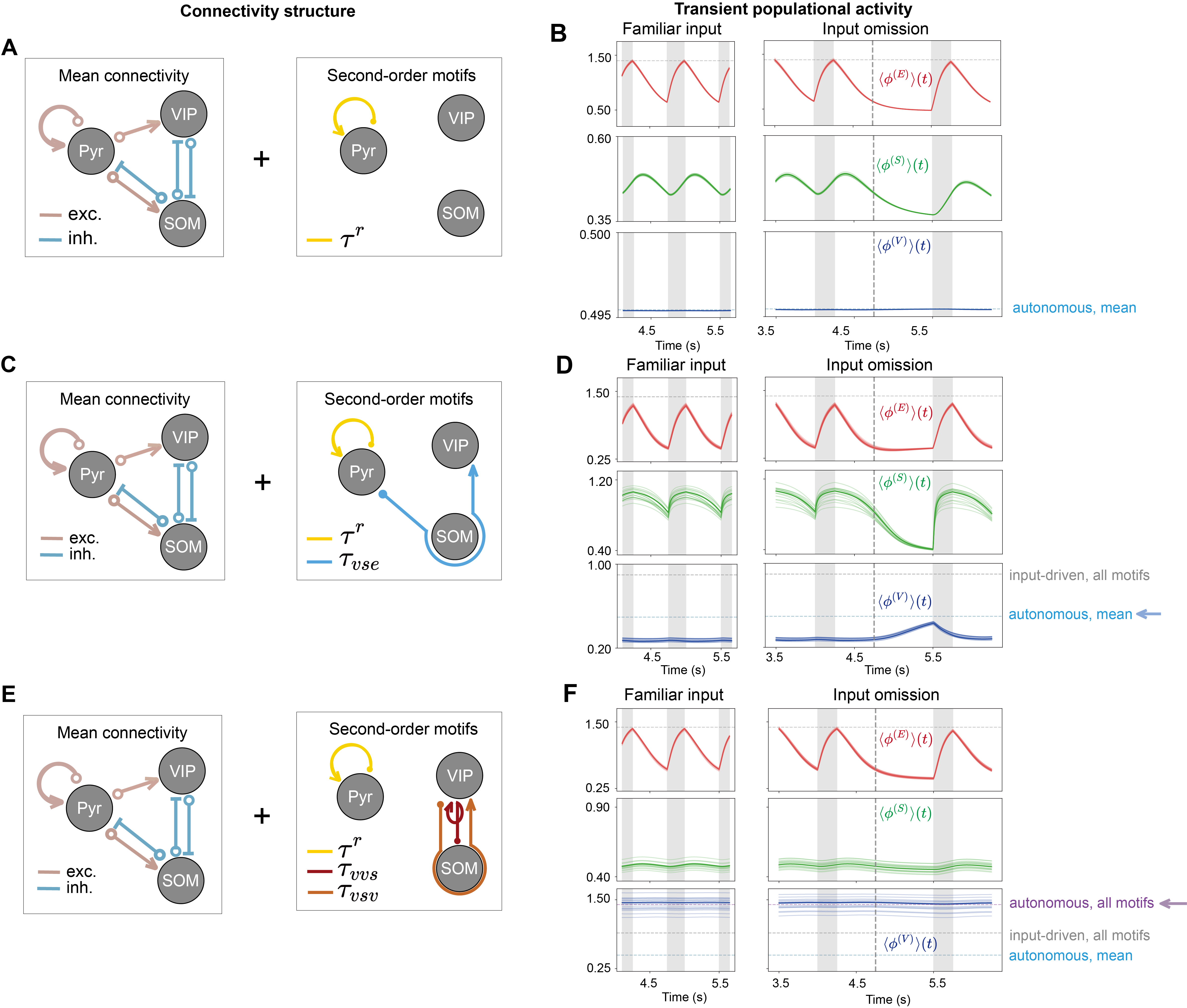}
\caption{{\bf Dissection of input integration pathways via motif ablation.} (A, C, E) Schematic depicting circuit connectivity following different motif ablation manipulations. (B, D, F) Transient activity profiles evoked by familiar stimuli
and input omission under each ablation condition.
The purple dashed line represents the autonomous steady state with mean connectivity and heterogeneous motifs, and the gray dashed line represents the input-driven steady state, with input parallel to $\boldsymbol{u}^{(1)}$. The blue dashed line represents the autonomous steady state with mean connectivity only.
(B) Transient activity of networks with reciprocal motifs within Pyr neuron population $\tau^r$ only. (D) Transient activity of networks with reciprocal motifs within Pyr neuron population $\tau^r$, and chain motifs $\tau_{vse}$. (F) Transient activity of networks with reciprocal motifs within Pyr neuron population $\tau^r$, and chain motifs $\tau_{vsv}$ and $\tau_{vvs}$. Parameter settings follow Fig.~\ref{fig:experience}C.
}
\label{fig:motif_ablation}
\end{figure*}

Next, we investigate the mechanisms underlying VIP self-excitation during the inter-stimulus interval.
Recalling that in homogeneous simple EI networks~(Sec.~\ref {subsec:homogeneousGauss}), autonomous self-excitation dynamics emerge following a saddle-node bifurcation when positive chain motifs are over-represented. 
We also observe that in the connectome dataset, although the averaged connectivity across inhibitory interneuron populations is weak, individual connections are highly correlated\cite{dahmen_strong_2026,pfeffer2013inhibition}~(Supp.~Fig.~\ref{suppfig:pop3_transient_mean}A).
We propose that the strong chain motifs found between two interneuron populations, SOM and VIP cells, carry out this function (Figs.~\ref{fig:experience}A and~\ref{fig:motif_ablation}E,~F).
Specifically targeting the VIP cells, we assume significant positive chain-motif correlations \(\tau_{vsv}\) and \(\tau_{vvs}\), which contribute to the mean population activity of VIP cells \(K_3\) through the terms \(N_S\sigma^2\tau_{vsv}\langle\phi'^{(S)}\rangle\kappa_3\) and \(N_V\sigma^2\tau_{vvs}\langle\phi'^{(V)}\rangle\kappa_2\), respectively, as shown in the latent dynamics
\begin{equation}
\begin{aligned}
\dot \kappa_3 &= -\kappa_3 \underline{+\alpha_V \langle\phi(K_3,\Delta^{(V)})\rangle}, \\
\dot K_3 &= -K_3 + \bar K_V  - N_S\sigma^2|\tau_{vse}|\langle\phi'^{(S)}\rangle\kappa_1\\
&\underline{+N_S\sigma^2\tau_{vsv}\langle\phi'^{(S)}\rangle\kappa_3}\underline{+N_V\sigma^2\tau_{vvs}\langle\phi'^{(V)}\rangle\kappa_2}.
\end{aligned}
\end{equation}
The underlined terms form a positive feedback loop: the inhibitory neuron variability~(\(\kappa_3\) for SOM and \(\kappa_2\) for VIP), directly influences VIP population mean activity \(K_3\)~(Eq.~\eqref{eq:mu_delta2_pop3}),  thereby sustaining autonomous VIP self-excitation.

Building on the analysis in Sec.~\ref{subsubsec:tonic_familiar} and in this section, we now elucidate the computational principles behind VIP cell ramping activity during unexpected stimulus omissions.
Two \emph{competing pathways} govern VIP cell dynamics: first, strong chain motifs within triplets of inhibitory interneurons drive autonomous self-excitation in VIP cells, subsequently, the familiar stimulus-induced variability in Pyr activity propagates through negative feedforward chain motifs, thus suppressing VIP mean activity. 
These two pathways function as competing opposites, \emph{waxing and waning} as the external environment and internal dynamic landscape evolve.
During omissions in familiar stimulus sessions, the expected feedforward Pyr suppressions fail to arrive~(variability \(\kappa_1\)), disrupting the rhythmic ramping-and-decay population activity pattern
\begin{equation}
\begin{aligned}
\dot K_3 &= -K_3 + \bar K_V  \cancel{- N_S\sigma^2|\tau_{vse}|\langle\phi'^{(S)}\rangle\kappa_1}\\
&+N_S\sigma^2\tau_{vsv}\langle\phi'^{(S)}\rangle\kappa_3+N_V\sigma^2\tau_{vvs}\langle\phi'^{(V)}\rangle\kappa_2,
\end{aligned}
\end{equation}
leaving the autonomous ramping to dominate~(Fig.\ref{fig:experience}B and C, the right columns).

This autonomous ramping, normally emerging concurrently with stimulus-suppressed responses, is mechanistically distinct from the naive ramping activity induced by external input signals (Supp.~Fig.~\ref{suppfig:pop3_transient_mean}B and C). 
We modulate the balance between orthogonally arbitrary random input and correlated motif-driven input using the weight parameter $W$,  
characterized by the ratio $\sqrt{1-W^2}/W$.  Theoretical investigations and numerical simulations demonstrate that as $W\to0.0$, the system approaches a simple uniform-stimulus-driven ramping pattern~(Fig.~\ref{fig:experience}D-F), closely approximating the empirical responses observed for novel images. However as \(W\to 1.0\), self-excitation and stimulus-suppression emerge \cite{garrett2020experience,garrett2023stimulus} (See Appendix \ref{app:pyr-som-vip}4 for details).

\subsubsection{A dynamical system description}
The dual success of cell-type-specific motifs in reproducing empirical observations and the mean-variability framework in providing intuitive interpretations motivates a detailed, mechanistic dynamical systems analysis of the underlying principles.

Titrating the temporal response of the VIP neuron population is largely determined by the combinatorial effects of internal neuronal and synaptic properties and external structured inputs.
The VIP cells have a higher population mean activity as the strengths of the interneuron chain motifs (\(\tau_{vsv},~\tau_{vvs}\)) increase, across all biophysical neuronal parameters \(\theta\) examined (\(\theta\) in \(\phi(x)=b+\tanh(x-\theta)\)).
When \(\theta\) is sufficiently large~(e.g., \(\theta=1.6\)), VIP mean firing rate demonstrates the hysteresis phenomenon, similar to what we observed in networks with homogeneous chain-motif strength (Fig.~\ref{fig:experience}G, Supp.~Fig.~\ref{suppfig:thres_cusp_latent_phi}).

External random input parallel to $\boldsymbol{u}^{(1)}$ also influences the steady-state dynamics. 
When \(\theta\) is large and the autonomous steady-state dynamic \(\langle \phi^{(V)}\rangle\) is of a low magnitude, a rise in the intensity of the parallel random input, therefore elevating dynamic variability $\kappa_1$, leads to an increase in $\langle\phi^{(V)}\rangle$.
Conversely, increasing \(\delta_{I_{ext}}\) leads to a reduction in $\langle\phi^{(V)}\rangle$.
The latter scenario, in which VIP mean activity declines upon receiving external structured input, aligns with experimental findings thus validating the selection of a low value for \(\theta\) (Fig.~\ref{fig:experience}H, Supp.~Fig.~\ref{suppfig:response_kappa1_thres_latent_phi}).
We further analyze the combined effect of the feedforward chain motifs \(\tau_{vse}\) and the external random input \(\delta_{I_{ext}}\), under the selected \(\theta=0.7\). 
In the presence of the negative feedforward chain \(\tau_{vse}\), the external structured input suppresses VIP population activation, as expected. The stronger the negative feedforward chain motifs, the stronger the suppression of VIP mean activity (Fig.~\ref{fig:experience}I, Supp.~Fig.~\ref{suppfig:response_kappa1_tauvse_latent_phi}).

We subsequently outline the temporal velocity fields in \(\mu^{(V)}(K_3)-\kappa_2\) space under time-varying input circumstances, thereafter examining neuronal population dynamics on a shorter timescale.
The selection of the two-dimensional space accounts for the fact that VIP mean activity and variability are directly defined by \(K_3\) and \(\kappa_2\), respectively.
The flow of the velocity field changes with fluctuations in within-trial input conditions (Fig.~\ref{fig:experience}J). 
In the absence of external input (e.g., inter-stimulus interval), the upper region displays slower flow, creating an attractive basin that corresponds to a high activity state.
Upon receiving structured input, the lower region slows down, resulting in a distinct attractive basin corresponding to a relatively lower activity state.
This switching provides a clear dynamical explanation for the emergence of heterogeneous neuronal responses.

Together, network simulations and mathematical analysis support the biological relevance of our theoretical framework.
Multiscale  activity patterns across multiple populations can be captured by recurrent networks incorporating cell-type-specific synaptic motifs, demonstrating the sufficiency of the structural mechanisms. 
Furthermore, the theoretical framework reveals how mean-variability interactions can mediate experience-dependent neural heterogeneity and information flows, through the interplay between external stimuli and intrinsic dynamics.

\section{\label{sec:discussion}Discussion}
How microscopic connectivity structures contribute to macroscopic neural dynamics and neural computations remains a long-standing question\cite{dahmen2025heterogeneity}.
Here we show that locally correlated synaptic motifs -- specifically, chain motifs -- serve as the critical substrate that regulates neural coordination patterns underlying information integration, transformation, and ultimately, functional computations. 
We developed a motif-based low-rank representation of recurrent connectivity, which simultaneously captures correlated in-/out-degree statistics, and integrated it with a mean-field framework of nonlinear neural network activity. The resulting mathematical characterization of temporal dynamics spans scales from single-unit variability through population-averaged excitability to the coordinated activity of multiple neuronal populations underlying task-related computations. 
Interestingly, we demonstrated  that the latent dynamics separate into variables associated with population-averaged activity and variables controlling structured within-population variability, allowing microscopic connectivity to influence population activity means through nonlinear gain.
Network simulations confirmed the existence of a rich and diverse repertoire of dynamical states, and our theoretical framework captures the transitions among these states and their dynamical evolutions.
Our findings are consistent with recent work on spiking neural networks (SNNs), Ref.~\cite{biswas2025inhibitory}, which demonstrates that EI population–specific chain motifs and in-/out-degree distributions can strongly regulate synchronization. 
This convergence suggests that the dynamical consequences of chain-motif structure may extend across different neural models and implementation schemes. Furthermore, our theoretical framework provides a mechanistic basis for analytically examining whether the mean-variability mechanism developed here persists in spiking networks.

The extensive repertoire of cortical dynamics is widely hypothesized to emerge from the interplay among cortical connectivity, the intrinsic dynamical properties of individual neurons, and external inputs.
Established theoretical frameworks have focused on two complementary approaches. Random-network models emphasize statistical connectivity structure and provide analytical access to collective dynamics, whereas low-rank recurrent networks emphasize a small number of interpretable global modes associated with population activity and computation. 
Our framework bridges these perspectives through the lens of synaptic motifs, enabling bidirectional inquiry: forward prediction of heterogeneous dynamical states emerging from specific connectivity structures and, conversely, constrained construction of network connectivity consistent with empirically observed neural activity. In the mouse V1 example, we use experimentally observed motif statistics to construct a Pyr–SOM–VIP network in which distinct motif pathways reproduce key qualitative features of familiar-stimulus and omission responses. This demonstration of sufficiency suggests that a fruitful future direction is to incorporate richer connectome constraints to further constrain the effective connectivity and structural features underlying more complex population dynamics and task-related computation.

Matrix decompositions are generally not unique, and the same is true for higher-order tensor decompositions \cite{ballard1945tensor}. Our construction therefore should not be interpreted as recovering a unique decomposition of the connectivity matrix. Rather, it provides a statistically consistent and dynamically interpretable representation in which population mean activity and structured within-population variability correspond to distinct latent coordinates.
We exploit this flexibility in decomposition choice to construct a low-rank representation requiring only the orthogonalization of left connectivity vectors thus separating the latent dynamical directions. In this representation, population mean-activity and structured within-in population variability correspond to two distinct connectivity vectors, and consequently, to two sets of latent dynamic quantities \(\boldsymbol{\kappa}\) and \(\boldsymbol{K}\).
More generally, this connects two important and complementary organizational principles for computation in heterogeneous networks: computations distributed across functionally distinct neuronal populations, and computations distributed across low-dimensional dynamical modes.

These computational conveniences and biological interpretability motivate future extensions of the framework to cortical information processing across multiple scales and functions.
Recent work \cite{hadaeghi2025computational} has identified a no-strong-loop principle in empirical directed connectomes supporting sequential information integration, complementing classical symmetry principles for memory storage in hippocampal networks. Our framework is currently developed at a single-modular scale. Extending it to multi-modular networks while incorporating multi-scale, region-specific, and computational-demand-specific connectivity principles would be an important next step for understanding how local motif structures contribute to interactions across cortical regions\cite{kusmierz2025hierarchy}.

Finally, we consider learning. 
In both biological and artificial neural networks, learning occurs via changes in connectivity. 
While learning can be directed by strategically modifying specific connectivity components to selectively recruit specific neural populations \cite{thibeault2024low}, this approach faces an inherent constraint: increasing learning complexity demands ever more computational populations, which cannot scale indefinitely.
We demonstrate that this limitation can be circumvented by imposing learning directly in the latent dynamical variables \cite{clark2026structure}. Therefore, instead of recruiting additional populations, this approach reuses existing ones for different computations or across different learning phases by reshaping their coordination patterns, an idea similar to mixed selectivity \cite{tye2024mixed}. 
Within the present framework, a natural next step is therefore to ask how learning reshapes motif statistics and the associated mean–variability dynamics. Incorporating explicit synaptic plasticity would provide a route toward connecting changes in microscopic connectivity with the evolution of latent dynamical modes across diverse contexts 
\cite{schulz2021generation,wu2022regulation}.
In this view, learning may reshape not only mean connectivity but also the motif statistics that couple within-population variability to population-mean activity. Therefore, the mean–variability mechanism developed here provides a natural framework for studying how such microscopic synaptic reorganization changes low-dimensional network dynamics.

\begin{acknowledgments}
The authors would like to thank David Dahmen, Stefano Recanatesi, Eric Shea-Brown, and Srdjan Ostojic for valuable discussions during the early stages of this work. 
Y.S. was supported by funding from the French Government and implemented by the Agence Nationale de la Recherche (ANR) (Grant No. 26CPJ01Z6RCHX), the National Natural Science Foundation of China (Grant No. 32400936), and the NSF AccelNet IN-BIC program (Grant No. OISE-2019976).
We additionally acknowledge partial support of the National Science and Technology Innovation STI2030-Major Project No. 2022ZD0204600 and the Natural Science Foundation of China through Grant No. 31771147 (M.Z., J.Y., L.T.).
\end{acknowledgments}

% \clearpage
% \newpage
\bibliography{apssamp}% Produces the bibliography via BibTeX.

@article{dubreuil2022role,
  title={The role of population structure in computations through neural dynamics},
  author={Dubreuil, Alexis and Valente, Adrian and Beiran, Manuel and Mastrogiuseppe, Francesca and Ostojic, Srdjan},
  journal={Nature neuroscience},
  volume={25},
  number={6},
  pages={783--794},
  year={2022},
  publisher={Nature Publishing Group US New York}
}

@article{garrett2020experience,
  title={Experience shapes activity dynamics and stimulus coding of VIP inhibitory cells},
  author={Garrett, Marina and Manavi, Sahar and Roll, Kate and Ollerenshaw, Douglas R and Groblewski, Peter A and Ponvert, Nicholas D and Kiggins, Justin T and Casal, Linzy and Mace, Kyla and Williford, Ali and others},
  journal={elife},
  volume={9},
  pages={e50340},
  year={2020},
  publisher={eLife Sciences Publications, Ltd}
}

@article{shao2023relating,
  title={Relating local connectivity and global dynamics in recurrent excitatory-inhibitory networks},
  author={Shao, Yuxiu and Ostojic, Srdjan},
  journal={PLOS Computational Biology},
  volume={19},
  number={1},
  pages={e1010855},
  year={2023},
  publisher={Public Library of Science San Francisco, CA USA}
}

@article{shao2025impact,
  title={Impact of Local Connectivity Patterns on Excitatory-Inhibitory Network Dynamics},
  author={Shao, Yuxiu and Dahmen, David and Recanatesi, Stefano and Shea-Brown, Eric and Ostojic, Srdjan},
  journal={PRX Life},
  volume={3},
  number={2},
  pages={023008},
  year={2025},
  publisher={APS}
}

@article{miller2020generalized,
  title={Generalized paradoxical effects in excitatory/inhibitory networks},
  author={Miller, Kenneth D and Palmigiano, Agostina},
  journal={BioRxiv},
  pages={2020--10},
  year={2020},
  publisher={Cold Spring Harbor Laboratory}
}

@article{hu2018feedback,
  title={Feedback through graph motifs relates structure and function in complex networks},
  author={Hu, Yu and Brunton, Steven L and Cain, Nicholas and Mihalas, Stefan and Kutz, J Nathan and Shea-Brown, Eric},
  journal={Physical Review E},
  volume={98},
  number={6},
  pages={062312},
  year={2018},
  publisher={APS}
}

@article{dahmen_strong_2026,
	title = {Strong and localized recurrence controls the dimensionality of neural activity across brain areas},
	issn = {1546-1726},
	url = {https://doi.org/10.1038/s41593-026-02395-w},
	doi = {10.1038/s41593-026-02395-w},
	journal = {Nature Neuroscience},
	author = {Dahmen, David and Recanatesi, Stefano and Jia, Xiaoxuan and Ocker, Gabriel K. and Lahiji, Nilufar and Musall, Simon and Campagnola, Luke and Seeman, Stephanie and Jarsky, Tim and Helias, Moritz and Shea-Brown, Eric},
	month = aug,
	year = {2026},
}

@article{schuessler2020dynamics,
  title={Dynamics of random recurrent networks with correlated low-rank structure},
  author={Schuessler, Friedrich and Dubreuil, Alexis and Mastrogiuseppe, Francesca and Ostojic, Srdjan and Barak, Omri},
  journal={Physical Review Research},
  volume={2},
  number={1},
  pages={013111},
  year={2020},
  publisher={APS}
}

@article{ostojic2024computational,
  title={Computational role of structure in neural activity and connectivity},
  author={Ostojic, Srdjan and Fusi, Stefano},
  journal={Trends in Cognitive Sciences},
  volume={28},
  number={7},
  pages={677--690},
  year={2024},
  publisher={Elsevier}
}

@article{brunel2000dynamics,
  title={Dynamics of sparsely connected networks of excitatory and inhibitory spiking neurons},
  author={Brunel, Nicolas},
  journal={Journal of computational neuroscience},
  volume={8},
  number={3},
  pages={183--208},
  year={2000},
  publisher={Springer}
}

@article{horton2024excitatory,
  title={Excitatory and inhibitory synapses show a tight subcellular correlation that weakens over development},
  author={Horton, Sally and Mastrolia, Vincenzo and Jackson, Rachel E and Kemlo, Sarah and Machado, Pedro M Pereira and Carbajal, Maria Alejandra and Hindges, Robert and Fleck, Roland A and Aguiar, Paulo and Neves, Guilherme and others},
  journal={Cell Reports},
  volume={43},
  number={7},
  year={2024},
  publisher={Elsevier}
}

@article{biswas2025inhibitory,
  title={Inhibitory Motifs Quench Synchrony Induced by Excitatory Motifs in Biological Neuronal Networks},
  author={Biswas, Archishman and Kumar, Arvind},
  journal={bioRxiv},
  pages={2025--11},
  year={2025},
  publisher={Cold Spring Harbor Laboratory}
}

@article{pi2013cortical,
  title={Cortical interneurons that specialize in disinhibitory control},
  author={Pi, Hyun-Jae and Hangya, Bal{\'a}zs and Kvitsiani, Duda and Sanders, Joshua I and Huang, Z Josh and Kepecs, Adam},
  journal={Nature},
  volume={503},
  number={7477},
  pages={521--524},
  year={2013},
  publisher={Nature Publishing Group UK London}
}

@article{lee2013disinhibitory,
  title={A disinhibitory circuit mediates motor integration in the somatosensory cortex},
  author={Lee, Soohyun and Kruglikov, Illya and Huang, Z Josh and Fishell, Gord and Rudy, Bernardo},
  journal={Nature neuroscience},
  volume={16},
  number={11},
  pages={1662--1670},
  year={2013},
  publisher={Nature Publishing Group US New York}
}

@article{fishell2020interneuron,
  title={Interneuron types as attractors and controllers},
  author={Fishell, Gord and Kepecs, Adam},
  journal={Annual review of neuroscience},
  volume={43},
  number={1},
  pages={1--30},
  year={2020},
  publisher={Annual Reviews}
}

@article{tsodyks1997paradoxical,
  title={Paradoxical effects of external modulation of inhibitory interneurons},
  author={Tsodyks, Misha V and Skaggs, William E and Sejnowski, Terrence J and McNaughton, Bruce L},
  journal={Journal of neuroscience},
  volume={17},
  number={11},
  pages={4382--4388},
  year={1997},
  publisher={Society for Neuroscience}
}

@article{campagnola2022local,
  title={Local connectivity and synaptic dynamics in mouse and human neocortex},
  author={Campagnola, Luke and Seeman, Stephanie C and Chartrand, Thomas and Kim, Lisa and Hoggarth, Alex and Gamlin, Clare and Ito, Shinya and Trinh, Jessica and Davoudian, Pasha and Radaelli, Cristina and others},
  journal={Science},
  volume={375},
  number={6585},
  pages={eabj5861},
  year={2022},
  publisher={American Association for the Advancement of Science}
}

@article{hadaeghi2025computational,
  title={A computational perspective on the no-strong-loops principle in brain networks},
  author={Hadaeghi, Fatemeh and Fakhar, Kayson and Khajehnejad, Moein and Hilgetag, Claus C},
  journal={bioRxiv},
  pages={2025--09},
  year={2025},
  publisher={Cold Spring Harbor Laboratory}
}

@article{chau2025exact,
author = {Ho Yin Chau  and Kenneth D. Miller  and Agostina Palmigiano },
title = {Exact linear theory of perturbation response in a space- and feature-dependent cortical circuit model},
journal = {Proceedings of the National Academy of Sciences},
volume = {122},
number = {31},
pages = {e2426758122},
year = {2025},
doi = {10.1073/pnas.2426758122},
URL = {https://www.pnas.org/doi/abs/10.1073/pnas.2426758122},
eprint = {https://www.pnas.org/doi/pdf/10.1073/pnas.2426758122}
}

@article{krotov2016dense,
  title={Dense associative memory for pattern recognition},
  author={Krotov, Dmitry and Hopfield, John J},
  journal={Advances in neural information processing systems},
  volume={29},
  year={2016}
}

@article{whittington2025tale,
  title={A tale of two algorithms: Structured slots explain prefrontal sequence memory and are unified with hippocampal cognitive maps},
  author={Whittington, James CR and Dorrell, William and Behrens, Timothy EJ and Ganguli, Surya and El-Gaby, Mohamady},
  journal={Neuron},
  volume={113},
  number={2},
  pages={321--333},
  year={2025},
  publisher={Elsevier}
}

@article{clark2025transient,
  title={Transient dynamics of associative memory models},
  author={Clark, David G},
  journal={arXiv preprint arXiv:2506.05303},
  year={2025}
}

@article{shan2025graph,
  title={Graph embeddings for identifying symmetries in connectomes},
  author={Shan, Haozhe and Litwin-Kumar, Ashok},
  journal={bioRxiv},
  pages={2025--12},
  year={2025},
  publisher={Cold Spring Harbor Laboratory}
}

@article{saxena2025enriched,
  title={Enriched experience increases reciprocal synaptic connectivity and coding sparsity in higher-order cortex},
  author={Saxena, Rajat and Shobe, Justin L and Andujo, Aida M and Ning, Wing and Anaclet, Christelle and McNaughton, Bruce L},
  journal={bioRxiv},
  year={2025}
}

@article{sherf2025complexity,
  title={Complexity and dynamics of partially symmetric random neural networks},
  author={Sherf, Nimrod and Tang, Si and Hafner, Dylan and Touboul, Jonathan D and Pitkow, Xaq and Bassler, Kevin E and Josi{\'c}, Kre{\v{s}}imir},
  journal={arXiv preprint arXiv:2512.24439},
  year={2025}
}

@article{sompolinsky1988chaos,
  title={Chaos in random neural networks},
  author={Sompolinsky, Haim and Crisanti, Andrea and Sommers, Hans-Jurgen},
  journal={Physical review letters},
  volume={61},
  number={3},
  pages={259},
  year={1988},
  publisher={APS}
}

@article{kadmon2015transition,
  title={Transition to chaos in random neuronal networks},
  author={Kadmon, Jonathan and Sompolinsky, Haim},
  journal={Physical Review X},
  volume={5},
  number={4},
  pages={041030},
  year={2015},
  publisher={APS}
}

@article{clark2024theory,
  title={Theory of coupled neuronal-synaptic dynamics},
  author={Clark, David G and Abbott, LF},
  journal={Physical Review X},
  volume={14},
  number={2},
  pages={021001},
  year={2024},
  publisher={APS}
}

@article{clark2025connectivity,
  title={Connectivity structure and dynamics of nonlinear recurrent neural networks},
  author={Clark, David G and Marschall, Owen and Van Meegen, Alexander and Litwin-Kumar, Ashok},
  journal={Physical Review X},
  volume={15},
  number={4},
  pages={041019},
  year={2025},
  publisher={APS}
}

@article{kusmierz2020edge,
  title={Edge of chaos and avalanches in neural networks with heavy-tailed synaptic weight distribution},
  author={Ku{\'s}mierz, {\L}ukasz and Ogawa, Shun and Toyoizumi, Taro},
  journal={Physical Review Letters},
  volume={125},
  number={2},
  pages={028101},
  year={2020},
  publisher={APS}
}

@article{liu2024connectivity,
  title={How connectivity structure shapes rich and lazy learning in neural circuits},
  author={Liu, Yuhan Helena and Baratin, Aristide and Cornford, Jonathan and Mihalas, Stefan and Shea-Brown, Eric and Lajoie, Guillaume},
  journal={ArXiv},
  pages={arXiv--2310},
  year={2024}
}

@article{mastrovito2024transition,
  title={Transition to chaos separates learning regimes and relates to measure of consciousness in recurrent neural networks},
  author={Mastrovito, Dana and Liu, Yuhan Helena and Kusmierz, Lukasz and Shea-Brown, Eric and Koch, Christof and Mihalas, Stefan},
  journal={bioRxiv},
  year={2024}
}

@article{stubenrauch2025fixed,
  title={Fixed point geometry in chaotic neural networks},
  author={Stubenrauch, Jakob and Keup, Christian and Kurth, Anno C and Helias, Moritz and van Meegen, Alexander},
  journal={Physical Review Research},
  volume={7},
  number={2},
  pages={023203},
  year={2025},
  publisher={APS}
}

@article{van2021large,
  title={Large-deviation approach to random recurrent neuronal networks: Parameter inference and fluctuation-induced transitions},
  author={van Meegen, Alexander and K{\"u}hn, Tobias and Helias, Moritz},
  journal={Physical review letters},
  volume={127},
  number={15},
  pages={158302},
  year={2021},
  publisher={APS}
}

@article{steinmetz2021neuropixels,
  title={Neuropixels 2.0: A miniaturized high-density probe for stable, long-term brain recordings},
  author={Steinmetz, Nicholas A and Aydin, Cagatay and Lebedeva, Anna and Okun, Michael and Pachitariu, Marius and Bauza, Marius and Beau, Maxime and Bhagat, Jai and B{\"o}hm, Claudia and Broux, Martijn and others},
  journal={Science},
  volume={372},
  number={6539},
  pages={eabf4588},
  year={2021},
  publisher={American Association for the Advancement of Science}
}

@article{hong2019novel,
  title={Novel electrode technologies for neural recordings},
  author={Hong, Guosong and Lieber, Charles M},
  journal={Nature Reviews Neuroscience},
  volume={20},
  number={6},
  pages={330--345},
  year={2019},
  publisher={Nature Publishing Group UK London}
}

@article{steinmetz2018challenges,
  title={Challenges and opportunities for large-scale electrophysiology with Neuropixels probes},
  author={Steinmetz, Nicholas A and Koch, Christof and Harris, Kenneth D and Carandini, Matteo},
  journal={Current opinion in neurobiology},
  volume={50},
  pages={92--100},
  year={2018},
  publisher={Elsevier}
}

@article{perich2025neural,
  title={A neural manifold view of the brain},
  author={Perich, Matthew G and Narain, Devika and Gallego, Juan A},
  journal={Nature Neuroscience},
  volume={28},
  number={8},
  pages={1582--1597},
  year={2025},
  publisher={Nature Publishing Group US New York}
}

@article{langdon2023unifying,
  title={A unifying perspective on neural manifolds and circuits for cognition},
  author={Langdon, Christopher and Genkin, Mikhail and Engel, Tatiana A},
  journal={Nature Reviews Neuroscience},
  volume={24},
  number={6},
  pages={363--377},
  year={2023},
  publisher={Nature Publishing Group UK London}
}

@article{gallego2017neural,
  title={Neural manifolds for the control of movement},
  author={Gallego, Juan A and Perich, Matthew G and Miller, Lee E and Solla, Sara A},
  journal={Neuron},
  volume={94},
  number={5},
  pages={978--984},
  year={2017},
  publisher={Elsevier}
}

@article{mastrogiuseppe2018linking,
  title={Linking connectivity, dynamics, and computations in low-rank recurrent neural networks},
  author={Mastrogiuseppe, Francesca and Ostojic, Srdjan},
  journal={Neuron},
  volume={99},
  number={3},
  pages={609--623},
  year={2018},
  publisher={Elsevier}
}

@article{beiran2021shaping,
  title={Shaping dynamics with multiple populations in low-rank recurrent networks},
  author={Beiran, Manuel and Dubreuil, Alexis and Valente, Adrian and Mastrogiuseppe, Francesca and Ostojic, Srdjan},
  journal={Neural computation},
  volume={33},
  number={6},
  pages={1572--1615},
  year={2021},
  publisher={MIT Press One Rogers Street, Cambridge, MA 02142-1209, USA journals-info~…}
}

@article{beiran2023parametric,
  title={Parametric control of flexible timing through low-dimensional neural manifolds},
  author={Beiran, Manuel and Meirhaeghe, Nicolas and Sohn, Hansem and Jazayeri, Mehrdad and Ostojic, Srdjan},
  journal={Neuron},
  volume={111},
  number={5},
  pages={739--753},
  year={2023},
  publisher={Elsevier}
}

@article{pals2024trained,
  title={Trained recurrent neural networks develop phase-locked limit cycles in a working memory task},
  author={Pals, Matthijs and Macke, Jakob H and Barak, Omri},
  journal={PLOS Computational Biology},
  volume={20},
  number={2},
  pages={e1011852},
  year={2024},
  publisher={Public Library of Science San Francisco, CA USA}
}

@article{valente2022extracting,
  title={Extracting computational mechanisms from neural data using low-rank RNNs},
  author={Valente, Adrian and Pillow, Jonathan W and Ostojic, Srdjan},
  journal={Advances in Neural Information Processing Systems},
  volume={35},
  pages={24072--24086},
  year={2022}
}

@article{pellegrino2023low,
  title={Low tensor rank learning of neural dynamics},
  author={Pellegrino, Arthur and Cayco Gajic, N Alex and Chadwick, Angus},
  journal={Advances in Neural Information Processing Systems},
  volume={36},
  pages={11674--11702},
  year={2023}
}

@article{hertag2020learning,
  title={Learning prediction error neurons in a canonical interneuron circuit},
  author={Hert{\"a}g, Loreen and Sprekeler, Henning},
  journal={Elife},
  volume={9},
  pages={e57541},
  year={2020},
  publisher={eLife Sciences Publications, Ltd}
}

@article{palmigiano2020common,
  title={Common rules underlying optogenetic and behavioral modulation of responses in multi-cell-type V1 circuits},
  author={Palmigiano, Agostina and Fumarola, Francesco and Mossing, Daniel P and Kraynyukova, Nataliya and Adesnik, Hillel and Miller, Kenneth D},
  journal={bioRxiv},
  pages={2020--11},
  year={2020},
  publisher={Cold Spring Harbor Laboratory}
}

@article{murphy2009balanced,
  title={Balanced amplification: a new mechanism of selective amplification of neural activity patterns},
  author={Murphy, Brendan K and Miller, Kenneth D},
  journal={Neuron},
  volume={61},
  number={4},
  pages={635--648},
  year={2009},
  publisher={Elsevier}
}

@article{reimann2024specific,
  title={Specific inhibition and disinhibition in the higher-order structure of a cortical connectome},
  author={Reimann, Michael W and Egas Santander, Daniela and Ecker, Andr{\'a}s and Muller, Eilif B},
  journal={Cerebral Cortex},
  volume={34},
  number={11},
  pages={bhae433},
  year={2024},
  publisher={Oxford University Press}
}

@article{gal2017rich,
  title={Rich cell-type-specific network topology in neocortical microcircuitry},
  author={Gal, Eyal and London, Michael and Globerson, Amir and Ramaswamy, Srikanth and Reimann, Michael W and Muller, Eilif and Markram, Henry and Segev, Idan},
  journal={Nature neuroscience},
  volume={20},
  number={7},
  pages={1004--1013},
  year={2017},
  publisher={Nature Publishing Group US New York}
}

@article{perin2011synaptic,
  title={A synaptic organizing principle for cortical neuronal groups},
  author={Perin, Rodrigo and Berger, Thomas K and Markram, Henry},
  journal={Proceedings of the National Academy of Sciences},
  volume={108},
  number={13},
  pages={5419--5424},
  year={2011},
  publisher={National Academy of Sciences}
}

@article{rieubland2014structured,
  title={Structured connectivity in cerebellar inhibitory networks},
  author={Rieubland, Sarah and Roth, Arnd and H{\"a}usser, Michael},
  journal={Neuron},
  volume={81},
  number={4},
  pages={913--929},
  year={2014},
  publisher={Elsevier}
}

@article{song2005highly,
  title={Highly nonrandom features of synaptic connectivity in local cortical circuits},
  author={Song, Sen and Sj{\"o}str{\"o}m, Per Jesper and Reigl, Markus and Nelson, Sacha and Chklovskii, Dmitri B},
  journal={PLoS biology},
  volume={3},
  number={3},
  pages={e68},
  year={2005},
  publisher={Public Library of Science San Francisco, USA}
}

@article{trautmann2025large,
  title={Large-scale high-density brain-wide neural recording in nonhuman primates},
  author={Trautmann, Eric M and Hesse, Janis K and Stine, Gabriel M and Xia, Ruobing and Zhu, Shude and O’Shea, Daniel J and Karsh, Bill and Colonell, Jennifer and Lanfranchi, Frank F and Vyas, Saurabh and others},
  journal={Nature Neuroscience},
  pages={1--14},
  year={2025},
  publisher={Nature Publishing Group US New York}
}

@article{international2025brain,
  title={A brain-wide map of neural activity during complex behaviour},
  author={Angelaki, Dora and Benson, Brandon and Benson, Julius and Birman, Daniel and Bonacchi, Niccol{\`o} and Bougrova, Kc{\'e}nia and Bruijns, Sebastian A and Carandini, Matteo and Catarino, Joana A and others},
  journal={Nature},
  volume={645},
  number={8079},
  pages={177--191},
  year={2025},
  publisher={Nature Publishing Group UK London}
}

@article{schneider2023learnable,
  title={Learnable latent embeddings for joint behavioural and neural analysis},
  author={Schneider, Steffen and Lee, Jin Hwa and Mathis, Mackenzie Weygandt},
  journal={Nature},
  volume={617},
  number={7960},
  pages={360--368},
  year={2023},
  publisher={Nature Publishing Group UK London}
}

@article{gosztolai2025marble,
  title={MARBLE: interpretable representations of neural population dynamics using geometric deep learning},
  author={Gosztolai, Adam and Peach, Robert L and Arnaudon, Alexis and Barahona, Mauricio and Vandergheynst, Pierre},
  journal={Nature Methods},
  pages={1--9},
  year={2025},
  publisher={Nature Publishing Group US New York}
}

@article{panichello2021shared,
  title={Shared mechanisms underlie the control of working memory and attention},
  author={Panichello, Matthew F and Buschman, Timothy J},
  journal={Nature},
  volume={592},
  number={7855},
  pages={601--605},
  year={2021},
  publisher={Nature Publishing Group UK London}
}

@article{luo2025transitions,
  title={Transitions in dynamical regime and neural mode during perceptual decisions},
  author={Luo, Thomas Zhihao and Kim, Timothy Doyeon and Gupta, Diksha and Bondy, Adrian G and Kopec, Charles D and Elliott, Verity A and DePasquale, Brian and Brody, Carlos D},
  journal={Nature},
  volume={646},
  number={8087},
  pages={1156--1166},
  year={2025},
  publisher={Nature Publishing Group UK London}
}

@article{meissner2025computational,
  title={Computational functions of precisely balanced neuronal microcircuits in an olfactory memory network},
  author={Meissner-Bernard, Claire and Jenkins, Bethan and Rupprecht, Peter and Bouldoires, Estelle Arn and Zenke, Friedemann and Friedrich, Rainer W and Frank, Thomas},
  journal={Cell Reports},
  volume={44},
  number={3},
  year={2025},
  publisher={Elsevier}
}

@article{dahmen2025heterogeneity,
  title={How heterogeneity shapes dynamics and computation in the brain},
  author={Dahmen, David and Hutt, Axel and Indiveri, Giacomo and Kennedy, Ann and Lefebvre, Jeremie and Mazzucato, Luca and Motter, Adilson E and Narayanan, Rishikesh and Payvand, Melika and Planert, Henrike and others},
  journal={Neuron},
  year={2025},
  publisher={Elsevier}
}

@article{barbosa2023early,
  title={Early selection of task-relevant features through population gating},
  author={Barbosa, Joao and Proville, R{\'e}mi and Rodgers, Chris C and DeWeese, Michael R and Ostojic, Srdjan and Boubenec, Yves},
  journal={Nature communications},
  volume={14},
  number={1},
  pages={6837},
  year={2023},
  publisher={Nature Publishing Group UK London}
}

@article{nykamp2017mean,
  title={Mean-field equations for neuronal networks with arbitrary degree distributions},
  author={Nykamp, Duane Q and Friedman, Daniel and Shaker, Sammy and Shinn, Maxwell and Vella, Michael and Compte, Albert and Roxin, Alex},
  journal={Physical Review E},
  volume={95},
  number={4},
  pages={042323},
  year={2017},
  publisher={APS}
}

@article{pfeffer2013inhibition,
  title={Inhibition of inhibition in visual cortex: the logic of connections between molecularly distinct interneurons},
  author={Pfeffer, Carsten K and Xue, Mingshan and He, Miao and Huang, Z Josh and Scanziani, Massimo},
  journal={Nature neuroscience},
  volume={16},
  number={8},
  pages={1068--1076},
  year={2013},
  publisher={Nature Publishing Group US New York}
}

@article{kolda2009tensor,
  title={Tensor decompositions and applications},
  author={Kolda, Tamara G and Bader, Brett W},
  journal={SIAM review},
  volume={51},
  number={3},
  pages={455--500},
  year={2009},
  publisher={SIAM}
}

@article{ballard1945tensor,
  title={Tensor Decompositions for Data Science},
  author={Ballard, Grey and Kolda, Tamara G},
  journal={American history},
  volume={1861},
  number={1900},
  year={1945}
}

@article{clark2026structure,
  title={Structure, disorder, and dynamics in task-trained recurrent neural circuits},
  author={Clark, David G and Bordelon, Blake and Zavatone-Veth, Jacob A and Pehlevan, Cengiz},
  journal={bioRxiv},
  pages={2026--03},
  year={2026},
  publisher={Cold Spring Harbor Laboratory}
}

@article{pachitariu2026critical,
  title={A critical initialization for biological neural networks},
  author={Pachitariu, Marius and Zhong, Lin and Gracias, Alexa and Minisi, Amanda and Lopez, Crystall and Stringer, Carsen},
  journal={Nature},
  pages={1--7},
  year={2026},
  publisher={Nature Publishing Group UK London}
}

@article{thibeault2024low,
  title={The low-rank hypothesis of complex systems},
  author={Thibeault, Vincent and Allard, Antoine and Desrosiers, Patrick},
  journal={Nature Physics},
  volume={20},
  number={2},
  pages={294--302},
  year={2024},
  publisher={Nature Publishing Group UK London}
}

@article{tye2024mixed,
  title={Mixed selectivity: Cellular computations for complexity},
  author={Tye, Kay M and Miller, Earl K and Taschbach, Felix H and Benna, Marcus K and Rigotti, Mattia and Fusi, Stefano},
  journal={Neuron},
  volume={112},
  number={14},
  pages={2289--2303},
  year={2024},
  publisher={Elsevier}
}

@article{ito2024coordinated,
  title={Coordinated changes in a cortical circuit sculpt effects of novelty on neural dynamics},
  author={Ito, Shinya and Piet, Alex and Bennett, Corbett and Durand, S{\'e}verine and Belski, Hannah and Garrett, Marina and Olsen, Shawn R and Arkhipov, Anton},
  journal={Cell reports},
  volume={43},
  number={9},
  year={2024},
  publisher={Elsevier}
}

@article{garrett2023stimulus,
  title={Stimulus novelty uncovers coding diversity in visual cortical circuits},
  author={Garrett, Marina and Groblewski, Peter and Piet, Alex and Ollerenshaw, Doug and Najafi, Farzaneh and Yavorska, Iryna and Amster, Adam and Bennett, Corbett and Buice, Michael and Caldejon, Shiella and others},
  journal={BioRxiv},
  pages={2023--02},
  year={2023},
  publisher={Cold Spring Harbor Laboratory}
}

@article{miehl2023formation,
  title={Formation and computational implications of assemblies in neural circuits},
  author={Miehl, Christoph and Onasch, Sebastian and Festa, Dylan and Gjorgjieva, Julijana},
  journal={The Journal of Physiology},
  volume={601},
  number={15},
  pages={3071--3090},
  year={2023},
  publisher={Wiley Online Library}
}

@article{schulz2021generation,
  title={The generation of cortical novelty responses through inhibitory plasticity},
  author={Schulz, Auguste and Miehl, Christoph and Berry, Michael J and Gjorgjieva, Julijana},
  journal={Elife},
  volume={10},
  pages={e65309},
  year={2021},
  publisher={eLife Sciences Publications, Ltd}
}

@article{wu2022regulation,
  title={Regulation of circuit organization and function through inhibitory synaptic plasticity},
  author={Wu, Yue Kris and Miehl, Christoph and Gjorgjieva, Julijana},
  journal={Trends in Neurosciences},
  volume={45},
  number={12},
  pages={884--898},
  year={2022},
  publisher={Elsevier}
}

@article{clark2023dimension,
  title={Dimension of activity in random neural networks},
  author={Clark, David G and Abbott, LF and Litwin-Kumar, Ashok},
  journal={Physical Review Letters},
  volume={131},
  number={11},
  pages={118401},
  year={2023},
  publisher={APS}
}

@article{marti2018correlations,
  title={Correlations between synapses in pairs of neurons slow down dynamics in randomly connected neural networks},
  author={Mart{\'\i}, Daniel and Brunel, Nicolas and Ostojic, Srdjan},
  journal={Physical Review E},
  volume={97},
  number={6},
  pages={062314},
  year={2018},
  publisher={APS}
}

@article{kusmierz2025hierarchy,
  title={Hierarchy of chaotic dynamics in random modular networks},
  author={Ku{\'s}mierz, {\L}ukasz and Pereira-Obilinovic, Ulises and Lu, Zhixin and Mastrovito, Dana and Mihalas, Stefan},
  journal={Physical Review Letters},
  volume={134},
  number={14},
  pages={148402},
  year={2025},
  publisher={APS}
}
\clearpage
\newpage
\onecolumngrid
\appendix
\counterwithin{figure}{section}

\section{\label{app:appendix_neuralcircuit}Neural circuit model}
The recurrent connectivity matrix \(\mathbf{J}\in\mathbb{R}^{N\times N}\) in the circuit model is, in general, of full rank and decomposed as 
\begin{equation}\label{eq:conn_sum_mean_random}
\mathbf{J} = \mathbf{J^0}+\mathbf{Z}.
\end{equation}
Connectivity statistics of all \(N\times N\) synaptic weights in \(\mathbf{J}\) can be fully characterized by the joint distribution \(\mathcal{P}(\{J_{ij}\})\). 
Given our assumption of population-specific connectivity statistics, the marginal statistics of synaptic weights form low-rank block matrices, satisfying
\begin{equation}
\begin{aligned}
\mathcal{P}(J_{ij} &= J) = f^{pq}(J),\\
J^0_{pq} = \mathbb{E}[J_{ij}]_{f^{pq}},&\,\,\,\sigma_{z_{pq}}^2 = \mathbb{E}\left[\left(J_{ij}-\mathbb{E}[J_{ij}]\right)^2\right]_{f^{pq}},
\end{aligned}
\end{equation}
Furthermore, the strength of chain motifs is quantified by a three-dimensional low-rank tensor that characterizes the correlation coefficients
\begin{equation}\label{eq:definition_chain_stats}
\tau_{pqs}^{\rm chain} = \tau_{ijk}^{\rm chain} = \frac{[J_{ij} J_{jk}]-[J_{ij}][J_{jk}]}{\sqrt{[(J_{ij}-[J_{ij}])^2][(J_{jk}-[J_{jk}])^2]}}
= \frac{[z_{ij}z_{jk}]}{\sqrt{[z_{ij}^2][z_{jk}^2]}}
\end{equation}
where \(i\neq j\neq k\) and \(i\in p,\ j\in q,\ k\in s\). 
The symbol \([\cdot]\) represents the average over realizations on the corresponding connectivity distribution, where the numerator (on variable \(z\)) averages according to \(\mathcal{P}(\{J_{ij}\})\), and those in the denominator average according to \(f^{pq}\).

% Dynamics
The dynamics of the pre-activation of neuron \(i\), denoted by \(x_i(t)\), are governed by the following ordinary differential equation (ODE):
\begin{equation}\label{eq:rate_neural_network}
\dot{x}_i(t) = -x_i(t) + \sum_{j=1}^N J_{ij}\phi(x_j(t)) + I_i^{ext}u(t),
\end{equation}
where $\phi(x_i(t))$ denotes neuron \(i\)'s activation obtained via the positive nonlinearity $\phi(x)=b+\text{tanh}(x-\theta)$. The time-varying input is denoted by $I_i^{ext}u(t)$, which is neglected when studying spontaneous dynamics.
The solution to Eq.~\eqref{eq:rate_neural_network} for the pre-activation \(x_i(t)\) is expressed as 
\begin{equation}\label{eq:temporal_preactivation}
x_i(t) = e^{-t}x_{i,0} + \int_0^te^{-(t-\tau)}\left[\sum_{j=1}^NJ_{ij}\phi(x_j(\tau))+I_i^{ext}u(\tau)\right]d\tau,
\end{equation}
where \(x_{i,0}=x_i(0)\) is the initial activity of the network.
For time-invariant external input, when the network system reaches the steady state~(\(\lim t\to +\infty\)), the pre-activation $\{x_i^*\}_{i=1\dots N}$ satisfies
\begin{equation}
x_i^* = \sum_{j=1}^NJ_{ij}\phi(x_j^*) + I_i^{ext}u^*.
\end{equation}

We characterize the network's response to external input perturbations. An input perturbation \(\delta_h\boldsymbol{h}\) (s.t. \(\Vert\boldsymbol{h}\Vert^2=1\)) induces a deviation in the pre-activation of neuron  \(i\) from its equilibrium as \(\delta x_i\). Therefore, the response function relating the pre-activation of neuron \(i\) to \(\boldsymbol{h}\) is 
\begin{equation}\label{eq:response_full_sum}
\chi_{ih}^x = \frac{\delta x_i}{\delta_h}= \sum_{j=1}^N \left[(\mathbf{I}-\mathbf{J}\mathbf{\Phi}')^{-1}\right]_{ij}h_j.
\end{equation}
When the patterned perturbation is applied only on neuron population \(q\) (with \(h_i\neq 0,~i\in q\)), and is uncorrelated with the recurrent connectivity \(\mathbf{J}\) (covariance between terms \([\cdot]_{ij}\) and \(h_j\) is zero),  the average value \(\langle\chi_{ih}^x\rangle_{i\in p}\) represents the mean response of neurons in population \(p\) to uniform input to neurons in population \(q\).

\section{\label{app:appendix_lrRNN}Low-rank neural network}
The low-rank connectivity matrix with rank \(R\) is expressed as the sum of \(R\) unit rank outerproducts
\begin{equation}\label{eq:gen_synapse_lowrank}
\mathbf{J} = \frac{1}{N}\sum_{r=1}^R\boldsymbol{m}^{(r)}\boldsymbol{n}^{(r)}.
\end{equation}
We further assume that the connectivity vectors \(\{\boldsymbol{m}^{(r)}\}_{r=1\dots R}\) constitute an orthogonal basis, as \(\langle \boldsymbol{m}^{(r)},\boldsymbol{m}^{(r')}\rangle =0\) for \(r\neq r'\).
To characterize the first- and second-order statistics of the low-rank connectivity structures, we have 
\begin{equation}\label{eq:lrconn_statistics}
\begin{aligned}
a_{x_r}^{(p)} =E\left[x_i^{(r)}\right],\,\,\,\,
\sigma_{x_ry_{r'}}^{(p)} = E\left[\left(x_i^{(r)}-a_{x_r}^{(p)}\right)\left(y_i^{(r')}-a_{y_{r'}}^{(p)}\right)\right],
\end{aligned}
\end{equation}
where the neuron with index \(i\) belongs to the population \(p\), and variables \(x,y\) represent the connectivity vectors (\(m,n\)).

Substituting this low-rank approximation for the full-rank \(\mathbf{J}\)~(Eq.~\eqref{eq:rate_neural_network}), the dynamics evolve as
\begin{equation}\label{eq:lowrank_evolution}
\dot{x}_i(t) = -x_i(t) +\frac{1}{N}\sum_{r=1}^{R}\sum_{j=1}^N m_i^{(r)}n_j^{(r)}\phi(x_j(t))+I_i^{ext}u(t),
\end{equation}
and 
\begin{equation}\label{eq:lowrank_temporal_dynamics}
x_i(t) \approx\int_0^t \left[\frac{1}{N}\sum_{r=1}^{R}\sum_{j=1}^N m_i^{(r)}n_j^{(r)}\phi(x_j(\tau))+I_i^{ext}u(\tau)\right]e^{-(t-\tau)}d\tau.
\end{equation}
The above equation ~\eqref{eq:lowrank_temporal_dynamics} shows that under the low-rank connectivity assumption, after a sufficiently long time evolution, the network dynamics eventually become constrained within the subspace spanned by the connectivity vectors \(\{\boldsymbol{m}^{(r)}\}_{r=1\ldots R}\) and the external input vector.
Therefore, the population neural activity \(\boldsymbol{x}(t)\) can be expressed as 
\begin{equation}\label{eq:lowdim_subspace}
\boldsymbol{x}(t) = \sum_{r=1}^R\kappa_r(t)\boldsymbol{m}^{(r)} + \nu(t)\boldsymbol{I}^{ext}_{\perp},
\end{equation}
where \(\boldsymbol{I}^{ext}_{\perp}\) is the orthogonalized input vector compared to the recurrent basis \(\{\boldsymbol{m}^{(r)}\}_{r=1\ldots R}\) (i.e. using Gram-Schmidt algorithm).
Substituting Eq.~\eqref{eq:lowdim_subspace} into Eq.~\eqref{eq:lowrank_evolution} leads to the temporal evolution of the latent quantities 
\begin{equation}\label{eq:general_latent_dynamics}
\begin{aligned}
\dot{\kappa}_r(t) &= -\kappa_r(t) 
+ \sum_{p}\alpha_p\Bigg[a_{n_r}^{(p)}\langle\phi^{(p)}\rangle
+\bigg(\sigma^{(p)}_{n_rI_{\perp}^{ext}}\nu(t)
+\sum_{s=1}^R\sigma_{n_rm_s}^{(p)}\kappa_s(t)\bigg)\langle\phi'^{(p)}\rangle\Bigg] 
+\hat{\boldsymbol{m}}^{(r)\intercal}\boldsymbol{I}^{ext}u(t),\\
\dot{\nu}(t) &= -\nu(t) + \hat{\boldsymbol{I}}^{ext\intercal}_{\perp}\boldsymbol{I}^{ext}u(t),
\end{aligned}
\end{equation}
where \(\alpha_p = N_p/N\),
\(\hat{\boldsymbol{x}} = \boldsymbol{x}/\Vert\boldsymbol{x}\Vert^2\),
and \(\langle f^{(p)}\rangle\) represents the population-averaged quantity
\begin{equation}
\langle f^{(p)}\rangle = \int_{z\in D}p(z)f(z)\mathcal{D}z
\end{equation}
with respect to the population-specific activity distribution 
\begin{equation}
    D = \mathcal{N}(\mu^{(p)},\Delta^{(p)}),
\end{equation} 
where \(\mu^{(p)}\) and \(\Delta^{(p)}\) are the mean and variance of the neural activity \(x_i(t)\) within population \(p\) (\(i\in p\)), and they are associated with the connectivity vectors' statistics~(Eq.~\eqref{eq:lrconn_statistics}) as 
\begin{equation}\label{eq:mean and variance}
\begin{aligned}
\mu^{(p)}(t) = \sum_{r=1}^Ra_{m_r}^{(p)}\kappa_r(t) + a_{I_{\perp}^{ext}}^{(p)}\nu(t),\,\,\,\,
\Delta^{(p)}(t) = \sum_{r=1}^R \sigma_{m_r^2}^{(p)}(\kappa_r(t))^2+\sigma_{I_{\perp}^{ext}{}^2}^{(p)}(\nu(t))^2
\end{aligned}
\end{equation}
here \(a_{I_{\perp}^{ext}}^{(p)}\) and \(\sigma_{I_{\perp}^{ext}{}^2}^{(p)}\) share the definitions given in Eq.~\eqref{eq:lrconn_statistics}.
The steady-state of the latent dynamic quantities under a time-invariant \(u^*\) is calculated through the self-consistent equation 
\begin{equation}\label{eq:latent_dyn_equilibrium}
\begin{aligned}
\kappa_r^* &=  \sum_{p}\alpha_p\Bigg[a_{n_r}^{(p)}\langle\phi^{(p)}\rangle
+\bigg(\sigma^{(p)}_{n_rI_{\perp}^{ext}}\nu^*
+\sum_{s=1}^R\sigma_{n_rm_s}^{(p)}\kappa_s^*\bigg)\langle\phi'^{(p)}\rangle\Bigg] +\hat{\boldsymbol{m}}^{(r)\intercal}\boldsymbol{I}^{ext}u^*,\\
\nu^* &= \hat{\boldsymbol{I}}_{\perp}^{ext\intercal}\boldsymbol{I}^{ext}u^*.
\end{aligned}
\end{equation}
Combining Eq.~\eqref{eq:mean and variance} and Eq.~\eqref{eq:latent_dyn_equilibrium} gives the equilibria \(\mu^{(p)} = \sum_{r=1}^R \kappa_r^* a_{m_r}^{(p)} + \nu^* a_{I^{ext}_{\perp}}^{(p)}\) and \(\Delta^{(p)} =  \sum_{r=1}^R \sigma_{m_r^2}^{(p)}(\kappa_r^*)^2+\sigma_{I_{\perp}^{ext}{}^2}^{(p)}(\nu^*)^2\).
We let the new equilibria after applying input deviation \(\delta_h \boldsymbol{h}\) be \(\tilde{\cdot}\), so we have 
\begin{equation}\label{eq:kappa_perturbed}
\begin{aligned}
\tilde{\boldsymbol{\kappa}} =& \sum_{p}\alpha_p\Big[\boldsymbol{a}_n^{(p)}\langle\phi(\tilde{\mu}^{(p)},\tilde{\Delta}^{(p)})\rangle 
+\Big(\boldsymbol{\sigma}_{nI_{\perp}^{ext}}^{(p)}\tilde{\nu}
+\Sigma_{nm}^{(p)}\tilde{\boldsymbol{\kappa}}
+\boldsymbol{\sigma}_{nh_{\perp}}^{(p)}\tilde{\omega}\Big)\langle\phi'(\tilde{\mu}^{(p)},\tilde{\Delta}^{(p)})\rangle\Big] +\hat{\mathbf{M}}^{\intercal}(\boldsymbol{I}^{ext}u^* + \boldsymbol{h}\delta_h),\\
\tilde{\nu} =& \hat{\boldsymbol{I}}_{\perp}^{ext\intercal}(\boldsymbol{I}u^*+\boldsymbol{h}\delta_h),\\
\tilde{\omega} =& \hat{\boldsymbol{h}}_{\perp}^{\intercal}\boldsymbol{h}\delta_h,
\end{aligned}
\end{equation}
where \(\hat{\mathbf{M}}\) represents the scaled matrix with the \(r^{\rm th}\) column being \(\boldsymbol{m}^{(r)}/\Vert\boldsymbol{m}^{(r)}\Vert^2\), and 
\begin{equation}
\begin{aligned}
\tilde{\mu}^{(p)} = \sum_{r=1}^R\tilde{\kappa}_ra_{m_r}^{(p)} + \tilde{\nu}a_{I_{\perp}^{ext}}^{(p)}+ \tilde{\omega}a_{h_{\perp}}^{(p)},\,\,\,\,
\tilde{\Delta}^{(p)} = \sum_{r=1}^R(\tilde{\kappa}_r)^2\sigma_{m_r^2}^{(p)} + (\tilde{\nu})^2\sigma_{I_{\perp}^{ext}{}^2}^{(p)}+ (\tilde{\omega})^2\sigma_{h_{\perp}^2}^{(p)}.
\end{aligned}
\end{equation}
Comparing Eq.~\eqref{eq:kappa_perturbed} with Eq.~\eqref{eq:latent_dyn_equilibrium} we obtain the latent response function regarding \(\{\kappa_r\}\) as 
\begin{equation}
\chi_{rh}^{\kappa} = \frac{\tilde{\kappa}_r - \kappa_r^*}{\delta_h}.
\end{equation}
An explicit analytical solution for this latent response function is difficult to obtain due to the nonlinearity, we provide a heuristic derivation using linearization (Appendix~\ref{app:response_func}).
However, qualitatively, we observe that beyond the deviations of the latent variables \(\tilde{\boldsymbol{\kappa}}\) and \(\tilde{\nu}\) from their equilibrium values, an emergent term \(\tilde{\omega}\) associated with the input pattern \(\boldsymbol{h}\) significantly influences the response of the recurrent latent dynamics.
This emergent latent quantity is directly governed by the covariance \(\sigma_{n_rh_{\perp}}\) between the connectivity basis vectors \({\boldsymbol{n}^{(r)}}\) and the orthogonal input component \(\boldsymbol{h}_{\perp}\), suggesting that the integration of external inputs and the following computation can be manipulated through targeted design of the input pattern.

\section{\label{app:response_func}Latent response function through linearization}
We examine sufficiently minor perturbations \(\delta_h\) that allow for the linearization of latent dynamical variables around their respective steady-states, starting from Eq.~\eqref{eq:latent_dyn_equilibrium} and Eq.~\eqref{eq:kappa_perturbed}. Same as Eq.~\eqref{eq:latent_dyn_equilibrium}, the unperturbed low-rank system with input $\boldsymbol{I}^{ext}$ is expressed as
\begin{equation}\label{eq:steady_latent_unpertb}
\begin{aligned}
\boldsymbol{\kappa}^{*} &= \sum_{p} \alpha_p \left[ \boldsymbol{a}_n^{(p)} \langle\phi^{(p)}\rangle^{*} + \left( \mathbf{\sigma}_{nI_{\perp}^{ext}}^{(p)}\nu^* + \mathbf{\sigma}_{nm}^{(p)}\boldsymbol{\kappa}^* +\sigma_{nh}^{(p)}\times 0\right)\langle\phi^{\prime(p)}\rangle^{*}\right] + \hat{\mathbf{M}}^{\intercal} \boldsymbol{I}^{ext} u^* , \\
\nu^* &= \hat{I}_{\perp}^{ext \intercal} \boldsymbol{I}^{ext} u^*, 
\end{aligned}
\end{equation}
where \(*\) denotes the steady-state value of latent quantities. The average and variance of the Gaussian integral of the nonlinearity $\langle\phi^{(p)}\rangle$, $\langle\phi^{\prime(p)}\rangle$ are expressed as
\begin{equation}
\mu^{(p)} = \boldsymbol{a}_m^{(p)\intercal}\boldsymbol{\kappa} + a_{I^{ext}_{\perp}}^{(p)} \nu, \ \ \ \Delta^{(p)} = {(\boldsymbol{\sigma}_{m^2}^{(p)} \odot \boldsymbol{\kappa})}^{\intercal} \boldsymbol{\kappa} + \sigma_{I_{\perp}^{ext}{}^2}^{(p)} \nu^2 + \sigma_{h^2_{\perp}}^{(p)} \omega^2,
\end{equation}
where \(\odot\) represents the Hadamard~(elementwise) product.
Introducing the perturbative input $\delta_h \boldsymbol{h}$, which is probably not orthogonal to the original input $\boldsymbol{I}^{ext}$, the steady-state self-consistent equations are expressed as
\begin{equation}\label{eq:steady_latent_pertb}
\begin{aligned}
\tilde{\boldsymbol{\kappa}} &= \sum_{p} \alpha_p \left[ \boldsymbol{a}_n^{(p)} \langle\tilde\phi^{(p)}\rangle + \left( \mathbf{\sigma}_{nI_{\perp}^{ext}}^{(p)}\tilde\nu + \mathbf{\sigma}_{nm}^{(p)}\tilde{\boldsymbol{\kappa}} + \mathbf{\sigma}_{nh_{\perp}}^{(p)} \tilde{\omega} \right)\langle\tilde\phi^{\prime(p)}\rangle\right] + \hat{\mathbf{M}}^{\intercal} (\boldsymbol{I}^{ext}u^* + \delta_{h} \boldsymbol{h}) , \\
\tilde{\nu} &= \hat{\boldsymbol{I}}^{ext \intercal}_{\perp} (\boldsymbol{I}^{ext}u^* + \delta_{h} \boldsymbol{h}) , \\
\tilde{\omega} &= \hat{\boldsymbol{h}}_{\perp}^{\intercal} \boldsymbol{h} \delta_h, 
\end{aligned}
\end{equation}
same as Eq.\eqref{eq:kappa_perturbed}.
The average and variance of the Gaussian integral for population \(p\) are expressed as
\begin{equation}
\tilde{\mu}^{(p)} = \boldsymbol{a}_m^{(p)\intercal}\tilde{\boldsymbol{\kappa}} + a_{I^{ext}_{\perp}}^{(p)} \tilde\nu, \ \ \ \tilde{\Delta}^{(p)} = {(\boldsymbol{\sigma}_{m^2}^{(p)} \odot \tilde{\boldsymbol{\kappa}})}^{\intercal} \tilde{\boldsymbol{\kappa}} + \sigma_{I^{ext}_{\perp}{}^2}^{(p)} \tilde{\nu}^2 + \sigma_{h^2_{\perp}}^{(p)} \tilde{\omega}^2. 
\end{equation}
Notating the steady-state change of latent quantities $\delta{\boldsymbol{\kappa}} = \tilde{\boldsymbol{\kappa}} -\boldsymbol{\kappa}^*, \delta \nu = \tilde\nu - \nu^*, \delta\omega = \tilde\omega$, and notating that the average and variance of population pre-activations $\delta\mu^{(p)} = \tilde{\mu}^{(p)} - \mu^{(p)}{}^*, \ \delta\Delta^{(p)} = \tilde{\Delta}^{(p)} - \Delta^{(p)}{}^*$, 
\begin{equation}
\delta\mu^{(p)} = \boldsymbol{a}_m^{(p)\intercal} \delta\boldsymbol{\kappa} + a_{I_{\perp}^{ext}}^{(p)}\delta\nu + a_{h_{\perp}}^{(p)}\delta\omega , \ \ \delta\Delta^{(p)} = 2\left(( \boldsymbol{\sigma}_{m^2}^{(p)} \odot \boldsymbol{\kappa}^* )^{\intercal}\delta\boldsymbol{\kappa} + \sigma_{I_{\perp}^{ext}{}^2}^{(p)}\nu^*\delta\nu +0\times \sigma_{h_{\perp}^2}^{(p)}\delta\omega \right), 
\end{equation}
which contribute to the steady-state change of population-specific Gaussian integral $\delta \langle\phi^{(p)}\rangle = \langle\tilde{\phi}^{(p)}\rangle - \langle\phi^{(p)}\rangle^*$, and so as $\langle\phi^{\prime(p)}\rangle$. The steady-state changes of the population-specific Gaussian integral are expressed as
\begin{equation}
\begin{aligned}
\delta\langle\phi^{(p)}\rangle = &\frac{\partial \langle\phi^{(p)}\rangle}{\partial\mu^{(p)}}\delta \mu^{(p)} + \frac{\partial \langle\phi^{(p)}\rangle}{\partial\Delta^{(p)}}\delta \Delta_p 
= \left( \langle\phi^{\prime(p)}\rangle^*\boldsymbol{a}_m^{(p)\intercal} + \langle\phi^{\prime\prime(p)}\rangle^*{(\boldsymbol{\sigma}_{m^2}^{(p)}\odot\boldsymbol{\kappa}^*)}^{\intercal} \right) \delta\boldsymbol{\kappa} \\
&+ \left( \langle\phi^{\prime(p)}\rangle^* a_{I_{\perp}^{ext}}^{(p)} +  \langle\phi^{\prime\prime(p)}\rangle^* (\sigma_{I_{\perp}^{ext}{}^2}^{(p)}\nu^*) \right) \delta\nu
+ \left( \langle\phi^{\prime(p)}\rangle^* a_{h_{\perp}}^{(p)} \right) \delta\omega.
\end{aligned}
\end{equation}
From the above two steady-state equations (Eq.\eqref{eq:steady_latent_unpertb} and Eq.\eqref{eq:steady_latent_pertb}), ignoring the second-order intersection $\delta$ term, the perturbation-induced change of $\boldsymbol{\kappa}$ satisfies
\begin{equation}
\begin{aligned}
\delta\boldsymbol{\kappa} = &\sum_p\alpha_p\left[ \boldsymbol{a}_n^{(p)}\delta\langle\phi^{(p)}\rangle 
+ \boldsymbol{\sigma}_{nI_{\perp}^{ext}}^{(p)}\left(\nu^*\delta\langle\phi^{\prime(p)}\rangle + \langle\phi^{\prime(p)}\rangle^*\delta\nu \right) 
+ \mathbf{\sigma}_{nm}^{(p)} \left( \boldsymbol{\kappa}^*\delta\langle\phi^{\prime(p)}\rangle + \langle\phi^{\prime(p)}\rangle^*\delta\boldsymbol{\kappa} \right)
+ \boldsymbol{\sigma}_{nh_{\perp}}^{(p)} \langle\phi^{\prime(p)}\rangle^*\delta{\omega} \right] \\
&+ \hat{\mathbf{M}}^{\intercal} \boldsymbol{h} \delta_h
\end{aligned}
\end{equation}
then, the perturbation-induced change of $\boldsymbol{\kappa}$, 
\begin{equation}\label{eq:general_response_latent_form}
\begin{aligned}
\delta\boldsymbol{\kappa} = \tilde{\boldsymbol{\kappa}} - \boldsymbol{\kappa}^* 
=\mathbf{A}_{\kappa}\delta\boldsymbol{\kappa} + \boldsymbol{A}_{\nu} \delta\nu + \boldsymbol{A}_{\omega} \delta\omega + \boldsymbol{A}_{h}\delta_h, 
\end{aligned}
\end{equation}
where the matrix $\mathbf{A}_{\kappa}$
\begin{equation}
\begin{aligned}
\mathbf{A}_{\kappa} = \sum_{p}\alpha_p \bigg\{
&\langle\phi'\rangle_p^* \left(\boldsymbol{a}_n^{(p)}\boldsymbol{a}_m^{(p)\top} + \mathbf{\Sigma}_{nm}^{(p)}\right) +\langle\phi''\rangle_p^* \left(\boldsymbol{a}_n^{(p)}\left(\boldsymbol{\sigma}_{m^2}^{(p)}\odot\boldsymbol{\kappa}^*\right)^{\top} + \left(\boldsymbol{\sigma}_{nI^{ext}_{\perp}}^{(p)}\nu^* + \mathbf{\Sigma}_{nm}^{(p)}\boldsymbol{\kappa}^*\right)\boldsymbol{a}_m^{(p)\top}\right) \\
&+\langle\phi'''\rangle_p^* \left(\boldsymbol{\sigma}_{nI_{\perp}^{ext}}^{(p)}\nu^* + \boldsymbol{\Sigma}_{nm}^{(p)}\boldsymbol{\kappa}^*\right) \left(\boldsymbol{\sigma}_{m^2}^{(p)}\odot\boldsymbol{\kappa}^*\right)^{\top}\bigg\} .
\end{aligned}
\end{equation}
The first term in the bracket corresponds to the response under a linear response model. 
Analogous to \(\mathbf{A}_{\kappa}\), the coefficient-vector \(\boldsymbol{A}_{\nu}\) is
\begin{equation}
\begin{aligned}
\boldsymbol{A}_{\nu} =& \sum_p\alpha_p\bigg\{\langle\phi'\rangle^*\left(\boldsymbol{a}_n^{(p)}a_{I_{\perp}^{ext}}^{(p)}+\boldsymbol{\sigma}_{nI_{\perp}^{ext}}^{(p)}\right)
+\langle\phi''\rangle_p^* \left(\boldsymbol{a}_n^{(p)}\left(\sigma_{I^{ext}_{\perp}{}^2}^{(p)}\nu^*\right) + \left(\boldsymbol{\sigma}_{nI^{ext}_{\perp}}^{(p)}\nu^* + \mathbf{\Sigma}_{nm}^{(p)}\boldsymbol{\kappa}^*\right)a_{I_{\perp}^{ext}}^{(p)}\right) \\
&+\langle\phi'''\rangle_p^* \left(\boldsymbol{\sigma}_{nI_{\perp}^{ext}}^{(p)}\nu^* + \boldsymbol{\Sigma}_{nm}^{(p)}\boldsymbol{\kappa}^*\right) \left(\sigma_{I^{ext}_{\perp}{}^2}^{(p)}\nu^*\right)\bigg\},
\end{aligned}
\end{equation}
and \(\boldsymbol{A}_{\omega}\) is
\begin{equation}
\begin{aligned}
\boldsymbol{A}_{\omega} =& \sum_p\alpha_p\bigg\{\langle\phi^{\prime(p)}\rangle^*\left(\boldsymbol{a}_n^{(p)}a_{h_{\perp}}^{(p)}+\boldsymbol{\sigma}_{nh_{\perp}}^{(p)}\right)
+\langle\phi''\rangle_p^* \left(\boldsymbol{\sigma}_{nI^{ext}_{\perp}}^{(p)}\nu^* + \mathbf{\Sigma}_{nm}^{(p)}\boldsymbol{\kappa}^*\right)a_{I_{\perp}^{ext}}^{(p)}\bigg\}.
\end{aligned}
\end{equation}

\section{\label{app:outliers}Eigenvalue outliers}
In the recurrent neural network with a linear transfer function, the dynamical behavior of the network's steady-state is determined by the eigenvalues of the connectivity matrix $\mathbf{J}$ \cite{sompolinsky1988chaos}. When the maximum eigenvalue of the connectivity matrix $\mathbf{J}$ crosses 1, the quiescent steady-state loses stability. When extended to nonlinear networks, such a system may have non-trivial attractors. Here, the eigenvalue outliers of the connectivity matrix $\mathbf{J}$ are calculated, where $\mathbf{J}$ takes the form of Eq.~\eqref{eq:conn_sum_mean_random} and elements of the second-order correlation matrix $\mathbf{Z}$ are determined by the population-specific chain motifs (Eq.~\eqref{eq:definition_chain_stats}).  The theoretical derivation is similar to \cite{shao2023relating,shao2025impact}. 

The eigenvalues of the random matrix $\mathbf{J}$ are determined by the characteristic equation
\begin{equation}
{\rm det}(\mathbf{J}-\mathbf{I}\lambda)=0
\end{equation}
where the mean part of the connectivity $\mathbf{J}^0$ generally takes the form of $\frac{1}{N}\mathbf{M}_0\mathbf{N}_0^{\top}$. Following the matrix determinant lemma 
\begin{equation}
\begin{aligned}
&{\rm det} (\mathbf{M}_0\mathbf{N}_0^{\top}/N+(\mathbf{Z}-\mathbf{I}\lambda))  \\
&= {\rm det}(\mathbf{Z}-\mathbf{I}\lambda){\rm det}(\mathbf{I}+\mathbf{N}_0^{\top}(\mathbf{Z}-\mathbf{I}\lambda)^{-1}\mathbf{M}_0/N),
\end{aligned}
\end{equation}
the characteristic equation is expressed as
\begin{equation}
\frac{1}{\lambda^N}{\rm det}(\mathbf{Z}-\mathbf{I}\lambda){\rm det}\left(\lambda\mathbf{I}-\frac{1}{N}\mathbf{N}_0^{\top}\left(\mathbf{I}-\frac{\mathbf{Z}}{\lambda}\right)^{-1}\mathbf{M}_0\right) = 0
\end{equation}
Note that \({\rm det}(\mathbf{Z}-\mathbf{I}\lambda)\neq 0\), the eigenvalue outliers are the solution of \({\rm det}(\lambda\mathbf{I}-\mathbf{N}_0^{\top}(\mathbf{I}-\mathbf{Z}/\lambda)^{-1}\mathbf{M}_0 / N)=0\).
We therefore define a new matrix \(\mathbf{Q}\), which is expressed as 
\begin{equation}
\mathbf{Q} = \frac{1}{N}\mathbf{N}_0^{\top}\left(\mathbf{I}-\frac{\mathbf{Z}}{\lambda}\right)^{-1}\mathbf{M}_0 = \frac{1}{N}\sum_{k=0}^{\infty}\mathbf{N}_0^{\top}\left(\frac{\mathbf{Z}}{\lambda}\right)^k\mathbf{M}_0,
\end{equation}
The eigenvalue outliers are determined by the polynomial equation 
\begin{equation}\label{eq:polynomial_eigenvalue}
{\rm det}(\lambda\mathbf{I}-\mathbf{Q}) = 0.
\end{equation}
In networks with only second-order motifs, \([\mathbf{Z}^k]=0\) for odd \(k\), thereby \([\mathbf{Q}]=\sum_{l=0}^{\infty}\mathbf{N}_0^{\top}[\mathbf{Z}^{2l}]/\mathbf{M}_0/(N\lambda^{2l})\).
After some linear algebra and applying geometric series summation, we get the following
\begin{equation}\label{eq:heterogeneous_polynomial}
[\mathbf{Q}]=\frac{1}{N}\mathbf{N}_0^{\top}\left(\mathbf{I}-\frac{[\mathbf{Z}^2]}{\lambda^2}\right)^{-1}\mathbf{M}_0.
\end{equation}
Then, an important step is to calculate the matrix \([\mathbf{Z}^2]\) associated with the second-order statistics. The matrix element $[\mathbf{Z}^2]_{ik}$ with neuron $i,k$ in population $p$ and $s$ respectively, is expressed as 
\begin{equation}
\begin{aligned}
\left[\mathbf{Z}^2\right]_{ik} &= \left[\sum_{j=1}^N z_{ij}z_{jk}\right]
=\sum_{q}\sum_{j=1}^{N_q}\left[z_{ij}z_{jk}\right].
\end{aligned}
\end{equation}
From the definition of population-specific chain motifs (Eq.\eqref{eq:definition_chain_stats}), we have 
\begin{equation}
\left[z_{ij}z_{jk}\right] = \frac{\sigma_{pq}}{\sqrt{N}}\frac{\sigma_{qs}}{\sqrt{N}}\text{sgn}(\tau_{pqs})|\tau_{pqs}|, 
\end{equation}
then the element of \([\mathbf{Z}^2]\) is calculated as
\begin{equation}\label{eq:Z_second_order_moment_ap}
    \left[\mathbf{Z}^2\right]_{ik} = \sum_{q} \alpha_q\sigma_{pq}\sigma_{qs}{\rm sgn}(\tau_{pqs})|\tau_{pqs}|.
\end{equation}

In general, \([\mathbf{Z}^2]\) is the summation of \(P\) unit ranks, it thus can be expressed as the product of two $N \times P$ matrices $\mathbf{U}_{z2}$ and $\mathbf{V}_{z2}$, i.e., \([\mathbf{Z}^2] = \mathbf{U}_{z2}\mathbf{V}_{z2}^{\top}\).
Utilizing the Sherman-Morrison equation, we can compute the inverse of the \(N\times N\) matrix as 
\begin{equation}\label{eq:application_woodbury_inverse}
\left(\mathbf{I}-\frac{[\mathbf{Z}^2]}{\lambda^2}\right)^{-1} = \mathbf{I}_{\mathbb{R}^{N\times N}}+\mathbf{U}_{z2}\left(\mathbf{I}_{\mathbb{R}^{P\times P}}-\frac{\mathbf{V}_{z2}^{\top}\mathbf{U}_{z2}}{\lambda^2}\right)^{-1}\frac{\mathbf{V}_{z2}^{\top}}{\lambda^2}.
\end{equation}
Substituting Eq.~\eqref{eq:application_woodbury_inverse} into Eq.~\eqref{eq:heterogeneous_polynomial} and solving the resulting polynomial equation yields the theoretical outliers of $\mathbf{J}$.  For different population-specific chain-motif settings, the elements of the correlation matrix $\left[ \mathbf{Z}^2\right]$ take different values, thereby determining the eigenvalue outliers.

\subsection{End-neuron-specific chain motifs}
We consider the recurrent connectivity with end-neuron-specific chain motifs, that is \(\tau_E=\tau_{Epq}\) and \(\tau_I=\tau_{Ipq}\), where \(p,q\in\{E,I\}\).
Elements in the second-order moment matrix \([\mathbf{Z}^2]\) are computed as
\begin{equation}
\begin{aligned}
\left[\mathbf{Z}^2\right]_{ij} &= \sum_x^{E,I}\sum_{k\in x}\left[z_{ik}z_{kj}\right] = {\rm sgn}(\tau_{E\cdot})\sigma^2|\tau_{E\cdot}|\,\,\,i\in E\\
\left[\mathbf{Z}^2\right]_{ij} &= \sum_x^{E,I}\sum_{k\in x}\left[z_{ik}z_{kj}\right] = {\rm sgn}(\tau_{I\cdot})\sigma^2|\tau_{I\cdot}|\,\,\,i\in I.
\end{aligned}
\end{equation}
Therefore, this trial-averaged second-order moment matrix \([\mathbf{Z}^2]\) is of unit rank \([\mathbf{Z}^2]=\boldsymbol{U}_{z2}\boldsymbol{V}_{z2}^{\top}\), where \(\boldsymbol{U}_{z2} = [\sigma^2{\rm sgn}(\tau_{E\cdot})|\tau_{E\cdot}|\dots\sigma^2{\rm sgn}(\tau_{I\cdot})|\tau_{I\cdot}|\dots]^{\top}\) and \(\boldsymbol{V}_{z2} = \boldsymbol{e}\).
With these, we have 
\begin{equation}
\left(1-\frac{[\mathbf{Z}^2]}{\lambda^2}\right)^{-1} = \mathbf{I} +\boldsymbol{U}_{z2}\frac{\lambda^2}{\lambda^2-\sigma^2(N_E{\rm sgn}(\tau_{E\cdot})|\tau_{E\cdot}|+N_I{\rm sgn}(\tau_{I\cdot})|\tau_{I\cdot}|)}\frac{\boldsymbol{V}_{z2}^{\top}}{\lambda^2}. 
\end{equation}
Knowing that \(\mathbf{M}_0 = \boldsymbol{e}\) and \(\mathbf{N}_0 = [NJ_0\dots-gNJ_0\dots]^{\top}\), we get the equation 
\begin{equation}
\lambda = \lambda_0 + \frac{(1+\gamma)\sigma^2N_E^2({\rm sgn}(\tau_{E\cdot})|\tau_{E\cdot}|-g\gamma{\rm sgn}(\tau_{I\cdot})|\tau_{I\cdot}|)}{\lambda^2-\sigma^2N_E({\rm sgn}(\tau_{E\cdot})|\tau_{E\cdot}|+\gamma{\rm sgn}(\tau_{I\cdot})|\tau_{I\cdot}|)} J^0, 
\end{equation}
where $\lambda_0 = N_EJ^0(1-g\gamma)$ corresponds to the contribution of mean connectivity. The solutions of the above equation are the eigenvalue outliers under end-neuron-specific chain motifs in EI networks. When $\tau_E=\tau_I=\tau$, the characteristic equation is simplified as
\begin{equation}\label{eq:homogeneous_chain_outlier_characteristic}
\lambda(\lambda^2-\lambda_0\lambda-\sigma^2N\text{sgn}(\tau)|\tau|)=0,
\end{equation}
degenerating to the characteristic equation under homogeneous chain motifs \cite{shao2025impact}.

\subsection{I-to-I-to-I chain motifs and post-neuron-specific mean inhibition} %needed for $g_{II}$ decrease
Next, we consider the situation where the mean connectivity strength is postsynaptic neuron dependent, $g_{E}$ and $g_{I}$ are different. 
The nontrivial eigenvalues of the mean connectivity satisfy
\begin{equation}\label{eq:characteristic_eq_mean_gII}
  \lambda^2 + NJ^0 (\alpha_Ig_{I} - \alpha_E)\lambda + \alpha_E\alpha_I{(N J^0)}^2 (g_{E} - g_{I})=0, 
\end{equation}
giving the solutions
\begin{equation}
\begin{aligned}
\lambda_{1,2} &= \frac{N_E J^0}{2} \left(1- g_{I} \gamma \pm\sqrt{{(1-g_{I}\gamma)}^2 - 4 \gamma(g_{E} - g_{I})}\right) .
\end{aligned}
\end{equation}
The two eigenvalues $\lambda_{1,2}$ deviate from $\lambda_0 / 2$, and have conjugate imaginary part  when $g_E$ is sufficiently larger than $g_I$. When $g_{E} = g_{I}$, the two eigenvalues are $\lambda_0$ and zero, degenerating to the homogeneous case (Eq. (\ref{eq:homogeneous_chain_outlier_characteristic})). With a decrease in $g_{I}$, the two conjugate eigenvalues have a more positive real part.

Next, we add chain motifs, where only the nonzero $\tau_{III}$ is considered.
The second-order matrix $\left[\mathbf{Z}^2\right]$ (Eq.\eqref{eq:Z_second_order_moment_ap}) takes a simpler form
\begin{equation}
\begin{aligned}
  \left[\mathbf{Z}^2\right]_{ik} &= \sigma^2 \alpha_I \tau_{III},  \ \ i \in I, \ k \in I,   \\
  \left[\mathbf{Z}^2\right]_{ik} &= 0, \quad \quad \quad \quad \text{otherwise}.
\end{aligned}
\end{equation}
It is a unit-rank matrix, the factor matrices $\mathbf{U}_{z2}$ and $\mathbf{V}_{z2}$ are $\mathbf{U}_{z2} = {[0, ... \ , \sigma^2\alpha_I\tau_{III}, ...]}^{\intercal}$ and $\mathbf{V}_{z2} = {[0, ... \ , 1, ....]}^{\intercal}$. 
Substituting these vectors into Eq.~\eqref{eq:application_woodbury_inverse}, we have
\begin{equation}
{\left(1-\frac{\left[\mathbf{Z}^2\right]}{\lambda^2}\right)}^{-1} = \mathbf{I}_{\mathbb{R}^{N\times N}} + \mathbf{U}_{z2}\frac{\lambda^2}{\lambda^2 - \sigma^2N\alpha_I^2\tau_{III}}\frac{\mathbf{V}_{z2}^{\intercal}}{\lambda^2}.
\end{equation}
Substituting this into Eq.~\eqref{eq:heterogeneous_polynomial}, we have the matrix $\left[\mathbf{Q}\right]$, 
\begin{equation}
\begin{aligned}
\left[\mathbf{Q}\right] &= \frac{1}{N}\mathbf{N}_0^{\intercal}\mathbf{M}_0 + \frac{1}{N} \frac{\mathbf{N}_0^{\intercal}\mathbf{U}_{z2}\mathbf{V}_{z2}^{\intercal}\mathbf{M}_0}{\lambda^2 - \sigma^2N\alpha_I^2\tau_{III}} \\
&= NJ^0\left(\left(
\begin{matrix}
\alpha_E & -g_E\alpha_I \\
\alpha_E & -g_I\alpha_I
\end{matrix}\right) - \frac{\sigma^2N\alpha_I^3\tau_{III}}{\lambda^2 - \sigma^2N\alpha_I^2\tau_{III}}\left(\begin{matrix}
0 & g_E \\
0 & g_I
\end{matrix}\right)
\right).
\end{aligned}
\end{equation}
Let $A = \sigma^2 N \alpha_I^2\tau_{III}$, we rewrite the expression of matrix $\left[\mathbf{Q}\right]$ as
\begin{equation}
\left[\mathbf{Q}\right] = NJ^0 \begin{pmatrix}
    \alpha_E & -g_E\alpha_I\frac{\lambda^2}{\lambda^2 - A} \\
    \alpha_E & -g_I\alpha_I\frac{\lambda^2}{\lambda^2 - A}
\end{pmatrix}.
\end{equation}
Solving the polynomial equation, we obtain the characteristic equation of eigenvalue outliers,
\begin{equation}
  \lambda\left( (\lambda - \alpha_E N J^0) (\lambda^2 - A) + \alpha_I NJ^0 g_{I} \lambda^2 + \alpha_E\alpha_I{(N J^0)}^2 (g_{E} - g_{I}) \lambda\right)=0
\end{equation}
where the coefficient $A = \sigma^2 \alpha_I^2 \tau_{III}N$. When $\tau_{III}=0$, the above equation degenerates to Eq.~\eqref{eq:characteristic_eq_mean_gII}.

\section{\label{app:stability_analysis}Bifurcation and stability analysis for low-rank latent dynamics}
Here, we derived the Jacobian matrix of latent dynamics at steady-state under low-rank approximation. From the Jacobian matrix obtained, we obtained the phase transition line in parameter space by $\det(\mathcal{J})=0$. For the case of homogeneous chain motifs and end-neuron-specific chain motifs, we calculate the phase transition line for different nonlinear transfer functions.
\subsection{Homogeneous chain motifs}
Next, we consider the stability analysis of the fixed points.
We reformulate the dynamics of the latent quantities, $K$ and $\kappa$ as 
\begin{equation}
\begin{aligned}
\frac{dK}{dt} &= -K+\frac{1}{N}\boldsymbol{n}^{\top}\phi(\boldsymbol{x})=f(K,\kappa)\\
\frac{d\kappa}{dt} &= -\kappa+\frac{1}{N}\boldsymbol{v}^{\top}\phi(\boldsymbol{x})=g(K,\kappa).
\end{aligned}
\end{equation}
We further assume the fixed points $(K^0,\kappa^0)$ satisfying $f(K^0,\kappa^0) = g(K^0,\kappa^0) = 0$; the perturbations are $(K^1,\kappa^1)$, and define the first-order derivative \(\psi(\cdot) = \phi'(\cdot)\).
The dynamics of the perturbation $K^1$ can be expressed as 

\begin{equation}\label{eq:partialf_stabilityanalysis}
\begin{aligned}
\frac{\partial f}{\partial K} = \frac{\dot{K^1}}{K^1} =&-1 + \sum_{p=E,I}\alpha_p\langle n_i\psi_im_i\rangle_{i\in p}\\
 \approx&- 1 +\sum_{p=E,I} \alpha_p\bigg\{\underbrace{\langle\psi^{(p)}\rangle\left(a_{n}^{(p)}a_{m}^{(p)}+\sigma_{nm}^{(p)}\right)}_{{\rm Population-averaged~}\psi_{i\in P}}+ \underbrace{\langle\psi'^{(p)}\rangle\Big[a_{n}^{(p)}\left(\sigma_{m^2}^{(p)}K^0+\sigma_{mu}^{(p)}\kappa^0\right)+a_{m}^p\left(\sigma_{nm}^{(p)}K^0+\sigma_{nu}^{(p)}\kappa^0\right)\Big]}_{{\rm Fluctuation~in~}\psi_{i\in P}}\bigg\}\\
\frac{\partial f}{\partial \kappa} = \frac{\dot{K^1}}{\kappa^1} =&0 + \sum_{p=E,I}\alpha_p\langle n_i\psi_iu_i\rangle_{i\in p}\\
\approx& 0 + \sum_{p=E,I} \alpha_p\bigg\{\langle\psi^{(p)}\rangle\left(a_{n}^{(p)}a_{u}^{(p)}+\sigma_{nu}^{(p)}\right)+
 \langle\psi'^{(p)}\rangle\Big[a_{n}^{(p)}\left(\sigma_{mu}^{(p)}K^0+\sigma_{u^2}^{(p)}\kappa^0\right)
+a_{u}^{(p)}\left(\sigma_{nm}^{(p)}K^0+\sigma_{nu}^{(p)}\kappa^0\right)\Big]\bigg\}.
\end{aligned}
\end{equation}
% we further considered the derivative of $\phi(x_i)$ as $\psi(x_i)$. 
Utilizing Taylor expansion and Stein's lemma, both population-averaged components $\langle\psi^{(p)}\rangle$ and fluctuated components $\Delta \psi_i^{(p)} = \psi_i -\langle\psi^{(p)}\rangle$ influenced the dynamic evolution of the latent quantity perturbation.
For the fluctuated components $\Delta \psi_i^{(p)}$ that were correlated with fluctuating components in the structural connectivity, we retain terms up to $\psi'~(\phi'')$ for mathematical simplicity. 

Analogously, we have the dynamics of the perturbation $\kappa^1$ being expressed as
\begin{equation}\label{eq:partialg_stabilityanalysis}
\begin{aligned}
\frac{\partial g}{\partial K} = \frac{\dot{\kappa^1}}{K^1} =&0 + \sum_{p=E,I}\alpha_p\langle v_i\psi_im_i\rangle_{i\in p}\\
 \approx& 0 + \sum_{p=E,I}\alpha_p\bigg\{\langle\psi^{(p)}\rangle\left(a_{v}^{(p)}a_{m}^{(p)}+\sigma_{vm}^{(p)}\right)+
\langle\psi'^{(p)}\rangle\Big[a_{v}^{(p)}\left(\sigma_{m^2}^{(p)}K^0+\sigma_{mu}^{(p)}\kappa^0\right) + a_{m}^{(p)}\left(\sigma_{vm}^{(p)}K^0+\sigma_{vu}^{(p)}\kappa^0\right)\Big]\bigg\}\\
\frac{\partial g}{\partial \kappa} = \frac{\dot{\kappa^1}}{\kappa^1} =&-1 + \sum_{p=E,I}\alpha_p\langle v_i\psi_iu_i\rangle_{i\in p}\\
 \approx& -1 + \sum_{p=E,I}\alpha_p\bigg\{\langle\psi^{(p)}\rangle\left(a_{v}^{(p)}a_{u}^{(p)}+\sigma_{vu}^{(p)}\right)+
\langle\psi'^{(p)}\rangle\Big[a_{v}^{(p)}\left(\sigma_{mu}^{(p)}K^0+\sigma_{u^2}^{(p)}\kappa^0\right) + a_{u}^{(p)}\left(\sigma_{vm}^{(p)}K^0+\sigma_{vu}^{(p)}\kappa^0\right)\Big]\bigg\}.
\end{aligned}
\end{equation}

Combining the general expressions provided in Eqs.~\eqref{eq:partialf_stabilityanalysis},\eqref{eq:partialg_stabilityanalysis} with the connectivity statistics of the simple homogeneous EI network, we obtained the Jacobian matrix $\mathcal{J}\in\mathbb{R}^{2\times 2}$ for the specific fixed point $(K^0,\kappa^0)$.
We substituted the connectivity statistics from Table~\ref{tab:EI_connectivity_vectors} into the general expressions above. We obtain
\begin{equation}
\begin{aligned}
\frac{\partial f}{\partial K} = -1 + J^0(N_E-gN_I)\langle\psi\rangle + \sigma^2N\tau\kappa^0\langle\psi'\rangle, \,\,\,\,\,\,\frac{\partial f}{\partial \kappa} = \sigma^2N\tau\langle\psi\rangle + J^0(N_E-gN_I)\kappa^0\langle\psi'\rangle N\sigma^2|\tau|
\end{aligned}
\end{equation}
and
\begin{equation}
\begin{aligned}
\frac{\partial g}{\partial K} = \langle\psi\rangle, \,\,\,\,\,\,\frac{\partial g}{\partial \kappa} = -1+\kappa^0\langle\psi'\rangle N\sigma^2|\tau|.
\end{aligned}
\end{equation}

When one of the eigenvalues of the Jacobian matrix becomes zero, the stability of the steady-state changes from unstable to stable or vice versa. When this happens, the determinant of the Jacobian matrix equals zero. From the above derivations, we have
\begin{equation}\label{eq:Jacobian_Homo_Res}
\begin{aligned}
\det(\mathcal{J}) &= \frac{\partial f}{\partial K}\frac{\partial g}{\partial \kappa} - \frac{\partial f}{\partial\kappa}\frac{\partial g}{\partial K} \\
&= 1-C\langle\psi'\rangle\kappa^0-|C|\langle\psi'\rangle\kappa^0 + \text{sgn}(\tau)(C\langle\psi'\rangle \kappa^0)^2-\lambda_0\langle\psi\rangle-C(\langle\psi\rangle)^2
\end{aligned}
\end{equation}
where $\lambda_0 = J^0(N_E - gN_I)$ and $C = \sigma^2 N \tau$, which means the parameters for mean connectivity and for the second-order motif shape the dynamics, respectively. At the transition between different phases, one of the eigenvalues of the Jacobian matrix becomes zero, which means $\det(\mathcal{J})=0$ gives the phase transition line.
For nonlinearity $\phi(\cdot)=\text{tanh}(\cdot)$, the transition line is simplified as
\begin{equation}
1-\lambda_0-\sigma^2N|\tau|=0
\end{equation}
rewritten as 
\begin{equation}
J^0 = \frac{1-\sigma^2N|\tau|}{N\alpha_E(1-g\gamma)}.
\end{equation}
For other positive transfer functions where the integration of the nonlinear function and its derivative is not constant at the steady-state, there is no explicit solution for the transition line. Combining $\det(\mathcal{J})=0$ and the expression of latent quantities' steady-state (Eq.~\eqref{eq:homo_dynamics}), we have
\begin{equation}\label{eq_set:homo_phase}
\begin{aligned}
\kappa^0 &= \langle\phi\rangle^0 \\
K^0 &= \lambda_0\langle\phi\rangle^0+C\langle\psi\rangle^0\kappa^0\\
\lambda_0 &= \frac{1-C\langle\psi'\rangle\kappa^0-|C|\langle\psi'\rangle\kappa^0+\text{sgn}(C)(C\langle\psi'\rangle\kappa^0)^2-C{(\langle\psi\rangle)}^2}{\langle\psi\rangle}
\end{aligned}
\end{equation}
where \(\langle f\rangle^0\) represents averaging the nonlinear function \(f(\cdot)\) over a Gaussian-distributed steady-state variable with the mean \(K^0\) and the variance \((\kappa^0)^2N\sigma^2|\tau|\).
Therefore, with no extra variables introduced, the above condition constitutes a set of self-consistent equations regarding variables $(K^0, \kappa^0, \lambda_0)$. 
For a given $\tau$, two sets of solutions are obtained. The obtained
\begin{equation}
\lambda_0 = \lambda_0^+(\tau), \ \ \ \lambda_0 = \lambda_0^-(\tau)
\end{equation}
gave
\begin{equation}
J_0^+ = \frac{\lambda_0^+(\tau)}{N\alpha_E(1-g\gamma)}, \ \ \ J_0^- = \frac{\lambda_0^-(\tau)}{N\alpha_E(1-g\gamma)},
\end{equation}
which are, respectively, the transition lines from the quiescent phase to the bisteady phase and from the bisteady phase to the high activity phase, in the \(J^0-g\) plane.

\subsection{End-neuron-specific chain motifs}
Analogously to the above derivation for the homogeneous chain motifs scenario, we analyze the linear stability of the system with end-neuron-specific chain motifs and I-to-I-to-I chain motifs by calculating the Jacobian matrix with Stein's lemma and obtaining the phase transition line from its determinant. We first rewrite the dynamics under end-neuron-specific chain motifs based on the given 4 latent quantities
\begin{equation}
\begin{aligned}
\frac{dK_1}{dt} &= -K_1 + \frac{1}{N}\boldsymbol{n}^{(1)\top}\phi(\boldsymbol{x})=f_1(K_r, \kappa_r),\,\,\,\frac{dK_2}{dt} = -K_2 + \frac{1}{N}\boldsymbol{n}^{(2)\top}\phi(\boldsymbol{x})=f_2(K_r, \kappa_r),  \\
\frac{d\kappa_1}{dt} &= -\kappa_1 + \frac{1}{N}\boldsymbol{v}^{(1)\top}\phi(\boldsymbol{x})=g_1(K_r, \kappa_r),\,\,\,
\frac{d\kappa_2}{dt} = -\kappa_2 + \frac{1}{N}\boldsymbol{v}^{(2)\top}\phi(\boldsymbol{x})=g_2(K_r, \kappa_r).
\end{aligned}
\end{equation}
The dynamics of perturbation $K_1$ is given by a partial derivative
\begin{equation}
\begin{aligned}
\frac{\partial f_1}{\partial K_1} &= -1 + \sum_{p=E,I}\alpha_p\langle n^{(1)}_i\psi_im^{(1)}_i\rangle_{i\in p} \\
&\approx -1 + \sum_{p=E,I} \alpha_p\bigg\{\langle\psi^{(p)}\rangle\left(a_{n_1}^{(p)}a_{m_1}^{(p)}+\sigma_{n_1m_1}^{(p)}\right)+ \langle\psi'^{(p)}\rangle\Big[a_{n_1}^{(p)}\sum_{r=1,2}\left(\sigma_{m_1m_r}^{(p)}K_r^0+\sigma_{m_1u_r}^{(p)}\kappa_r^0\right)
+a_{m_1}^{(p)}\sum_{r=1,2}\left(\sigma_{n_1m_r}^{(p)}K_r^0+\sigma_{n_1u_r}^{(p)}\kappa_r^0\right)\Big]\bigg\}   
\end{aligned}
\end{equation}
\begin{equation}
\begin{aligned}
\frac{\partial f_1}{\partial K_2} &= 0 + \sum_{p=E,I}\alpha_p\langle n^{(1)}_i\psi_im^{(2)}_i\rangle_{i\in p} \\
&\approx 0 + \sum_{p=E,I} \alpha_p\bigg\{\langle\psi^{(p)}\rangle\left(a_{n_1}^{(p)}a_{m_2}^{(p)}+\sigma_{n_1m_2}^{(p)}\right)+ \langle\psi'^{(p)}\rangle\Big[a_{n_1}^{(p)}\sum_{r=1,2}\left(\sigma_{m_2m_r}^{(p)}K_r^0+\sigma_{m_2u_r}^{(p)}\kappa_r^0\right)
+a_{m_2}^{(p)}\sum_{r=1,2}\left(\sigma_{n_1m_r}^{(p)}K_r^0+\sigma_{n_1u_r}^{(p)}\kappa_r^0\right)\Big]\bigg\}   \\
\end{aligned}
\end{equation}
and
\begin{equation}
\begin{aligned}
\frac{\partial f_1}{\partial \kappa_1} &= 0 + \sum_{p=E,I}\alpha_p\langle n^{(1)}_i\psi_iu^{(1)}_i\rangle_{i\in p} \\
&\approx 0 + \sum_{p=E,I} \alpha_p\bigg\{\langle\psi^{(p)}\rangle\left(a_{n_1}^{(p)}a_{u_1}^{(p)}+\sigma_{n_1u_1}^{(p)}\right)+ \langle\psi'^{(p)}\rangle\Big[a_{n_1}^{(p)}\sum_{r=1,2}\left(\sigma_{u_1m_r}^{(p)}K_r^0+\sigma_{u_1u_r}^{(p)}\kappa_r^0\right)
+a_{u_1}^p\sum_{r=1,2}\left(\sigma_{n_1m_r}^{(p)}K_r^0+\sigma_{n_1u_r}^{(p)}\kappa_r^0\right)\Big]\bigg\}   
\end{aligned}
\end{equation}
\begin{equation}
\begin{aligned}
\frac{\partial f_1}{\partial \kappa_2} &= 0 + \sum_{p=E,I}\alpha_p\langle n^{(1)}_i\psi_iu^{(2)}_i\rangle_{i\in p} \\
&\approx 0 + \sum_{p=E,I} \alpha_p\bigg\{\langle\psi^{(p)}\rangle\left(a_{n_1}^{(p)}a_{u_2}^{(p)}+\sigma_{n_1u_2}^{(p)}\right)+ \langle\psi'^{(p)}\rangle\Big[a_{n_1}^{(p)}\sum_{r=1,2}\left(\sigma_{u_2m_r}^{(p)}K_r^0+\sigma_{u_2u_r}^{(p)}\kappa_r^0\right)
+a_{u_2}^{(p)}\sum_{r=1,2}\left(\sigma_{n_1m_r}^{(p)}K_r^0+\sigma_{n_1u_r}^{(p)}\kappa_r^0\right)\Big]\bigg\}  \\
\end{aligned}
\end{equation}
The dynamics of perturbation $K_2, \kappa_1$ and $\kappa_2$ are given by similar partial derivatives and approximations based on Stein’s lemma. According to the connectivity statistics of the heterogeneous chain motif scenario Table~\ref{tab:EI_connectivity_vectors}, the elements of the Jacobian matrix are obtained
\begin{equation}
\begin{aligned}
\frac{\partial f_1}{\partial K_1} &= -1 + \alpha_E \langle\psi^{(E)}\rangle NJ^0 + \sigma^2 N \tau_E \alpha_E \langle\psi^{\prime(E)}\rangle (\kappa_1^0 + \kappa_2^0), \\
\frac{\partial f_1}{\partial K_2} &= - \alpha_I\langle\psi^{(I)}\rangle gNJ^0 + \sigma^2 N \tau_E \alpha_I \langle\psi^{\prime(I)}\rangle (\kappa_1^0 + \kappa_2^0), \\  
\frac{\partial f_1}{\partial \kappa_1} &= \sigma^2 N \tau_E (\alpha_E \langle\psi^{(E)}\rangle + \alpha_I \langle\psi^{(I)}\rangle) + \sigma^2 N (|\tau_E|+|\tau_I|) NJ^0(\alpha_E \langle\psi^{\prime(E)}\rangle - g\alpha_I \langle\psi^{\prime(I)}\rangle)\kappa_1^0, \\
\frac{\partial f_1}{\partial \kappa_2} &= \sigma^2 N \tau_E (\alpha_E \langle\psi^{(E)}\rangle + \alpha_I \langle\phi^{\prime(I)}\rangle) + \sigma^2 N (|\tau_E|+|\tau_I|) NJ^0(\alpha_E \langle\psi^{\prime(E)}\rangle - g\alpha_I \langle\psi^{\prime(I)}\rangle)\kappa_2^0, 
\end{aligned}
\end{equation}
and
\begin{equation}
\begin{aligned}
\frac{\partial f_2}{\partial K_1} &= \alpha_E \langle\psi^{(E)}\rangle NJ^0 + \sigma^2 N \tau_I \alpha_E \langle\psi^{\prime(E)}\rangle (\kappa_1^0 + \kappa_2^0), \\
\frac{\partial f_2}{\partial K_2} &= -1-\alpha_I\langle\psi^{(I)}\rangle gNJ^0 + \sigma^2 N \tau_I \alpha_I \langle\psi^{\prime(I)}\rangle (\kappa_1^0 + \kappa_2^0), \\
\frac{\partial f_2}{\partial \kappa_1} &= \sigma^2 N \tau_I (\alpha_E \langle\psi^{(E)}\rangle + \alpha_I \langle\psi^{(I)}\rangle) + \sigma^2 N (|\tau_E|+|\tau_I|) NJ^0(\alpha_E \langle\psi^{\prime(E)}\rangle - g\alpha_I \langle\psi^{\prime(I)}\rangle)\kappa_1^0, \\
\frac{\partial f_2}{\partial \kappa_2} &= \sigma^2 N \tau_I (\alpha_E \langle\psi^{(E)}\rangle + \alpha_I \langle\psi^{(I)}\rangle) + \sigma^2 N (|\tau_E|+|\tau_I|) NJ^0(\alpha_E \langle\psi^{\prime(E)}\rangle - g\alpha_I \langle\psi^{\prime(I)}\rangle)\kappa_2^0, 
\end{aligned}
\end{equation}
and
\begin{equation}\label{eq:jacobian_element_kappa1}
\begin{aligned}
\frac{\partial g_1}{\partial K_1} &= \alpha_E\langle\psi^{(E)}\rangle,
&\frac{\partial g_1}{\partial K_2} &= 0, \\
\frac{\partial g_1}{\partial \kappa_1} &= -1+\alpha_E\langle\psi'^{(E)}\rangle N\sigma^2(|\tau_E|+|\tau_I|)\kappa_1^0,
&\frac{\partial g_1}{\partial \kappa_2} &= \alpha_E\langle\psi'^{(E)}\rangle N\sigma^2(|\tau_E|+|\tau_I|)\kappa_2^0, 
\end{aligned}
\end{equation}
and
\begin{equation}\label{eq:jacobian_element_kappa2}
\begin{aligned}
\frac{\partial g_2}{\partial K_1} &= 0,
&\frac{\partial g_2}{\partial K_2} &= \alpha_I\langle\psi^{(I)}\rangle, \\
\frac{\partial g_2}{\partial \kappa_1} &= \alpha_I\langle\psi'^{(I)}\rangle N\sigma^2(|\tau_E|+|\tau_I|)\kappa_1^0,
&\frac{\partial g_2}{\partial \kappa_2} &= -1+\alpha_I\langle\psi'^{(I)}\rangle N\sigma^2(|\tau_E|+|\tau_I|)\kappa_2^0.
\end{aligned}
\end{equation}
Setting the determinant of the Jacobian matrix to zero and combining it with the steady-state equations of the system (Eq.\eqref{eq:latents_heter}) yields a system of five coupled equations, which is similar to Eq.~\eqref{eq_set:homo_phase} at a homogeneous scenario. For a given value of the parameter $g$, this system determines five unknowns: the state variables $(\kappa_1^0, \kappa_2^0, K_1^0, K_2^0)$ and the mean coupling parameter $J^0$. The numerical solutions of this system give the phase transition lines in the plane of parameters $(g, J^0)$, shown in Fig.\ref{fig:hetero_auto_EI_fig}C.

\subsection{End-neuron-specific chain motifs and post-neuron-specific mean inhibition}

\begin{figure}[ht]
\centering
\includegraphics[width=0.95\linewidth]{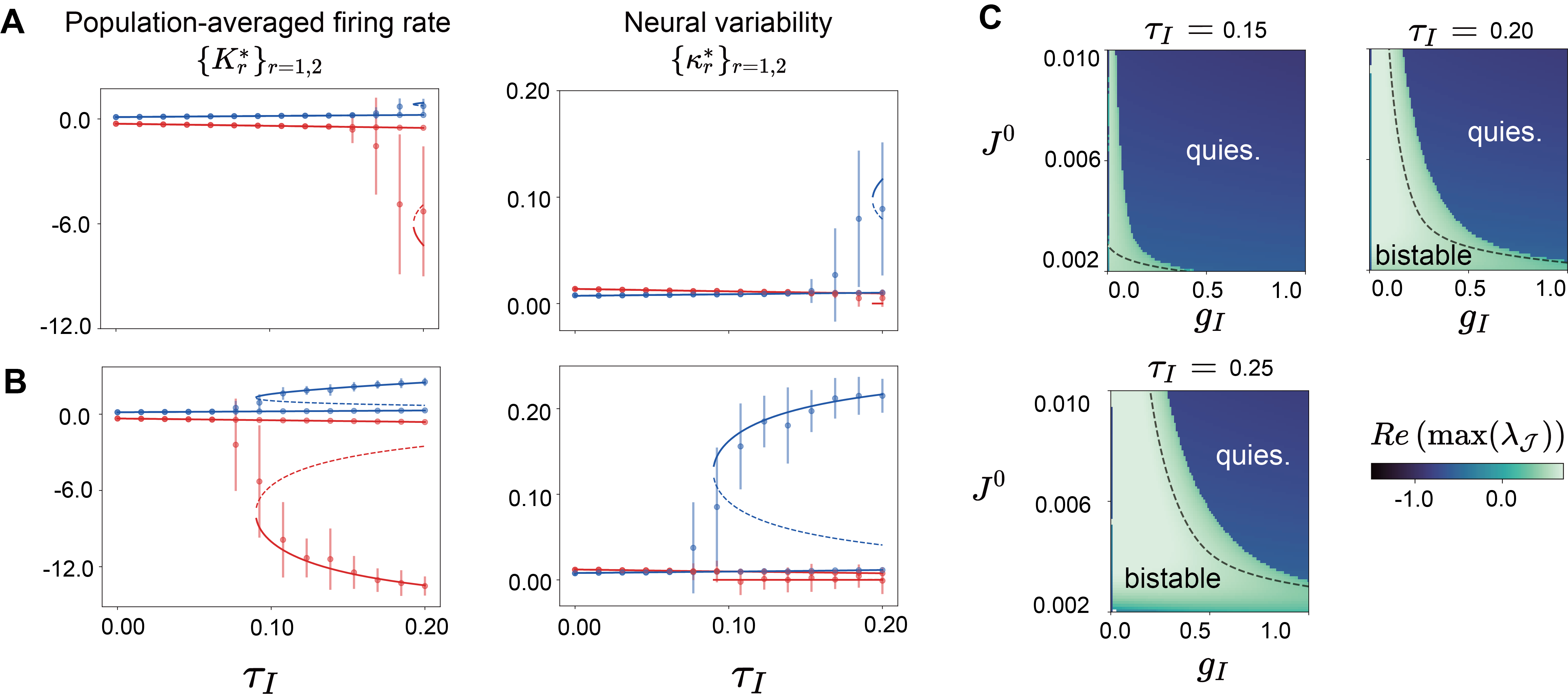}
\caption{{\bf Combined effects of chain motifs terminating on the inhibitory neuron population \(\tau_I\) and post-neuron-specific mean inhibition.} 
(A) Bistable transition of the latent quantities $\{K_r\}_{r=1,2}$.  
(B) Bistable transition of the latent quantities $\{\kappa_r\}_{r=1,2}$. 
(C) Phase diagram in mean connectivity plane $g_{I}-J^0$. Bistability exists within a range of relatively small $g_I$ values. 
Other figure descriptions follow the same conventions as in Fig.~\ref{fig:tau_iii_reponse_fig}A, B.
Parameters: $N_E=2000$, $N_I=500$, $\sigma=0.6$, $J^0=0.005$ in (A) and (B), $\theta=2.1$, $g_E=5.0$, $g_I=0.4$ in (A), $g_I=0.0$ in (B), $\tau_E=0.01$ in (C).}
\label{supp_fig:hetero_auto_tauI}
\end{figure}

Based on the above calculations for the end-neuron-specific scenario, we consider the post-neuron-specific mean inhibition when calculating the Jacobian matrix, in other words, $g_{E} \neq g_{I}$. The resulting adjustments to the Jacobian matrix entries are marginal. Specifically, the four elements determining the perturbation dynamics of $K_1$:
\begin{equation}
\frac{\partial f_1}{\partial K_1} = -1 + \alpha_E \langle\psi^{(E)}\rangle NJ^0 + \sigma^2 N \tau_E \alpha_E \langle\psi^{\prime(E)}\rangle (\kappa_1^0 + \kappa_2^0), 
\end{equation}
\begin{equation}
\frac{\partial f_1}{\partial K_2} = - \alpha_I\langle\psi^{(I)}\rangle g_{E}NJ^0 + \sigma^2 N \tau_E \alpha_I \langle\psi^{\prime(I)}\rangle (\kappa_1^0 + \kappa_2^0), 
\end{equation}
\begin{equation}
\frac{\partial f_1}{\partial \kappa_1} = \sigma^2 N \tau_E (\alpha_E \langle\phi^{\prime(E)}\rangle + \alpha_I \langle\psi^{(I)}\rangle) + \sigma^2 N (|\tau_E|+|\tau_I|) NJ^0(\alpha_E \langle\phi^{\prime\prime(E)}\rangle - g_{E}\alpha_I \langle\psi^{\prime(I)}\rangle)\kappa_1^0, 
\end{equation}
\begin{equation}
\frac{\partial f_1}{\partial \kappa_2} = \sigma^2 N \tau_E (\alpha_E \langle\psi^{(E)}\rangle + \alpha_I \langle\psi^{(I)}\rangle) + \sigma^2 N (|\tau_E|+|\tau_I|) NJ^0(\alpha_E \langle\psi^{\prime(E)}\rangle - g_{E}\alpha_I \langle\psi^{\prime(I)}\rangle)\kappa_2^0.
\end{equation}
The four elements determine the perturbation dynamics of $K_2$:
\begin{equation}
\frac{\partial f_2}{\partial K_1} =  \alpha_E \langle\psi^{(E)}\rangle NJ^0 + \sigma^2 N \tau_I \alpha_E \langle\psi^{\prime(E)}\rangle (\kappa_1^0 + \kappa_2^0), 
\end{equation}
\begin{equation}
\frac{\partial f_2}{\partial K_2} = -1-\alpha_I\langle\psi^{(I)}\rangle g_{I}NJ^0 + \sigma^2 N \tau_I \alpha_I \langle\psi^{\prime(I)}\rangle (\kappa_1^0 + \kappa_2^0), 
\end{equation}
\begin{equation}
\frac{\partial f_2}{\partial \kappa_1} = \sigma^2 N \tau_I (\alpha_E \langle\psi^{(E)}\rangle + \alpha_I \langle\psi^{(I)}\rangle) + \sigma^2 N (|\tau_E|+|\tau_I|) NJ^0(\alpha_E \langle\psi^{\prime(E)}\rangle - g_{I}\alpha_I \langle\psi^{\prime(I)}\rangle)\kappa_1^0, 
\end{equation}
\begin{equation}
\frac{\partial f_2}{\partial \kappa_2} = \sigma^2 N \tau_I (\alpha_E \langle\psi^{(E)}\rangle + \alpha_I \langle\psi^{(I)}\rangle) + \sigma^2 N (|\tau_E|+|\tau_I|) NJ^0(\alpha_E \langle\psi^{\prime(E)}\rangle - g_{I}\alpha_I \langle\psi^{\prime(I)}\rangle)\kappa_2^0.
\end{equation}
The elements determine the perturbation dynamics of $\kappa_1$ and $\kappa_2$ are the same as Eq.\eqref{eq:jacobian_element_kappa1} and Eq.\eqref{eq:jacobian_element_kappa2}.
The condition $\det(\mathcal{J})=0$ gives the phase diagram in plane $(g_{I}, \tau_{I})$, shown in Fig. \ref{supp_fig:hetero_auto_tauI}.

\subsection{I-to-I-to-I chain motifs and post-neuron-specific mean inhibition}
An analogous calculation for the Jacobian matrix, under mean connectivity $g_{E} \neq g_{I}$, also the second-order chain motifs $\tau_{III} \neq 0$ and $\tau_{pqx}=0$ otherwise. Rewriting the latent dynamics as
\begin{equation}
\begin{aligned}
\frac{dK_1}{dt} &= - K_1 + \frac{1}{N} \boldsymbol{n}^{(1)\top}\phi(\boldsymbol{x}) = f_1(K_r, \kappa_r) \\
\frac{dK_2}{dt} &= - K_2 + \frac{1}{N} \boldsymbol{n}^{(2)\top}\phi(\boldsymbol{x}) = f_2(K_r, \kappa_r) \\
\frac{d\kappa_2}{dt} &= - \kappa_2 + \frac{1}{N} \boldsymbol{v}^{(2)\top}\phi(\boldsymbol{x}) = g_2(K_r, \kappa_r),
\end{aligned}
\end{equation}
where \(u_i^{(1)}=0\), \(\forall i\). Thus only \(K_1,K_2\) and \(\kappa_2\) exhibit dynamics.
The detailed deviation by use of Stein's lemma is omitted here for simplicity, and the final results
\begin{equation}
\begin{aligned}
\frac{\partial f_1}{\partial K_1} = -1 + NJ^0 \alpha_E \langle\psi^{(E)}\rangle , \,\, 
\frac{\partial f_1}{\partial K_2} = - g_{E} NJ^0 \alpha_I \langle\psi^{(I)}\rangle , \,\,
\frac{\partial f_1}{\partial \kappa_2} = - g_{E} NJ^0 \alpha_I \langle\psi^{\prime(I)}\rangle \sigma^2 N |\tau_{III}| \kappa_2^0, 
\end{aligned}
\end{equation}
and
\begin{equation}
\begin{aligned}
\frac{\partial f_2}{\partial K_1} &= NJ^0 \alpha_E \langle\psi^{(E)}\rangle , \,\,
\frac{\partial f_2}{\partial K_2} = -1 - g_{I} NJ^0 \alpha_I \langle\psi^{(I)}\rangle + \alpha_I \langle\psi^{\prime(I)}\rangle \sigma^2 N \tau_{III}\kappa_2^0,  \\
\frac{\partial f_2}{\partial \kappa_2} &= \alpha_I \langle\psi^{(I)}\rangle \sigma^2 N \tau_{III} - g_{I} NJ^0 \alpha_I \langle\psi^{\prime(I)}\rangle \sigma^2 N |\tau_{III}| \kappa_2^0 ,
\end{aligned}
\end{equation}
and
\begin{equation}
\begin{aligned}
\frac{\partial g_2}{\partial K_1} = 0 ,\,\,
\frac{\partial g_2}{\partial K_2} = \alpha_I\langle\psi^{(I)}\rangle , \,\,
\frac{\partial g_2}{\partial \kappa_2} = -1 + \alpha_I\langle\psi^{\prime(I)}\rangle \sigma^2 N |\tau_{III}| \kappa_2^0.
\end{aligned}
\end{equation}
The condition $\det(\mathcal{J})=0$ gives the phase diagram in plane $(g_{I}, \tau_{III})$, shown in  Fig. \ref{fig:tau_iii_reponse_fig}B.

\section{\label{app:pyr-som-vip}Pyr-SOM-VIP network}
\subsection{Mean Connectivity Strength}
The mean synaptic coupling between Pyr, SOM and VIP is qualitatively based on the cell-type-specific connectivity probability matrix of VISp Layer 2 in \cite{campagnola2022local},
\begin{equation}\label{eq:mean_conn_param}
N\mathbf{J}^0 = N
\begin{bmatrix}
J_{EE}^0 & J_{ES}^0 & J_{EV}^0 \\
J_{SE}^0 & J_{SS}^0 & J_{SV}^0 \\
J_{VE}^0 & J_{VS}^0 & J_{VV}^0
\end{bmatrix}
=
\begin{bmatrix}
0.35 & -4.75 & 0\\
0.425 & 0 & -8.285 \\
0 & -0.0078 & 0
\end{bmatrix}.
\end{equation}
In addition, the ratio of each population: $\alpha_E = 0.8$, $\alpha_S = 0.1$, $\alpha_V = 0.1$ with $N = 2500$. 

\subsection{Mean-driven dynamics}  
The dynamics of the mean-driven (\(\delta_{I_{ext}}=0\)) Pyr-SOM-VIP network are shown in Eq.~\eqref{eq:mean_driven_syseqn_pop3}. It's a rank-3 system, each rank corresponds to a population. Without second-order correlations in connectivity, the three ranks correspond to the trial-to-trial variability $\kappa_1, \kappa_2$ and $\kappa_3$ degenerate. Similar to the classical mean-field model, the recurrent dynamics depend on mean coupling and deterministic nonlinearity
\begin{equation}\label{eq:mean_driven_syseqn_pop3}
\begin{aligned}
\dot{K_1} &= -K_1 + N \sum_{q_r=E,S,V} \alpha_{q_r} J_{E{q_r}}^0 \phi(K_r) + \mu_{I_{ext}},\\
\dot{K_2} &= -K_2 + N \sum_{q_r=E,S,V} \alpha_{q_r} J_{S{q_r}}^0 \phi(K_r), \\
\dot{K_3} &= -K_3 + N \sum_{q_r=E,S,V} \alpha_{q_r} J_{V{q_r}}^0 \phi(K_r),
\end{aligned}
\end{equation}
where \(q_r = E,S,V\) maps to \(r=1,2,3\) respectively. Here, the population-averaged firing rate $\langle\phi^{(q_r)}\rangle$ degenerates to the value of nonlinear transfer function of population-averaged pre-activation $K_r$.

\begin{figure}[ht]
\centering
\includegraphics[width=1.0\linewidth]{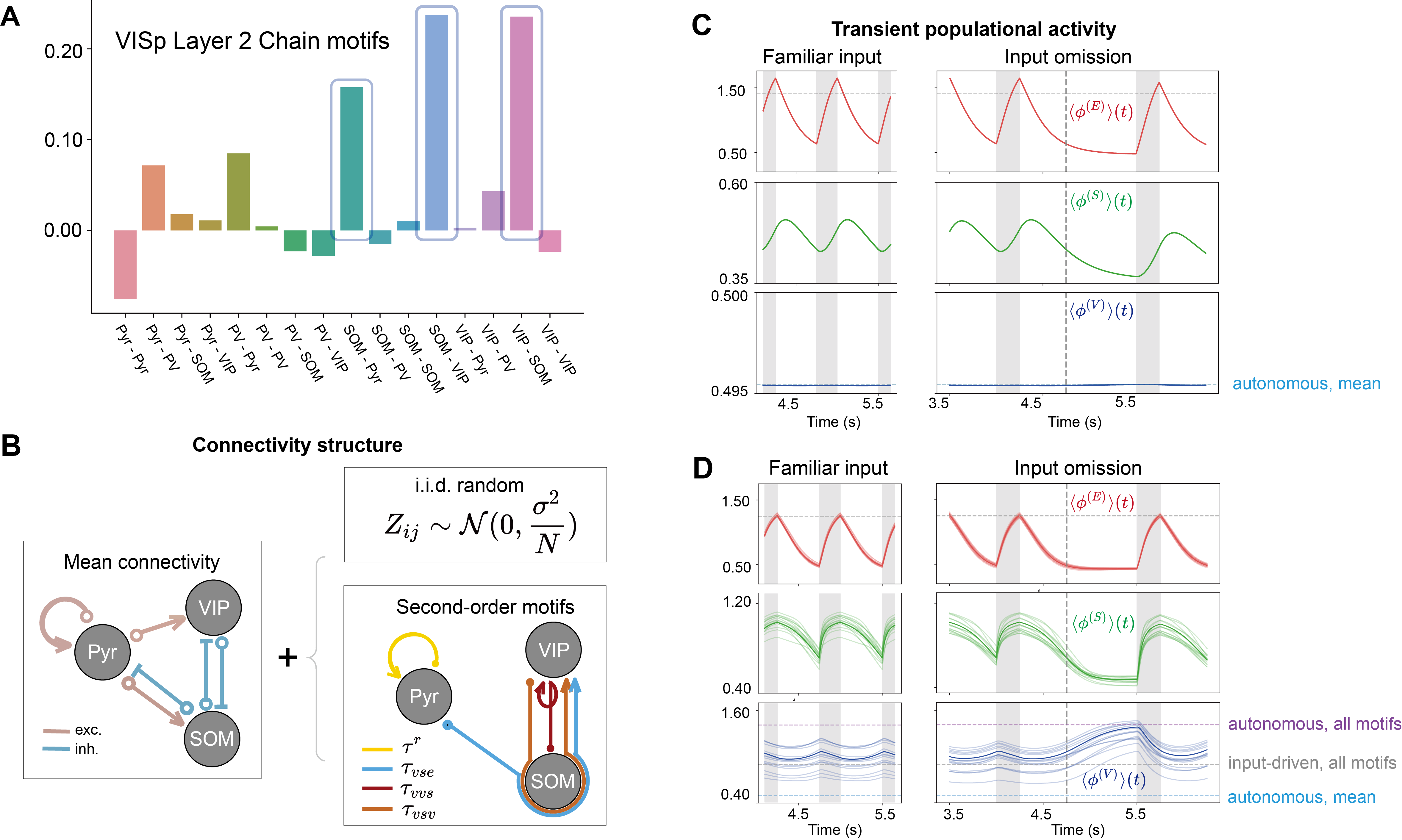}
\caption{{\bf Transient activity across Pyr, SOM and VIP neuron populations in networks with i.i.d. random connectivity and heterogeneous motifs.} 
(A) Strengths of chain motifs among four types of neurons in layer 2 of the mouse primary visual cortex. The index on the vertical axis is calculated as the ratio of the number of detected chain motifs to the total number of potential chain motifs across all tested neurons.\cite{dahmen_strong_2026} 
(B) Schematic of network connectivity decomposed into a mean component and a random component. The random component lacking chain motifs follows an independent and identically distributed (i.i.d.) zero-mean Gaussian distribution. Reconstructed motifs form the core integration pathways. 
(C) Transient network activity under i.i.d. random connectivity. 
(D) Transient network activity under random connectivity incorporating heterogeneous motifs (same as the results shown in Fig.~\ref{fig:experience}C). 
Other figure descriptions follow the same conventions as in Fig.~\ref{fig:experience}.
}
\label{suppfig:pop3_transient_mean}
\end{figure}

Next, we examine the population neuronal activities in response to external inputs. We consider the uniform input only for Pyr cells \(\langle I^{ext}_i\rangle_{i\in E}=\mu_{I_{ext}}\neq 0\). The transient activity under uniform input of the network with mean connectivity only is shown in Fig.\ref{suppfig:pop3_transient_mean}.

\subsection{Motif-considered dynamics} 
The dynamics of the network are shown in Eq. \eqref{eq:mean_motif_syseqn_pop3}. It's a rank-6 system, where the firing rate of each population $\langle\phi(\mu^{(E)}, \Delta^{(E)})\rangle$, $\langle\phi(\mu^{(S)}, \Delta^{(S)})\rangle$, $\langle\phi(\mu^{(V)}, \Delta^{(V)})\rangle$, and the mean-driven part of the recurrent activity $\overline{K}_p=N \sum_{q=E,S,V} \alpha_q J_{pq}^0 \langle\phi^{(q)}\rangle$ where \(p=E,S,V\):
\begin{equation}\label{eq:mean_motif_syseqn_pop3}
\begin{aligned}
\dot{\kappa}_1 &= -\kappa_1 + C_f\sigma^2 N \tau^r \alpha_E\langle\phi^{\prime(E)}\rangle \kappa_1 + \delta_{I_{ext}}/\Vert\boldsymbol{u}^{(1)}\Vert, \\
\dot{\kappa}_2 &= -\kappa_2 + \alpha_S\langle\phi^{(S)}\rangle, \\
\dot{\kappa}_3 &= -\kappa_3 + \alpha_V\langle\phi^{(V)}\rangle; \\
\dot{K}_1 &= - K_1 + \overline{K}_E + \mu_{I_{ext}}   , \\
\dot{K}_2 &= - K_2 + \overline{K}_S,\\
\dot{K}_3 &= - K_3 + \overline{K}_V - \sigma^2 N|\tau_{vse}|\alpha_S\langle\phi^{\prime(S)}\rangle\kappa_1 + \sigma^2 N\tau_{vsv}\alpha_S\langle\phi^{\prime(S)}\rangle\kappa_3 + \sigma^2 N\tau_{vvs}\alpha_V\langle\phi^{\prime(V)}\rangle\kappa_2.\\
\end{aligned}
\end{equation}
The norm of $\boldsymbol{u}^{(1)}$ is determined by second-order motifs, and $||\boldsymbol{u}^{(1)}|| = \sigma N \sqrt{\alpha_e|\tau^{r}|+\alpha_s|\tau_{vse}|}$ here. The population-averaged pre-activations and the neural variability within each population
\begin{equation}
\begin{aligned}\label{eq:mu_variance_three_population}
\mu^{(E)}&=K_1, \ \ \mu^{(S)}=K_2, \ \ \mu^{(V)}=K_3, \\
\Delta^{(E)}&=\sigma^2N\vert\tau^r\vert\kappa_1^2, \ \ \Delta^{(S)}=\sigma^2 N (|\tau_{vse}|\kappa_1^2 + |\tau_{vsv}|\kappa_3^2), \ \ \Delta^{(V)}=\sigma^2 N |\tau_{vvs}|\kappa_2^2.
\end{aligned}
\end{equation}\label{eq:mu_delta2_pop3}
Next, we examine the population neuronal activities in response to external inputs. Beyond considering the uniform input on pyramidal neurons, the structured input is considered on both Pyr and SOM cells \(\boldsymbol{I}^{ext}-\mu_{I_{ext}}\boldsymbol{m}^{(1)} = \delta_{I_{ext}}\boldsymbol{u}^{(1)}/\Vert\boldsymbol{u}^{(1)}\Vert \).
The pyramidal neuron population is driven by a uniform input while relaying information about structural random input. 
Two interneuron populations, SOM and VIP, exhibit both mean and random couplings, with the latter including second-order motifs. 
Latent quantities $\kappa_1$ and $\kappa_3$ determine the variance of SOM interneurons, and the latent quantity $\kappa_2$ determines the variance of VIP interneurons.

\subsection{\label{supp_subsection:familiarity}Familiarity-dependent population responses}
Three distinct connectivity sources regulate VIP population excitation. 
The average synaptic connectivity of Pyr-to-VIP, which responds to the uniform input, mediates a stimulus-driven regulation; 
chain motifs of Pyr-to-SOM-to-VIP (\(\tau_{vse}<0\)), which respond to a particular random input pattern, mediate a stimulus-suppressed regulation; 
and chain motifs in recurrent connectivity within interneurons (\(\tau_{vsv},\ \tau_{vvs}\)) form a self-excitation regulation, which contributes to the elevated VIP firing rates during inter-stimulus intervals.
Among all connectivity components, only second-order chain motifs (\(\tau_{vse}, \tau_{vvs}, \tau_{vsv}\)) substantially modulate population neural responses to familiar stimulus patterns.
We therefore examine how stimulus familiarity\cite{ito2024coordinated} drives the rewiring of recurrent connectivity through these chain motifs, thereby producing familiarity-dependent neural population responses.

We assume that the Pyr-SOM-VIP network receives environmental input
\begin{equation}
\boldsymbol{I}^{ext} = \mu_{I_{ext}}\boldsymbol{m}^{(1)} + \frac{\delta_{I_{ext}}}{||\boldsymbol{u}^{(1)}||}\left(W\boldsymbol{u}^{(1)}+ \sqrt{1-W^2}\boldsymbol{\xi}\right), 
\end{equation}
where $\mu_{I_{ext}}\boldsymbol{m}^{(1)}$ denotes the uniform input to Pyr neuron population, and $\delta_{I_{ext}}(W\boldsymbol{u}^{(1)}+ \sqrt{1-W^2}\boldsymbol{\xi})/\Vert\boldsymbol{u}^{(1)}\Vert $ denotes the random input with a specific pattern. 
Here we introduce a scalar variable $W$ regarding familiarity. The scalar $W \in [0,1]$, increasing $W$ from $0$ to $1$ corresponds to the network's familiarization of a given input pattern, that is,  $W=0$ corresponds to the completely novel input, and $W=1$ corresponds to the completely familiar input. 
The parameter $W$ is designed following a rule similar to that of the control parameter $\gamma$, which corresponds to the process whereby reservoir random connectivity transforms into structured connectivity for task-adaptation \cite{clark2026structure}. The random vector $\boldsymbol{\xi}\in\mathbb{R}^{N_E+N_S}$ is independent of any connectivity component $\boldsymbol{\eta}$. The 2-norm of $\boldsymbol{\xi}$ is kept the same as $\boldsymbol{u}^{(1)}$, which gives the variance $\Sigma^2$ of the elements in $\boldsymbol{\xi}$, 
\begin{equation}
\Sigma^2 = \sigma^2 N \frac{\alpha_e|\tau^r|+\alpha_s|\tau_{vse}|}{\alpha_e+\alpha_s}.
\end{equation}
Under this scenario, we have a new base vector $\boldsymbol{I}^{ext}_{\perp} = \boldsymbol{\xi}$, which is orthogonal to other base vectors $\{\boldsymbol{u}^{(r)}, \boldsymbol{m}^{(r)}\}_{r-1,2,3}$. Therefore, we have the dynamical equations for the latent quantities as 
\begin{equation}
\begin{aligned}
\dot{\kappa}_1 &= -\kappa_1 + C_f\sigma^2 N \tau^r \alpha_E\langle\phi^{\prime(E)}\rangle \kappa_1 + \delta_{I_{ext}}W/\Vert\boldsymbol{u}^{(1)}\Vert ,  \\
\dot{\kappa}_2 &= -\kappa_2 + \alpha_S\langle\phi^{(S)}\rangle, \\
\dot{\kappa}_3 &= -\kappa_3 + \alpha_V\langle\phi^{(V)}\rangle, \\
\dot{K}_1 &= - K_1 + \overline{K}_E + \mu_{I_{ext}}   , \\
\dot{K}_2 &= - K_2 + \overline{K}_S,\\
\dot{K}_3 &= - K_3 + \overline{K}_V - \sigma^2 N|\tau_{vse}|\alpha_S\langle\phi^{\prime(S)}\rangle\kappa_1 + \sigma^2 N\tau_{vsv}\alpha_S\langle\phi^{\prime(S)}\rangle\kappa_3 + \sigma^2 N\tau_{vvs}\alpha_V\langle\phi^{\prime(V)}\rangle\kappa_2,\\
\dot{\nu} & = -\nu + \delta_{I_{ext}}\sqrt{1-W^2}/\Vert\boldsymbol{u}^{(1)}\Vert,
\end{aligned}
\end{equation}
where the additional \(\nu\) corresponds to the new base vector \(\boldsymbol{I}_{\perp}^{ext}\).
The population-averaged pre-activations and the neural variability within each population are
\begin{equation}
\begin{aligned}\label{eq:mu_variance_three_population_familiarity}
\mu^{(E)}&=K_1, \ \ \mu^{(S)}=K_2, \ \ \mu^{(V)}=K_3, \\
\Delta^{(E)}&=\sigma^2N\vert\tau^r\vert\kappa_1^2 + \Sigma^2\nu^2, \\ 
\Delta^{(S)}&=\sigma^2 N \left(|\tau_{vse}|\kappa_1^2 + |\tau_{vsv}|\kappa_3^2\right) + \Sigma^2\nu^2, \\ 
\Delta^{(V)}&=\sigma^2 N |\tau_{vvs}|\kappa_2^2.
\end{aligned}
\end{equation}
\begin{comment}
Then, for the stimulus-driven pathway, we assume
\begin{equation}
J_{VE}^0 = (1-W)J^{0, base}_{VE}, 
\end{equation}
and 
\end{comment}
for the self-excitation within interneurons, we assume
\begin{equation}
\tau_{vsv} = \frac{1+W}{2}\tau_{vsv}^{(base)}, \ \ \ \tau_{vvs} = \frac{1+W}{2}\tau_{vvs}^{(base)}, 
\end{equation}
for the $\tau_{vse}$, which contributes to the familiar stimulus suppression, obeys
\begin{equation}
\tau_{vse} = W \tau_{vse}^{(base)},
\end{equation}
and for the reciprocal motifs among Pyr neurons
\begin{equation}
\tau^r = W\tau^{r(base)}.
\end{equation}
The specific parameter values: $\tau_{vsv}^{(base)}=0.38, \tau_{vvs}^{(base)}=0.38, \tau_{vse}^{(base)}=-0.1$ and  $\tau^{r(base)}=0.001$. %$J^{0(base)}_{VE}=0.0018$,
Results are shown in Fig.~\ref{fig:experience} D-F.

\clearpage
\newpage
\section{Supplementary Figures}

\begin{figure}[ht]
\centering
\includegraphics[width=0.91\linewidth]{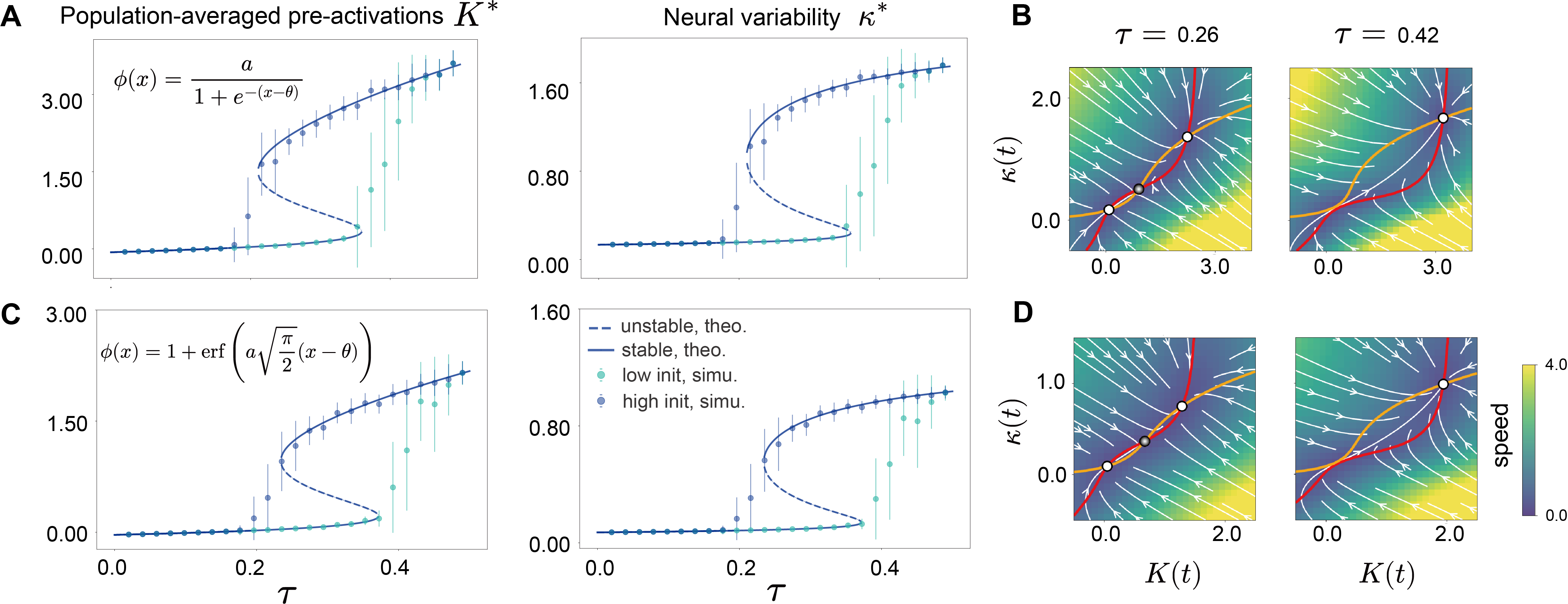}
\caption{{\bf Saddle-node bifurcation induced by gradually increasing the strength of homogeneous chain motifs under different positive saturating nonlinearity.} 
(A, C) Steady-state of latent quantities $K$ and $\kappa$ as homogeneous chain motif strength $\tau$ increases. (B, D) Change in velocity field $\kappa(t)-K(t)$ as $\tau$ increases. 
(A, B) under nonlinear transfer function $\phi(x) = a \cdot\text{sigmoid}(x-\theta) = \frac{a}{1+e^{-(x-\theta)}}$, with $a=3.3$, $\theta=3.1$. (C, D) under nonlinear transfer function $\phi(x) = 1+\text{erf}(a\cdot\sqrt{\frac{\pi}{2}}(x-\theta))$, with $a=0.5$, $\theta=2.0$. 
Other figure descriptions follow the same conventions as in Fig.~\ref{fig:homo_fig}.
Parameters: $N_E=2000$, $N_I=500$, $J^0=0.001$, $\sigma=0.1$, $g=5.0$. }
\label{supp_fig:homo_general_nonlinearity}
\end{figure}
\begin{figure}[ht]
\centering
\includegraphics[width=0.91\linewidth]{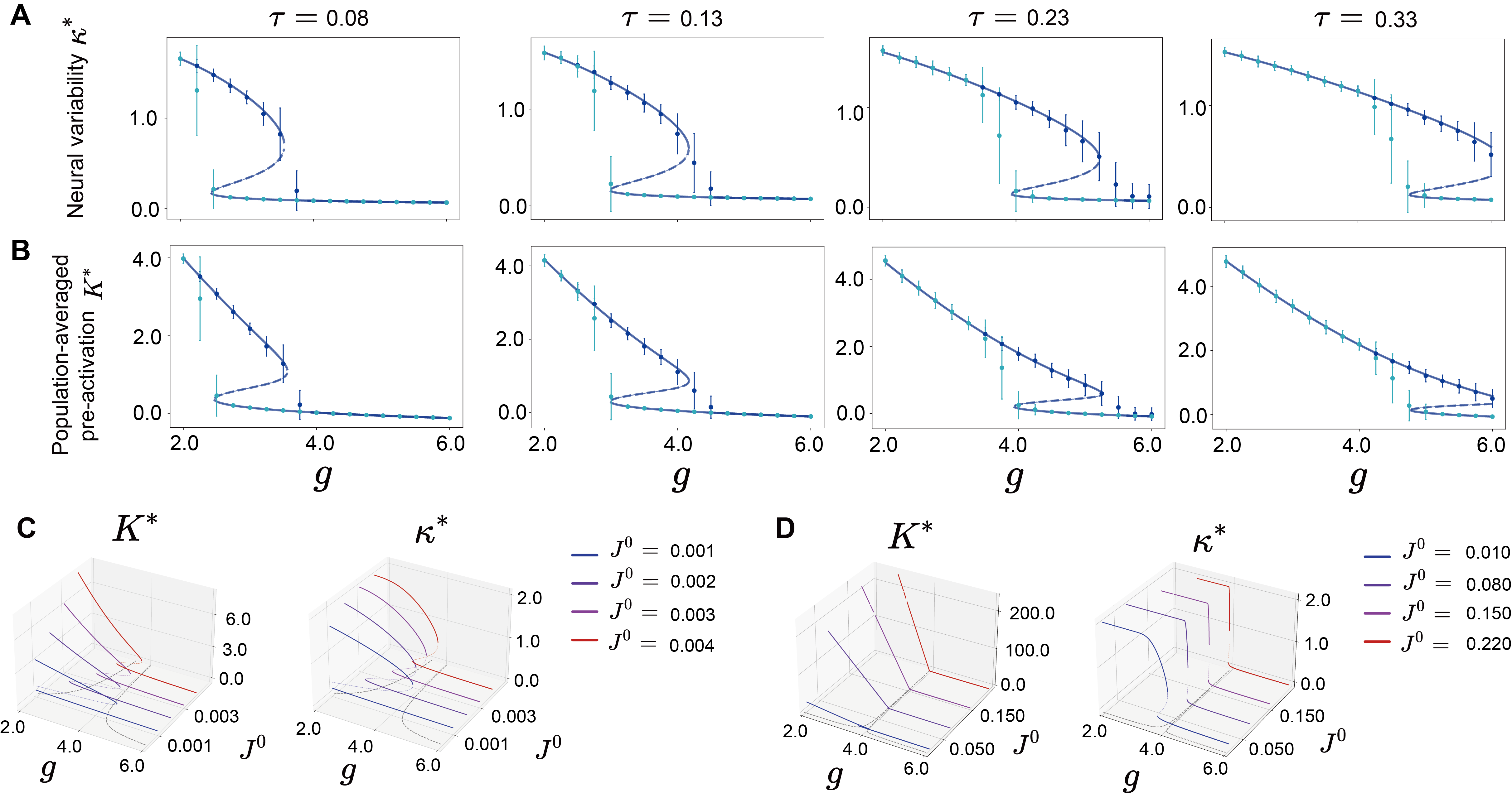}
\caption{{\bf Change in the steady-state values of latent quantities as a function of global inhibition with increasing strength of homogeneous chain motifs.} 
Steady state of the latent quantity $\kappa$ in (A) and $K$ in (B). 
(C) Bifurcation of $\kappa$ and $K$ as a function of inhibition parameter $g$, examined within a range of relatively small $J_0$ values. 
(D) Same as (C), but within a range of relatively large $J_0$ values. 
Other figure descriptions follow the same conventions as in Figs.~\ref{fig:homo_fig},~\ref{fig:hetero_auto_EI_fig}.
Parameters: $N_E=2000$, $N_I=500$, $\sigma=0.1$, $J^0=0.002$ in (A) and (B), $\theta=1.6$.}
\label{supp_fig:homo_gj0_saddle}
\end{figure}

\begin{figure}[ht]
\centering
\includegraphics[width=0.95\linewidth]{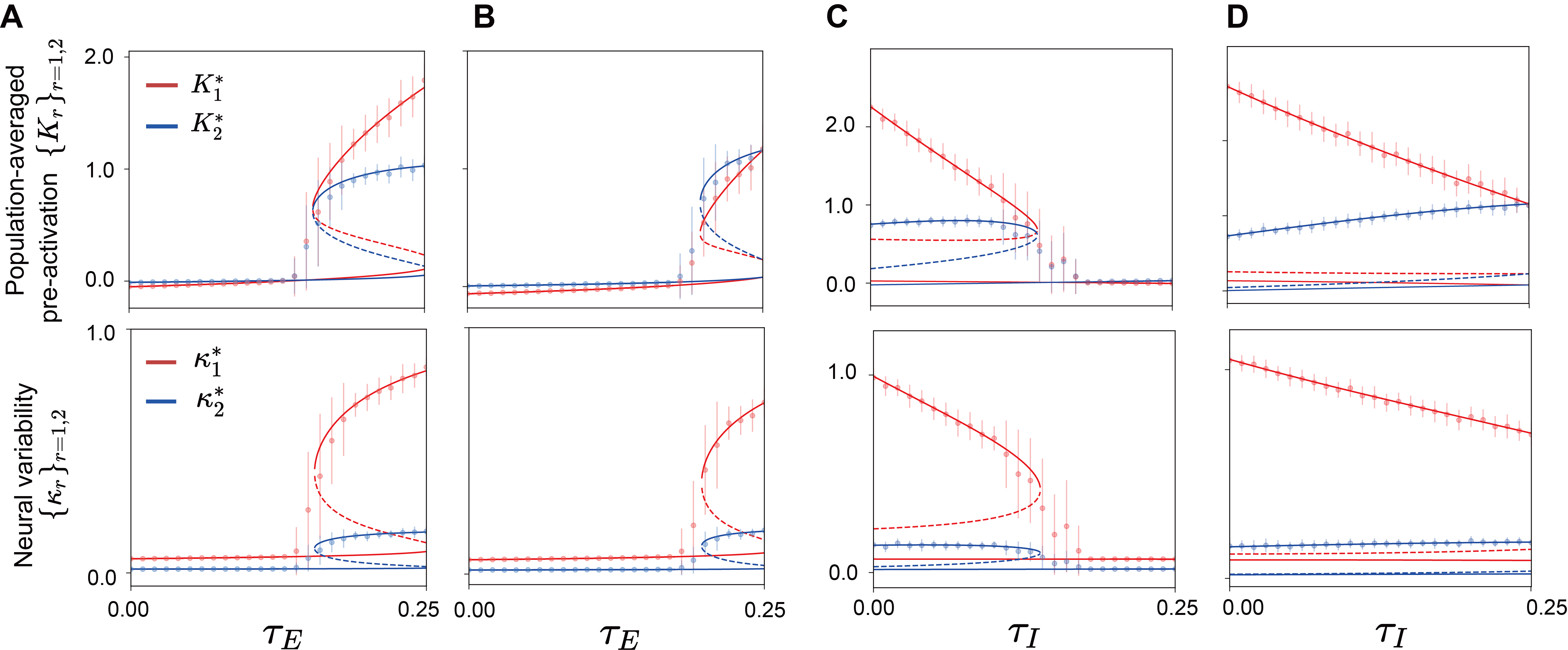}
\caption{{\bf Impact of end-neuron-specific chain motifs in EI networks with fixed mean connectivity parameters.} 
(A, B) Bistable transition of latent quantities $\{K_r\}$ and $\{\kappa_r\}$ (r=1, 2) induced by increased strength of $\tau_E$. 
(C, D) Transition from bistability to quiescence of latent quantities $\{K_r\}$ and $\{\kappa_r\}$ (r=1, 2) induced by increased strength $\tau_I$. 
Other figure descriptions follow the same conventions as in Fig.~\ref{fig:hetero_auto_EI_fig}.
Parameters: $N_E=2000$, $N_I=500$, $\sigma=0.1$, $J^0=0.001$, $g=5.0$, $\theta=1.6$. $\tau_I=0.15$ in (A), $\tau_I=0.25$ in (B), and $\tau_E=0.15$ in (C), $\tau_E=0.25$ in (D).}
\label{suppfig:supp_fig_hetero_end_type}
\end{figure}

\begin{figure}[ht]
\centering
\includegraphics[width=0.85\linewidth]{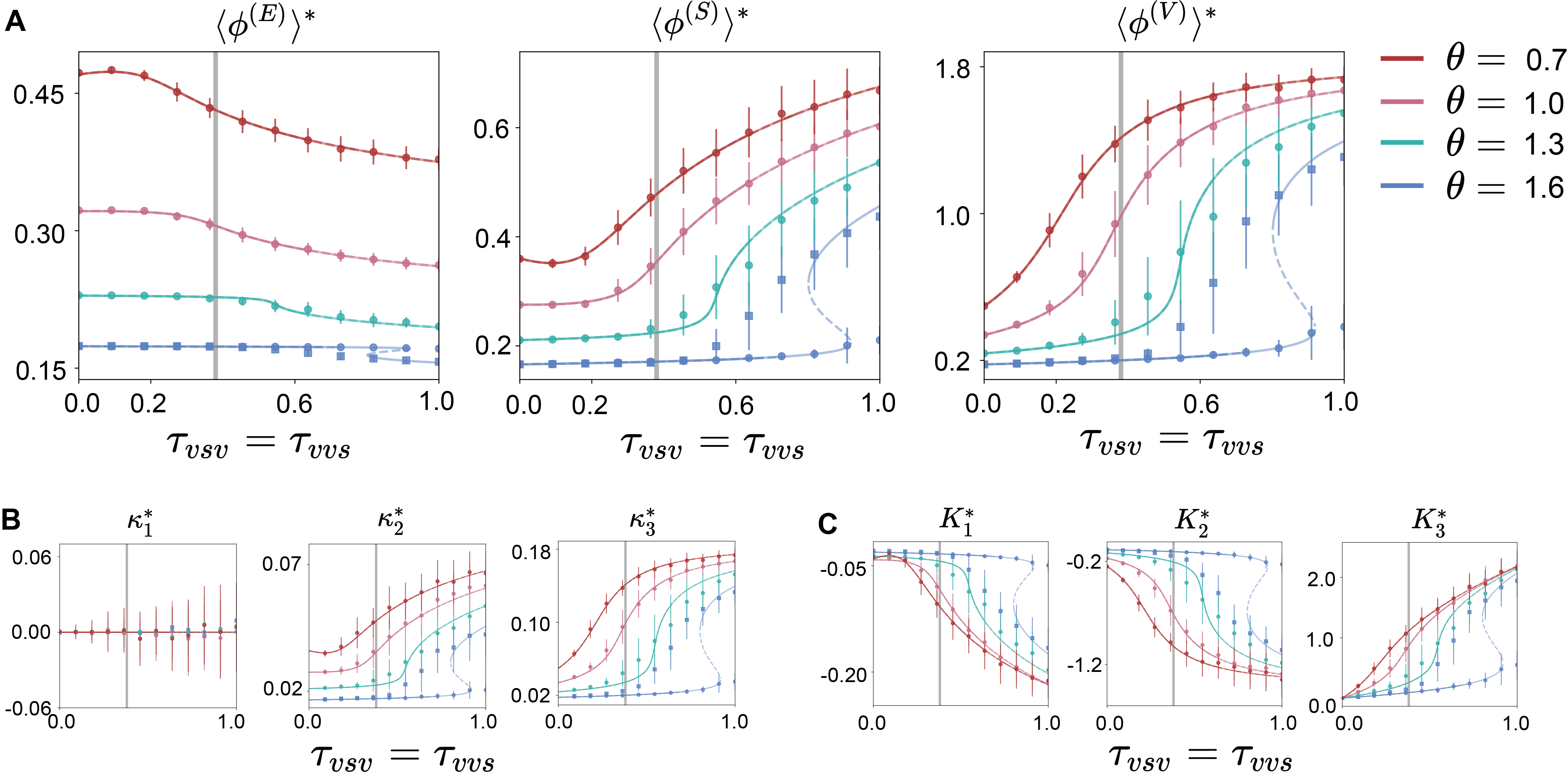}
\caption{{\bf Bifurcation of the steady states of population-averaged firing rates and latent quantities with simultaneous increases in $\tau_{vsv} = \tau_{vvs}$ analyzed under varying nonlinearity thresholds $\theta$.} 
Vertical gray lines: Values of $\tau_{vsv}$ and $\tau_{vvs}$ used in Fig. \ref{fig:experience}C. 
Other figure descriptions follow the same conventions as in Fig.~\ref{fig:experience}G.
Parameters: $\sigma_{vs}=\sigma_{se}=\sigma_{sv}=\sigma_{vv}=0.4$, $\tau_{vse}=-0.10$. 
}
\label{suppfig:thres_cusp_latent_phi}
\end{figure}

\begin{figure}[ht]
\centering
\includegraphics[width=0.9\linewidth]{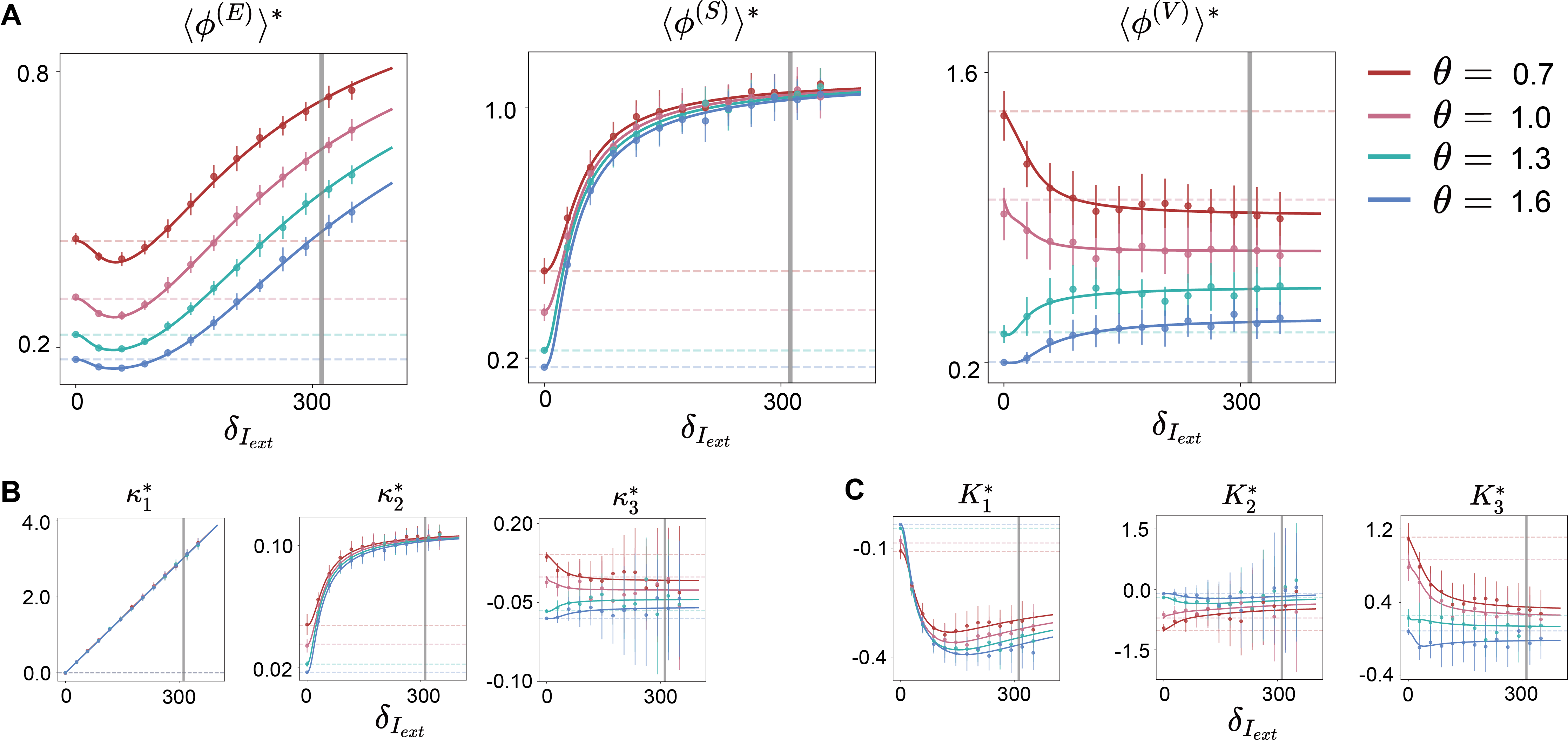}
\caption{{\bf Change in steady states of population-averaged firing rates and latent quantities as a function of increasing patterned input strength $\delta_{I_{ext}}$ of the patterned input, under varying nonlinearity thresholds $\theta$.} 
Vertical gray lines: value of random input strength $\delta_{I_{ext}}$ used in Fig. \ref{fig:experience}C.
Other figure descriptions follow the same conventions as in Fig.~\ref{fig:experience}H.
Parameters: $\sigma_{vs}=\sigma_{se}=\sigma_{sv}=\sigma_{vv}=0.4$, $\tau_{vse}=-0.1$, $\tau_{vsv}=\tau_{vvs}=0.38$, $\theta=0.7$.
}
\label{suppfig:response_kappa1_thres_latent_phi}
\end{figure}

\begin{figure}[ht]
\centering
\includegraphics[width=0.9\linewidth]{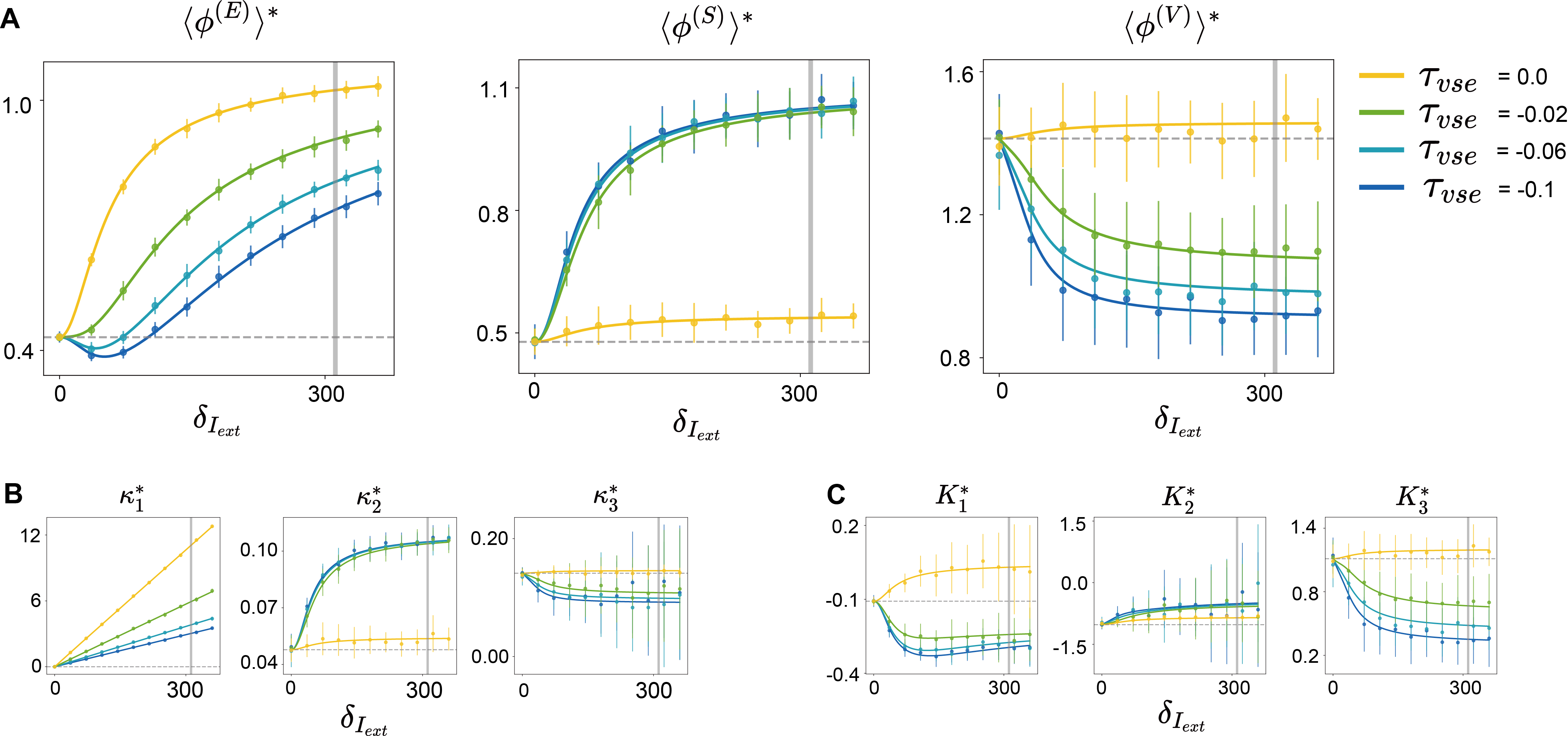}
\caption{{\bf Change in steady states of population-averaged firing rates and latent quantities as a function of increasing patterned input strength $\delta_{I_{ext}}$ of the patterned input, under varying values of $\tau_{vse}$.} 
Vertical gray lines: value of random input strength $\delta_{I_{ext}}$ used in Fig. \ref{fig:experience}C. 
Other figure descriptions follow the same conventions as in Fig.~\ref{fig:experience}I.
Parameters: $\sigma_{vs}=\sigma_{se}=\sigma_{sv}=\sigma_{vv}=0.4$, $\tau_{vse}=-0.1$, $\tau_{vsv}=\tau_{vvs}=0.38$, $\theta=0.7$.
}
\label{suppfig:response_kappa1_tauvse_latent_phi}
\end{figure}

\end{document}